\documentclass[aps,prf,superscriptaddress,longbibliography]{revtex4-1}

\usepackage{amsmath,rotating}
\usepackage{dashrule}

\usepackage{psfrag}
\usepackage{latexsym}
\usepackage{mathrsfs}
\usepackage{graphicx}
\usepackage{pifont}
\usepackage{amsmath}
\usepackage{subfigure}
\usepackage{mathtools}
\usepackage{mathcomp}
\usepackage{mathdots}
\usepackage{rotating}
\usepackage{array}
\usepackage{color}
\usepackage{algorithm}
\usepackage{algorithmic}
\usepackage[normalem]{ulem}
\usepackage{bm}
\usepackage{amssymb}
\usepackage[dvipsnames]{xcolor}

\newcommand{\bK}{\mathbf{K}}
\newcommand{\bP}{\mathbf{P}}

\newtheorem{remark}{Remark}

\newcommand{\enma}[1]   {\ensuremath{#1}}

\newcommand{\beq}{\begin{equation}}
\newcommand{\eeq}{\end{equation}}
\newcommand{\bseq}{\begin{subequations}}
\newcommand{\eseq}{\end{subequations}}
\newcommand{\beqn}{\begin{eqnarray}}
\newcommand{\eeqn}{\end{eqnarray}}
\newcommand{\ba}{\begin{array}}
\newcommand{\ea}{\end{array}}
\newcommand{\bct}{\begin{center}}
\newcommand{\ect}{\end{center}}
\newcommand{\btmz}{\begin{itemize}}
\newcommand{\etmz}{\end{itemize}}
\newcommand{\benum}{\begin{enumerate}}
\newcommand{\eenum}{\end{enumerate}}

\newcommand{\diag}      {\enma{\mathrm{diag}}}

\newcommand{\trace}     {\enma{\mathrm{trace}}}

\newcommand{\inner}[2]{\left\langle #1,#2 \right\rangle}

\newcommand{\bv}{{\bf v}}

\newcommand{\be}{\begin{equation}}
\newcommand{\ee}{\end{equation}}

\newcommand{\cplxs}{ C\kern -.35em \rule{0.03 em}{.7 ex}~   }

\def\complex{\hbox{C\kern -.45em \rule{0.03 em}{1.5 ex}}~}

\newcommand{\bi}{\begin{itemize}}
\newcommand{\ei}{\end{itemize}}

\newcommand{\bu}{{\bf u}}

\newcommand{\btab}{\begin{tabular}}
\newcommand{\etab}{\end{tabular}}
\newcommand{\bd}{{\bf d}}

\newcommand{\bpsi}{\mbox{\boldmath$\psi$}}

\newcommand{\non}{\nonumber}
\newcommand{\mrd}{\mathrm{d}}
\newcommand{\mre}{\mathrm{e}}
\newcommand{\mri}{\mathrm{i}}

\newcommand{\bA}{\mathbf{A}}
\newcommand{\bB}{\mathbf{B}}
\newcommand{\bC}{\mathbf{C}}

\newcommand{\bI}{\mathbf{I}}

\newcommand{\bF}{\mathbf{F}}
\newcommand{\bG}{\mathbf{G}}

\newcommand{\bQ}{\mathbf{Q}}

\newcommand{\bk}{\mathbf{k}}
\newcommand{\DefinedAs}[0]{\mathrel{\mathop:}=}

\definecolor{bgblue}{rgb}{0.04,0.19,0.53}
\definecolor{dblue1}{rgb}{0,0.3,0.7}
\definecolor{dred}{rgb}{0.4,0.2,0}

\begin{document}

\title{\LARGE \bf
Stochastic receptivity analysis of boundary layer flow
}

\author{Wei Ran}
\email[E-mail:]{wran@usc.edu}
\affiliation{Department of Aerospace and Mechanical Engineering, University of Southern California, Los Angeles, CA, 90089}
\author{Armin Zare}
\email[E-mail:]{armin.zare@usc.edu}
\affiliation{Ming Hsieh Department of Electrical and Computer Engineering, University of Southern California, Los Angeles, CA, 90089}
\author{M.\ J.\ Philipp Hack}
\email[E-mail:]{mjph@stanford.edu}
\affiliation{Center for Turbulence Research, Stanford University, Stanford, CA, 94305}
\author{Mihailo R.\ Jovanovi\'c}
\email[E-mail:]{mihailo@usc.edu}
\affiliation{Ming Hsieh Department of Electrical and Computer Engineering, University of Southern California, Los Angeles, CA, 90089}

    \begin{abstract}
        We utilize the externally forced linearized Navier-Stokes equations to study the receptivity of pre-transitional boundary layers to persistent {sources of stochastic excitation}. Stochastic forcing is used to model the effect of free-stream turbulence that enters at various wall-normal locations and the fluctuation dynamics are studied via linearized models that arise from locally parallel and global perspectives. In contrast to the widely used resolvent analysis that quantifies {the amplification of deterministic disturbances at a given temporal frequency,} our approach examines the steady-state response to stochastic excitation {that is uncorrelated in time}. In addition to stochastic forcing with identity covariance, we utilize the spatial spectrum of homogeneous isotropic turbulence to model the effect of free-stream turbulence. Even though {locally parallel analysis does not} account for the effect of the spatially evolving base flow, we demonstrate that {it captures the essential mechanisms and the} prevailing length-scales in stochastically forced boundary layer flows. On the other hand, global analysis, which accounts for the spatially evolving nature of the boundary layer flow, predicts the amplification of a cascade of streamwise scales throughout the streamwise domain. We show that the flow structures that can be extracted from a modal decomposition of the resulting velocity covariance matrix, can be closely captured by {conducting locally parallel analysis} at various streamwise locations and over {different wall-parallel wavenumber pairs. Our approach does not rely on costly stochastic simulations and it provides insight into mechanisms for perturbation growth including the interaction of the slowly varying base flow with streaks and Tollmien-Schlichting waves.}
    \end{abstract}

\maketitle

\section{Introduction}
\label{sec.intro}

Laminar-turbulent transition of fluid flows is important in many engineering applications. Predicting the point of transition requires an accurate understanding of the mechanisms that govern the physics of transitional flows. Since the 1990's, numerical simulations with various levels of fidelity have been used to uncover many essential features of the transition phenomenon. In spite of this progress, the complicated sequence of events that leads to transition and the inherent complexity of the Navier-Stokes (NS) equations have hindered the development of practical control strategies for delaying transition in boundary layer flows~\cite{morresher94,sarreeker02,kimbew07}.

It is generally accepted that the transition process can be divided into three stages; receptivity, instability growth, and breakdown~\cite{morresher94}. In the laminar boundary layer flow, disturbances that lead to transition are amplified either through modal, {i.e., exponential,} instability mechanisms or non-modal amplification, e.g., via transient growth mechanisms such as lift-up~\cite{lan75,lan80} and Orr mechanisms~\cite{orr1907,butfar92,hacmoi17}. An important aspect in both scenarios is the receptivity of the boundary layer flow to external excitation sources, e.g., free-stream turbulence and surface roughness. Such sources of excitation perturb the velocity field and give rise to initial disturbances within the shear that can grow to critical levels. Depending on the amplitude and frequency of excitation, initial disturbances can take different routes to transition. For example, low-amplitude excitation of the boundary layer flow {can cause} the growth of two-dimensional Tollmien-Schlichting (TS) waves, which {can trigger natural transition to turbulence}~\cite{kletidsar62,kaclev84,mac84,her88,sayhammoi13}. {On the other hand, sufficiently high levels of broad-band excitation can induce the growth of streamwise elongated streaks that play an important role in bypass transition}~\cite{sarreeker02}. The effect of free-stream turbulence on the growth of boundary layer streaks has been the subject of various experimental~\cite{matalf01,framatalf05,ricwalbrimce16}, numerical~\cite{jocdur01,braschhen04}, and theoretical~\cite{gol14,haczak14} studies. In particular, it has been shown that free-stream disturbances that penetrate into the boundary layer are elongated in the streamwise direction~\cite{jacdur98}. While nonlinear dynamical models that are based on the NS equations provide insight into receptivity mechanisms, their implementation typically involves a large number of degrees of freedom and it ultimately requires direct simulations. This motivates the development of low-complexity models that are better suited for comprehensive quantitative studies.

In recent years, increasingly accurate descriptions of coherent structures in wall-bounded shear flows, e.g.~\cite{smimckmar11,hacmoi18}, have inspired the development of reduced-order models. Such models are computationally tractable and can be trained to replicate statistical features that are estimated from experimentally or numerically generated data measurements. However, their data-driven nature is accompanied by a lack of robustness. Specifically, control actuation and sensing may significantly alter the identified modes which introduces nontrivial challenges for model-based control design~\cite{noamortad11}. In contrast, models that are based on the linearized NS equations are less prone to such uncertainties and are, at the same time, well-suited for analysis and synthesis using tools of modern robust control. {While the nonlinear terms in the NS equations play an important role in transition to turbulence and in sustaining the turbulent state, they are conservative and, as such, they do not contribute to the transfer of energy between the mean flow and velocity fluctuations but only transfer energy between different Fourier modes~\cite{mcc91,durrei11}. This feature has inspired modeling the effect of nonlinearity using additive stochastic forcing with early efforts focused on homogeneous isotropic turbulence (HIT)~\cite{ors70,kra71,monyag75}.
In the presence of stochastic excitation, the linearized NS equations have been used to model heat and momentum fluxes and spatio-temporal spectra in quasi-geostrophic turbulence~\cite{farioa93c,farioa94a,delfar95}. Moreover, they have been used to characterize the most detrimental stochastic forcing and determine scaling laws for energy amplification at subcritical Reynolds numbers~\cite{farioa93,bamdah01,jovbamJFM05}, and to replicate structural~\cite{hwacosJFM10a,hwacosJFM10b} and statistical~\cite{moajovJFM12,zarjovgeoJFM17} features of wall-bounded turbulent flows. In these studies, stochastic forcing has been commonly used to model the impact of exogenous excitation sources and initial conditions, or to capture the effect of nonlinearity in the NS equations.}

The linearized NS equations have been widely used for modal and non-modal stability analysis of both parallel and non-parallel flows~\cite{huemon90,schhen01,sch07}. In parallel flows, homogeneity in the streamwise and spanwise dimensions allows for the decoupling of the governing equations across streamwise and spanwise wavenumbers via Fourier transform, which results in significant computational advantages for analysis, optimization, and control. On the other hand, in the flat-plate boundary layer, streamwise and wall-normal inhomogeneity require discretization over two spatial directions and lead to models of significantly larger sizes. Conducting modal and non-modal analyses is thus more challenging than for locally parallel flows. However, due to the slowly varying nature of the boundary layer flow, parallel flow assumptions can still provide meaningful results. For example, primary disturbances can be {identified using the eigenvalue analysis} of the Orr-Sommerfeld and Squire equations~\cite{schhen01} and the secondary instabilities can be obtained via Floquet analysis~\cite{her84,her88}. Moreover, the NS equations can be parabolized to account for the downstream propagating nature of waves in slowly varying flows via spatial marching. This technique has enabled the analysis of transitional boundary layers and turbulent jet flows using various forms of the unsteady boundary-region equations~\cite{leiwungol99a,ricluowu11}, parabolized stability equations~\cite{her97,lozhacmoi18}, and the more recent one-way Euler equations~\cite{towcol15}. Furthermore, drawing on Floquet theory, the linear parabolized stability equations have also been extended to study interactions between different modes in slowly growing boundary layer flow~\cite{ranzarhacjovPRF19}.

While the parallel flow assumption offers significant computational advantages, it does not account for the effect of the spatially evolving base flow on the stability of the boundary layer. Global stability analysis addresses this issue by accounting for the spatially varying nature of the base flow and discretizing all inhomogeneous spatial directions. Previously, tools from sparse linear algebra in conjunction with iterative schemes have been employed to analyze the eigenspectrum of the governing equations and provide insight into the dynamics of transitional flows~\cite{ehrgal05,alirob07,niclel11,pargosthekim16,schtowcolcavjorbre17}. Efforts have also been made to conduct non-modal analysis of spatially evolving flows including transient growth~\cite{barblashe08,monakebrahen10} and resolvent{~\cite{brasippramar11,jeunicjovPOF16,schtowrigcolbre18,dwisidniccanjovJFM18} analyses. In particular, for the flat-plate boundary layer flow, the sensitivity of singular values of the resolvent operator to base-flow modifications and subsequent effects on the TS instability mechanism and streak amplification was investigated in~\cite{brasippramar11}.} However, previous studies did not incorporate information regarding the spatio-temporal spectrum and spatial localization of excitation sources. {The} widely used resolvent analysis~\cite{tretrereddri93,mj-phd04,mcksha10} is limited to monochromatic forcing, and as such, {may not fully capture} naturally occurring sources of excitation. {Furthermore, the evolution of exact optimal perturbations that are identified using resolvent analyses is seldom encountered in practical configurations~\cite{frabratalcos04}.}
		
The approach advanced in the present work {enables the study of} receptivity mechanisms in boundary layer flows subject to stochastic sources of excitation. We model the effect of free-stream turbulence as a persistent white-in-time stochastic forcing that enters at various wall-normal locations and {analyze the dynamics of velocity fluctuations around locally parallel and spatially evolving base flows using the solution to the algebraic Lyapunov equation}. Our simulation-free approach enables computationally efficient assessment of the energy spectrum of spatially evolving flows, without relying on a particular form of the inflow conditions or computation of the full spectrum of the linearized dynamical generator. {Moreover, the broad-band nature of our forcing model captures the aggregate effect of all time-scales without the need to integrate the frequency response over all energetically relevant frequencies.}

We compare and contrast results obtained under locally parallel flow assumption with those of global analysis. Coherent structures that emerge as the response to free-stream turbulence are extracted using the modal decomposition of the steady-state velocity covariance matrix. We demonstrate how parallel and global flow analyses can be used to quantify the amplification of streamwise elongated streaks and Tollmien-Schlichting (TS) waves, which are important in the laminar-turbulent transition of boundary layer flows. Our analysis shows that subordinate eigenmodes of the steady-state velocity covariance matrices that result from global flow analyses have nearly equal energetic contributions to that of the principal modes. This {observation} demonstrates that global covariance matrices cannot be well-approximated by low-rank representations. On the other hand, we show how locally parallel analysis, which breaks up the receptivity process of the boundary layer flow over various streamwise length-scales, can uncover certain flow structures that are difficult to observe in global analysis. We also demonstrate that modeling the effect of free-stream turbulence using the spectrum of HIT yields similar results as the analysis based on white-in-time stochastic excitation with identity covariance matrix. For the considered range of moderate Reynolds numbers, our results support the assumption of parallel flow in the low-complexity modeling and analysis of boundary layer flows.

The remainder of this paper is organized as follows. In Section~\ref{sec.formulation}, we introduce the stochastically forced linearized NS equations and describe the algebraic Lyapunov equation that we use to compute second-order statistics of velocity fluctuations, extract information about the energy amplification, and identify energetically dominant flow structures. In Section~\ref{sec.parallel-BL}, we study the receptivity to stochastic excitations of the velocity fluctuations around a locally parallel Blasius boundary layer profile. In Section~\ref{sec.global-BL}, we extend the receptivity analysis to stochastically forced non-parallel flows. We also discuss the effect of exponentially attenuated HIT on the amplification of streaks and TS waves. {In Section~\ref{sec.discussion}, we compare the results of locally parallel and global analyses and examine the spatio-temporal frequency response of the linearized dynamics.} We provide concluding remarks in Section~\ref{sec.conclusion}.

    \vspace*{-2ex}
\section{Stochastically forced linearized NS equations}
\label{sec.formulation}

\begin{figure}
\vspace{.3cm}
\begin{center}
         \includegraphics[width=8.5cm]{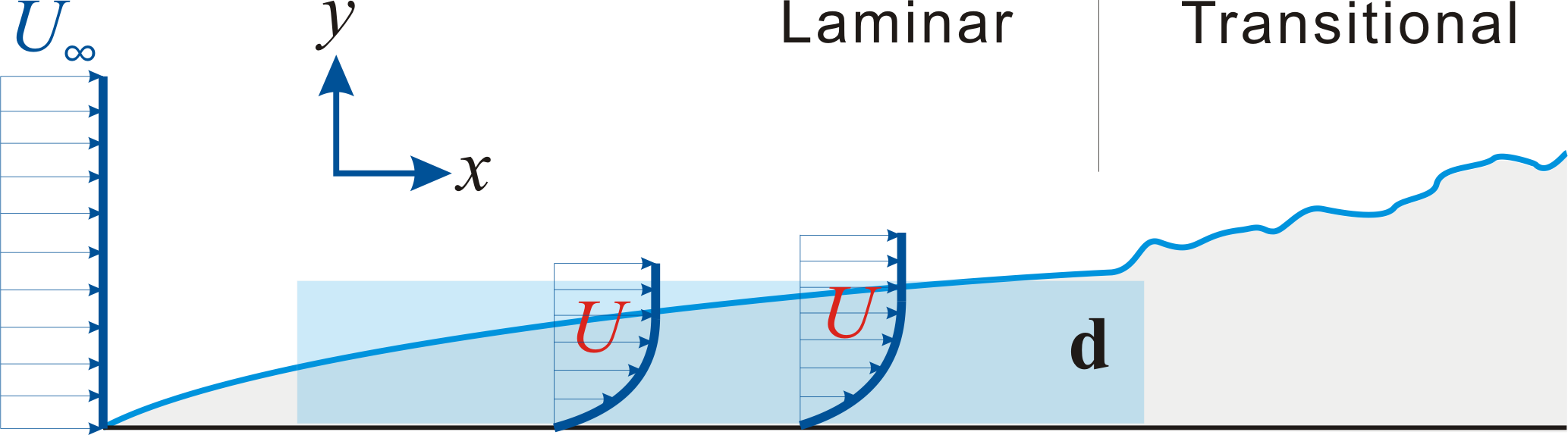}
\end{center}
\caption{Geometry of transitional boundary layer flow with stochastic excitation $\bd$ entering in the blue shaded region.}
\label{fig.forcedBL}
\end{figure}

In a flat-plate boundary layer, the linearized incompressible NS equations around the Blasius base flow profile $\bar{\bu} = [\,U(x,y)\,~V(x,y)\,~0\,]^T$ are given by
\begin{align}
	\label{eq.turblin}
	\ba{rcl}
    		\bv_t
    		&=&
    		-
    		\left( \nabla \cdot \bar{\bu} \right) \bv
     		\; - \;
    		\left( \nabla \cdot \bv \right) \bar{\bu}
    		\; - \;
    		\nabla p
    		\; + \;
    		\dfrac{1}{Re_0} \, \Delta \bv
    		\; + \;
    		{\bd},
    		\\[.15cm]
    		0
    		&=&
    		\nabla \cdot \bv,
	\ea
\end{align}
where $\bv = [\,u~v~w\,]^T$ is the vector of velocity fluctuations, $p$ denotes pressure fluctuations, $u$, $v$, and $w$ represent components of the fluctuating velocity field in the streamwise ($x$), wall-normal ($y$), and spanwise ($z$) directions, and $\bd$ denotes an additive zero-mean stochastic body forcing. The stochastic perturbation $\bd$ is used to model the effect of exogenous sources of excitation on the boundary layer flow and, as illustrated in Fig.~\ref{fig.forcedBL}, it can be introduced in various wall-normal regions. In Eqs.~\eqref{eq.turblin}, $Re_0 = U_\infty \delta_0/\nu$ is the Reynolds number based on the Blasius length-scale $\delta_0=\sqrt{\nu\, x_0/U_{\infty}}$, where the initial streamwise location $x_0$ denotes the distance from the leading edge, $U_\infty$ is the free-stream velocity, and $\nu$ is the kinematic viscosity. The local Reynolds number at distance $x$ to the starting position $x_0$ is thus given by
$
	Re
	=
	Re_0
	\sqrt{1+ x/x_0}.
$
The velocities are non-dimensionalized by $U_\infty$, time by $\delta_0/U_\infty$, and pressure by $\rho U_\infty^2$, where $\rho$ is the fluid density.

	\vspace*{-3ex}
\subsection{Evolution model}

Elimination of the pressure yields an evolution form of the linearized equations with the state variable $\bm{\varphi} = [\,v\,~ \eta\,]^T$, which contains the wall-normal velocity $v$ and vorticity $\eta = \partial_z u - \partial_x w$~\cite{schhen01}. In addition, homogeneity of the Blasius base flow in the spanwise direction allows a normal-mode representation with respect to $z$, yielding \mbox{the evolution model}
\begin{align}
	\label{eq.lnse}
	\ba{rcl}
		\partial_t
		\bm{\varphi}(x,y,k_z,t)
		& = &
		[\bA(k_z)\, \bm{\varphi} (\cdot,k_z,t)](x,y)
		\;+\;
		[\bB(k_z)\, \bd(\cdot,k_z,t)](x,y),
		\\[.15cm]
		\bv(x,y,k_z,t)
		&=&
		[\bC(k_z)\, \bm{\varphi}(\cdot,k_z,t)](x,y),
	\ea
\end{align}
which is parameterized by the spanwise wavenumber $k_z$. Definitions of the operators $\bA$, $\bB$, and $\bC$ are provided in Appendix~\ref{sec.appendix-operators}. We note that an additional wall-parallel base flow assumption that entails $\bar{\bu} = [\,U(y)\,~0\,~0\,]^T$ renders the coefficients in Eqs.~\eqref{eq.turblin} independent of $x$ and thus enables a normal-mode representation in that dimension as well.

We obtain finite-dimensional approximations of the operators in Eqs.~\eqref{eq.lnse} using a pseudospectral discretization scheme~\cite{weired00} in the spatially inhomogeneous directions. For streamwise-varying base flows we consider $N_x$ and $N_y$ Chebyshev collocation points in $x$ and $y$, and for streamwise invariant base flows we use $N_y$ points in $y$. Furthermore, a change of variables is employed to obtain a state-space representation in which the kinetic energy is determined by the Euclidean norm of the state vector; see Appendix~\ref{sec.appendix-coc}. We thus arrive at the state-space model
\be
	\label{eq.lnse1}
	\ba{rcl}
		\dot{\bpsi}(t)
		& = &
		A\, \bpsi(t)
		\;+\;
		B\, \bd(t),
		\\[.15cm]
		\bv(t)
		&=&
		C\, \bpsi(t),
	\ea
\ee
where $\bpsi(t)$ and $\bv(t)$ are vectors with $2 N_x N_y$ and $3 N_x N_y$ complex-valued components, respectively ($2 N_y$ and $3 N_y$ components, respectively, for parallel flows), and state-space matrices $A$, $B$, and $C$ incorporate the aforementioned change of variables and wavenumber parameterization over $k_z$ (over $(k_x,k_z)$ for parallel flows).

	\vspace*{-2ex}
\subsection{Second-order statistics of velocity fluctuations}
\label{sec.stats}

We next characterize the structural dependence between the second-order statistics of the state and forcing term in the linearized dynamics. We also describe how the energy amplification arising from persistent stochastic excitation and the energetically dominant flow structures can be computed from these flow statistics. All mathematical statements in the remainder of this section are parameterized over homogeneous directions.

In statistical steady-state, the covariance matrices
$
	\Phi
	=
	\lim_{t \, \to \, \infty} \left< \bv(t)\, \bv^*(t)\right>
$
of the velocity fluctuation vector and
$
	X
	=
	\lim_{t \, \to \, \infty} \left< \bpsi(t)\, \bpsi^*(t)\right>
$
of the state vector in Eq.~\eqref{eq.lnse1} are related by
\be
	\label{eq.outputcovariance}
	\Phi
	\;=\;
	C\,X\,C^*,
\ee
where $\left< \, \cdot \, \right>$ denotes the expectation and superscript $*$ denotes complex conjugate transpose. The matrix $\Phi$ contains information about all second-order statistics of the fluctuating velocity field {in statistical steady-state}, including the Reynolds stresses. We assume that the persistent source of excitation $\bd(t)$ in Eq.~\eqref{eq.lnse1} is zero-mean and white-in-time with spatial covariance matrix $W = W^*$,
\be
	\left< {\bd} (t_1) \, {\bd}^* (t_2) \right>
	\;=\;
	W
	\,
	\delta(t_1 - t_2),
	\label{eq.white_cov}
\ee
where $\delta$ is the Dirac delta function. {When the linearized dynamics~\eqref{eq.lnse1} are stable,} the steady-state covariance $X$ of the state $\bpsi(t)$ can be determined as the solution to the algebraic Lyapunov equation
\be
	A\, X
	\;+\;
	X\, A^*
	\;=\;
	-B\, W  B^*.
	\label{eq.standard_lyap}
\ee
The Lyapunov equation~\eqref{eq.standard_lyap} relates the statistics of white-in-time forcing, represented by $W$, to the {infinite-horizon} state covariance $X$ via system matrices $A$ and $B$. It can also be used to compute the energy spectrum of velocity fluctuations $\bv$,
	\be
	\label{eq.H2norm}
	E
	{~=~}
	{
	\trace \left( \Phi \right)
	}
	\; = \;
	\trace \left( C\, X\, C^*\right).
	\ee

{We note that the steady-state velocity covariance matrix $\Phi$ can be alternatively obtained from the spectral density matrix of velocity fluctuations $S_{\bv} (\omega)$ as~\cite{kwasiv72},
	\be
	\Phi 
	\; = \;
	\dfrac{1}{2 \pi}
	\int_{- \infty}^{\infty}
	S_{\bv} (\omega) \, \mrd \omega.
	\non
	\ee
For the linearized NS equations, we have
	\be
	S_{\bv}(\omega)
	\; \DefinedAs \;
	T_{\bv \bd} (\omega) \, W \, T^*_{\bv \bd} (\omega)
	\label{eq.psd}
	\ee
where the frequency response matrix
\begin{align}
\label{eq.resolvent}
	T_{\bv \bd} (\omega)
	~=~
	C \left( \mri \omega I \,-\, A \right)^{-1} B,
\end{align}
is obtained by applying the temporal Fourier transform on system~\eqref{eq.lnse1}. We note that the solution $X$ to the algebraic Lyapunov equation~\eqref{eq.standard_lyap} allows us to avoid integration over temporal frequencies and compute the energy spectrum $E$ using~\eqref{eq.H2norm}; see Section~\ref{sec.resolvent} for additional details.
}
	
Following the proper orthogonal decomposition of~\cite{baklum67,moimos89}, the velocity field can be decomposed into characteristic eddies by determining the spatial structure of fluctuations that contribute most to the energy amplification. For turbulent channel flow, it has been shown that the dominant characteristic eddy structures extracted from second-order statistics of the stochastically forced linearized model qualitatively agree with results obtained using eigenvalue decomposition of DNS-generated autocorrelation matrices; see Figs.~15 in~\cite{moajovJFM12} and~\cite{moimos89}. In addition to examining the energy spectrum of velocity fluctuations, we will use the eigenvectors of the covariance matrix $\Phi$ (defined in Eq.~\eqref{eq.outputcovariance}) to study dominant flow structures that are triggered by stochastic excitation.

{
\begin{remark}
\label{remark1}
Since linearized dynamics~\eqref{eq.lnse1} are globally stable even when the flow is convectively unstable~\cite{huemon90}, the Lyapunov-based approach can be used to conduct the steady-state analysis of the velocity fluctuations statistics for many flow configurations that are not stable from the perspective of local analysis.
\end{remark}}

\begin{figure}
	\begin{centering}
	\vspace{.15cm}
	\begin{tabular}{rc}
		\begin{tabular}{c}
		\vspace{.8cm}
		\rotatebox{90}{\normalsize $y$}
		\end{tabular}
		&
		\hspace{-.35cm}
		\begin{tabular}{c}
		\includegraphics[width=.40\textwidth]{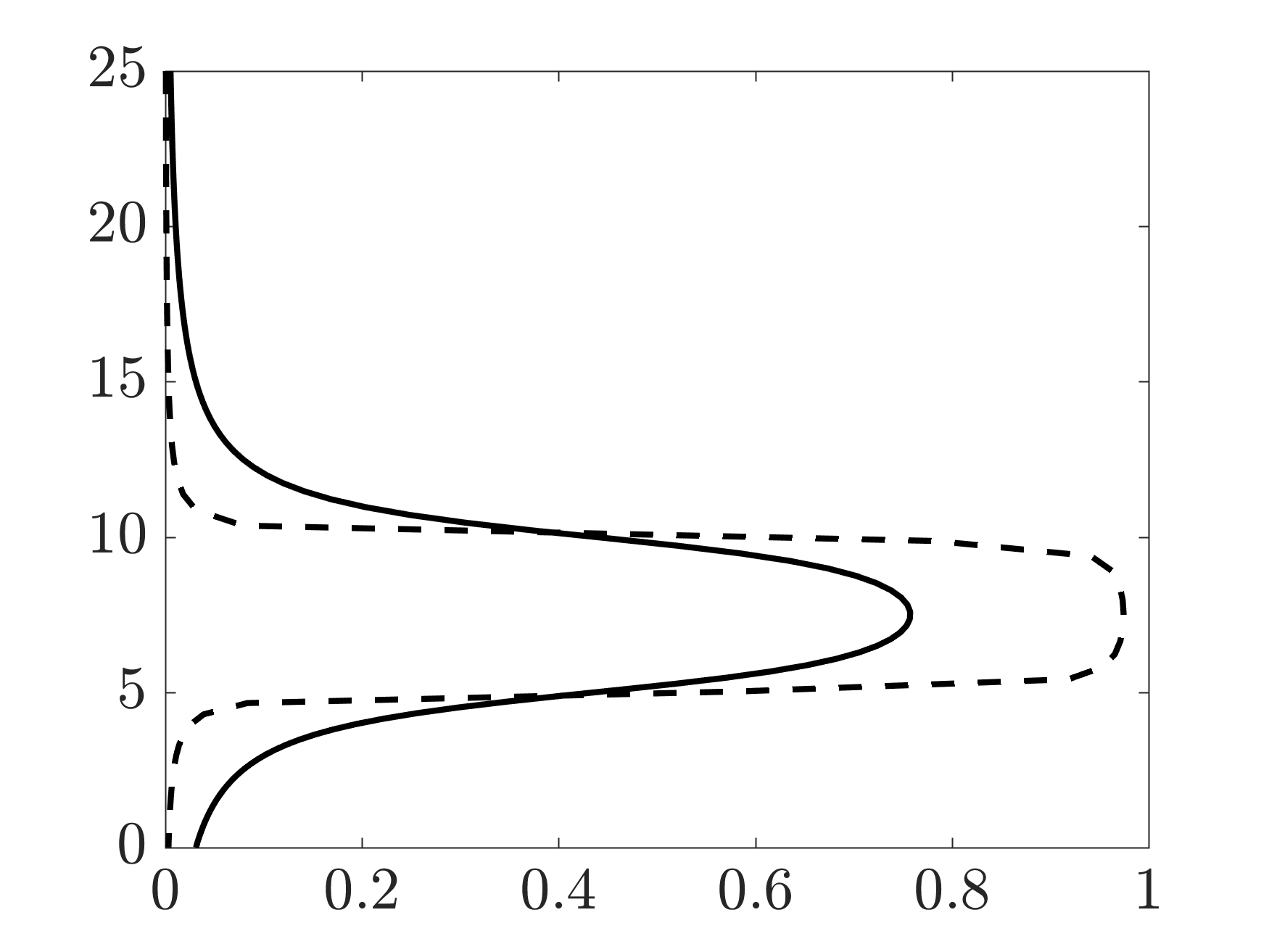}
		\\
		{\normalsize $f(y)$}
		\end{tabular}
	\end{tabular}
	\caption{The shape of the filter function $f(y)$ for $y_1=5$, $y_2=10$ {with $a=1$ ($-$) and $a=10$ ($- -$).}}
	\label{fig.fy}
	\end{centering}
\end{figure}

{
\vspace*{-3ex}
\subsection{Filtered excitation and receptivity coefficient}
\label{sec.filter-receptivitycoeff}

Let us specify the spatial region in which the forcing enters, by introducing
\begin{align}
\label{eq.forcing}
	{\bd} (x,y,z,t)
	\; \DefinedAs \;
	f(y)\, h(x)\, {\bd}_s (x,y,z,t),
\end{align}
where ${\bd}_s$ represents a white solenoidal forcing, $f(y)$ is a smooth filter function defined as
\begin{align}
	\label{eq.fy}
	f(y)
	\;\DefinedAs\;
	\dfrac{1}{\pi} \left( \mathrm{atan} \, (a\,(y\,-\,y_1)) \,-\, \mathrm{atan} \, ( a\,(y\,-\,y_2) ) \right),
\end{align}
and $h(x)$ is a filter function that determines the streamwise extent of the forcing. Here, $y_1$ and $y_2$ determine the wall-normal extent of $f(y)$ and $a$ specifies the roll-off rate; Fig.~\ref{fig.fy} shows $f(y)$ with $y_1=5$ and $y_2=10$, for two cases of $a=1$ and $a=10$. In Sections~\ref{sec.parallel-BL} and~\ref{sec.global-BL}, we study energy amplification arising from stochastic excitation that enters at various wall-normal locations; see Table~\ref{tab.forcing-cases}. For the near-wall forcing (with $y_1=0$ and $y_2=5$) with $a=1$, more than $96\%$ of the energy of the forcing is applied within the $\delta_{0.99}$ boundary layer thickness; on the other hand, for the outer-layer forcing (with $y_1=15$ and $y_2=20$) with $a=1$, less than $0.1\%$ is applied in that region. Our study mainly focuses on the forcing with $h(x)=1$; the effect of changing the function $h$ is considered in Section~\ref{sec.HIT}.

\begin{table}
\tabcolsep 1pt \caption{{Cases of stochastic excitation entering at various wall-normal regions}}
\begin{center}
{\rule{0.6\textwidth}{1pt}}
\begin{tabular*}{.6\textwidth}{@{\extracolsep{\fill}} l|c}
&
\\[-.3cm]
  ~~case number& wall-normal region of excitation; $[\,y_1,\, y_2\,]$ in Eq.~\eqref{eq.fy}~~~
  \\[-0.3cm] \\ \hline \\[-0.25cm]
  ~~1 (near-wall)~~ & $[\,0,\,5\,]$
  \\[0.1cm]
  ~~2 & $[\,5,\,10\,]$
  \\[0.1cm]
  ~~3 & $[\,10,\,15\,]$
  \\[0.1cm]
  ~~4 (outer-layer)~~ & $[\,15,\,20\,]$
  \\[0.1cm]
  \hline
\end{tabular*}
\end{center}
\label{tab.forcing-cases}
\end{table}

We quantify the receptivity of velocity fluctuations to stochastic forcing that enters at various wall-normal regions using the receptivity coefficient
\begin{align}
	\label{eq.recepcoeff}
	C_R
	~\DefinedAs~
	\dfrac{\lim_{t \, \to \, \infty} \left< (D_g \bv(t) )^* D_g \bv(t) \right>}{ \lim_{t \, \to \, \infty} \left< \bd^*(t)\, \bd(t) \right>}
	~=~
	\dfrac{\trace \left(D_g \Phi D_g^* \right)}{\trace \left(W \right)},
\end{align}
which determines the ratio of the energy of velocity fluctuations within the boundary layer to the energy of the forcing. Here, $D_g \DefinedAs g (x,y) I$, where the function $g(x,y)$ is a top-hat filter that extracts velocity fluctuations within the $\delta_{0.99}$ boundary layer thickness. In parallel flows, the function $g$ is invariant with respect to the streamwise direction.}

	\vspace*{-2ex}
\section{Receptivity analysis of locally parallel flow}
\label{sec.parallel-BL}

We first examine the dynamics of the stochastically forced Blasius boundary layer under the locally parallel flow assumption. In this case, the base flow only depends on the wall-normal coordinate $y$ and evolution model~\eqref{eq.lnse1} is parameterized by horizontal wavenumbers ($k_x,k_z$), which significantly reduces the computational complexity. We perform an input-output analysis to quantify the energy amplification of velocity fluctuations subject to \mbox{free-stream turbulence.}

We compute the energy spectrum of stochastically excited parallel Blasius boundary layer flow with $Re_0=232$ (the Blasius length-scale is $\delta_0=1$). Here, we consider a wall-normal region with {$L_y=35$} and discretize the differential operators in Eqs.~\eqref{eq.lnse} using $N_y = 100$ Chebyshev collocation points in $y$. In the wall-normal direction, homogenous Dirichlet boundary conditions are imposed on wall-normal vorticity,
$
	\eta(0) = \eta(L_y) = 0
$
and Dirichlet/Neumann boundary conditions are imposed on wall-normal velocity,
$
	v(0) = v(L_y) = 0
$,
$
	v_y(0) = v_y(L_y) = 0
$, where $v_y$ denotes the derivative of $v$ with respect to $y$. In the horizontal directions, we use $50 \times 51$ logarithmically spaced wavenumbers with $k_x \in [10^{-4}, 1]$ and $k_z \in [5\times 10^{-3}, 10]$ to parameterize the linearized model~\eqref{eq.lnse1}. Thus, for each pair ($k_x,k_z$), the state $\bpsi = [\,v^T \;\, \eta^T\,]^T$ is a complex-valued vector with ${2 N_y}$ components. Grid convergence has been verified by doubling the number of points used in the discretization of the differential operators in the wall-normal coordinate.

\begin{figure}[!ht]
        \begin{tabular}{cccc}
        \subfigure[]{\label{fig.Eturb_pBBL_Re500_fy_0_5}}
        &&
        \subfigure[]{\label{fig.Eturb_pBBL_Re232_fy_15_20}}
        &
        \\[-.5cm]
        \hspace{.2cm}
        \begin{tabular}{c}
                \vspace{.4cm}
                \rotatebox{90}{\normalsize $k_x$}
        \end{tabular}
        &
        \begin{tabular}{c}
                \includegraphics[width=0.41\textwidth]{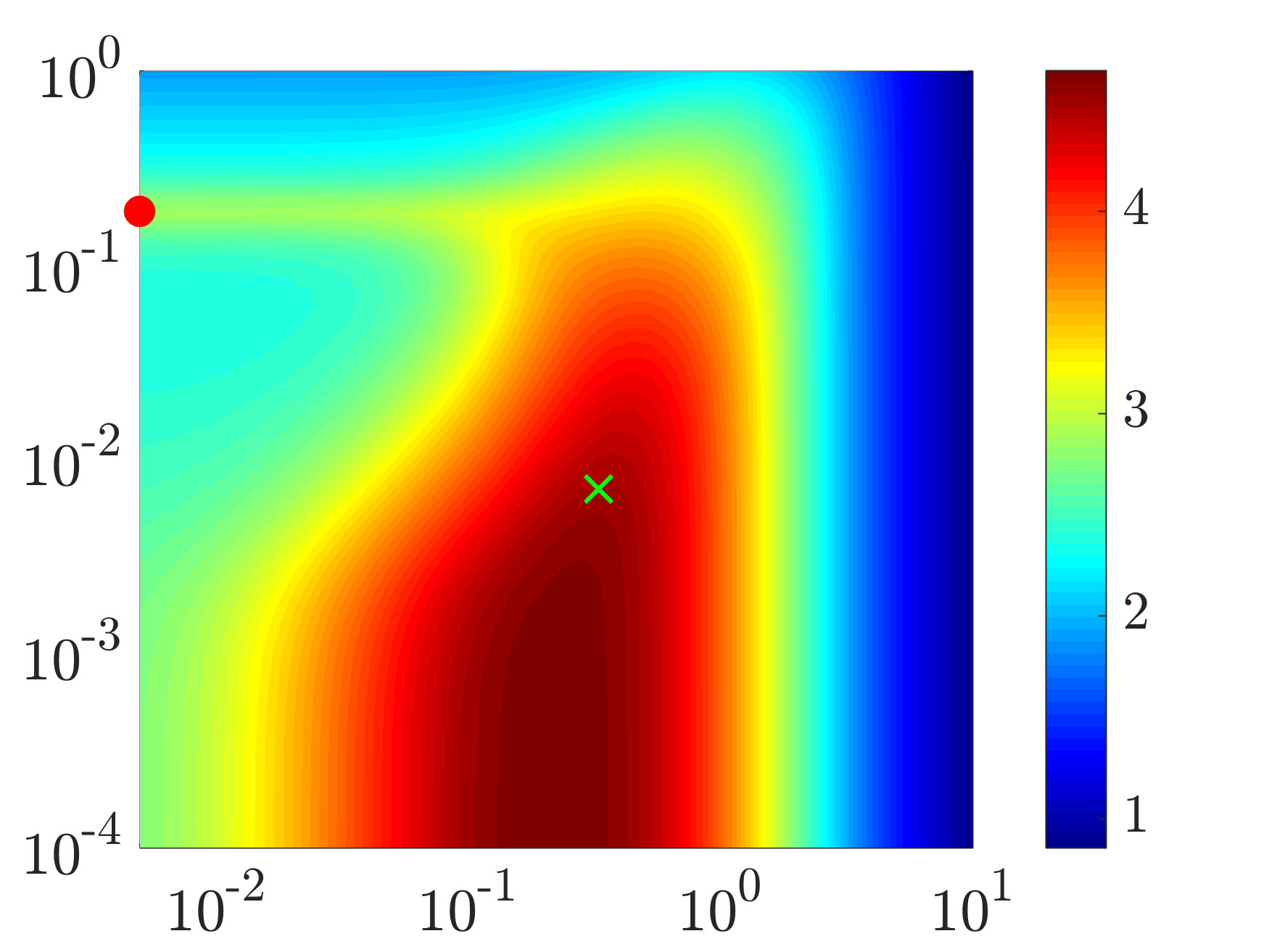}
        \end{tabular}
        &
        \hspace{.2cm}
        \begin{tabular}{c}
                \vspace{.4cm}
                \rotatebox{90}{\normalsize $k_x$}
        \end{tabular}
        &
        \begin{tabular}{c}
                \includegraphics[width=0.41\textwidth]{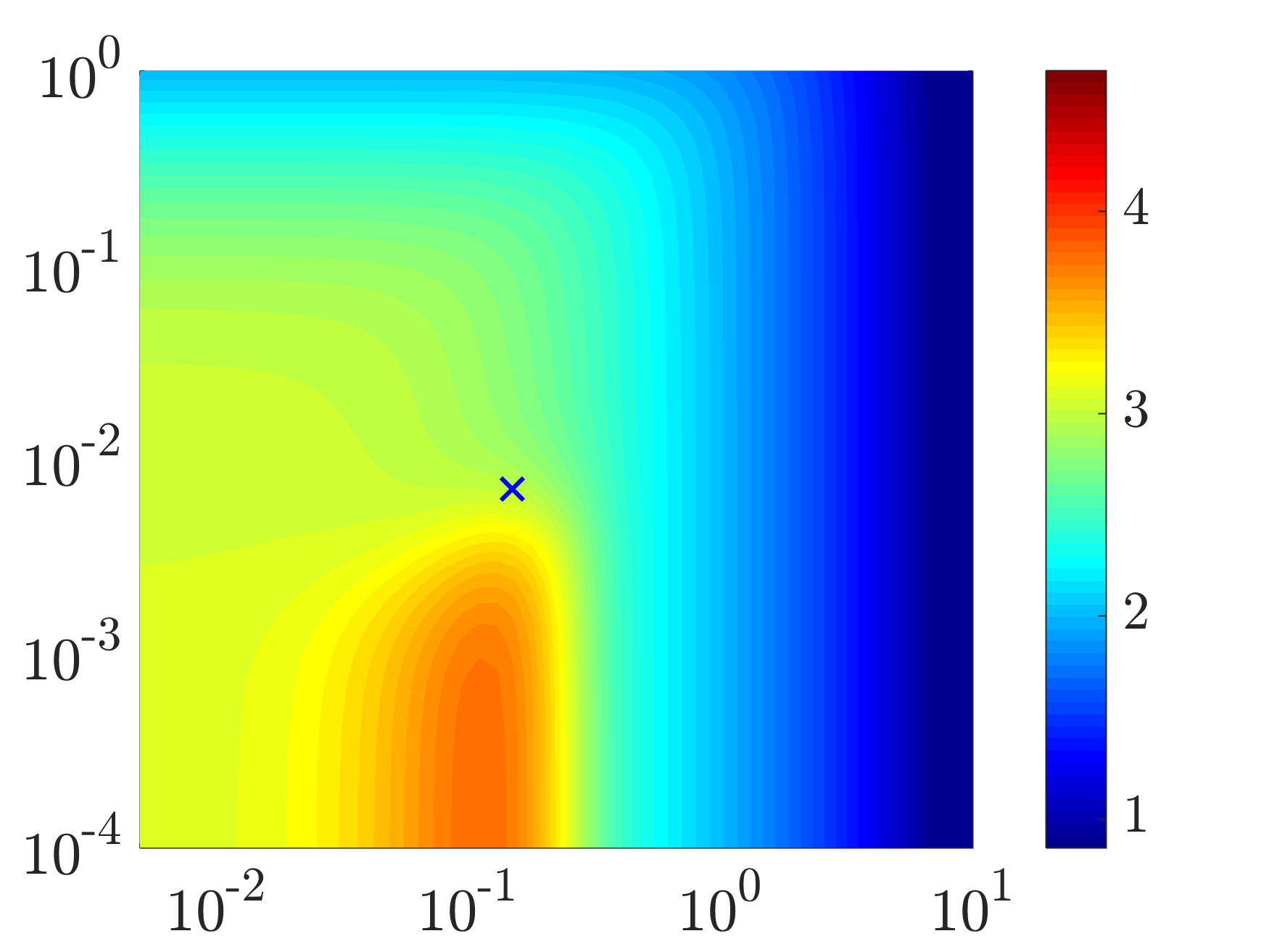}
        \end{tabular}
        \\[-.1cm]
        &
        \hspace{-1cm}
         {\normalsize $k_z$}
        &&
        \hspace{-1cm}
        {\normalsize $k_z$}
        \end{tabular}
        \caption{Plots of {$\log_{10}(C_R(k_x,k_z))$} in the parallel Blasius boundary layer flow with $Re_0=232$ subject to {(a) near-wall, and (b) outer-layer white-in-time stochastic excitation. The dot and crosses respectively mark the wavenumber pairs associated with TS waves and streaks that are closely examined in this paper.}}
        \label{fig.receptivity-coeff}
\end{figure}

We first consider a {streamwise invariant ($h(x)=1$)} solenoidal white-in-time excitation $\bd$ with covariance $W=I$ in the immediate vicinity of the wall {(case 1 in Table~\ref{tab.forcing-cases})}. Figure~\ref{fig.Eturb_pBBL_Re500_fy_0_5} shows {largest receptivity} at low streamwise wavenumbers ($k_x \approx 0$) with a global peak at $k_z \approx 0.25$. This indicates that streamwise elongated streaks are the dominant flow structures that result from persistent stochastic excitation of the boundary layer flow. Such streamwise elongated structures are reminiscent of energetically dominant streaks with spanwise wavenumbers {$k_z \approx 0.26$ (in Blasius length-scale)} that were identified {in analyses of} optimal disturbances~\cite{andberhen99,luc00}. Slightly smaller spanwise wavenumbers have been recorded from hot-wire signal correlations in the boundary layer subject to free-stream turbulence~\cite{matalf01}. {In addition to streaks,
Fig.~\ref{fig.Eturb_pBBL_Re500_fy_0_5}} also predicts the emergence of TS waves at $k_x \approx 0.19$. {For outer-layer forcing,}
the amplification of streamwise elongated structures persists while the amplification of the TS waves weakens; see Fig.~\ref{fig.Eturb_pBBL_Re232_fy_15_20}. It is also observed that as the region of excitation moves away from the wall, energy amplification becomes weaker and the peak of the {receptivity coefficient}
shifts to lower values of $k_z$. As we demonstrate in Section~\ref{sec.global-BL}, these observations are in agreement with the global receptivity analysis of stochastically excited boundary layer flow.

\begin{figure}
	\begin{centering}
	\vspace{.15cm}
	\begin{tabular}{rc}
		\begin{tabular}{c}
		\vspace{.8cm}
		\rotatebox{90}{\normalsize {$\log_{10}(C_R)$}}
		\end{tabular}
		&
		\hspace{-0.3cm}
		\begin{tabular}{c}
		\includegraphics[width=0.41\textwidth]{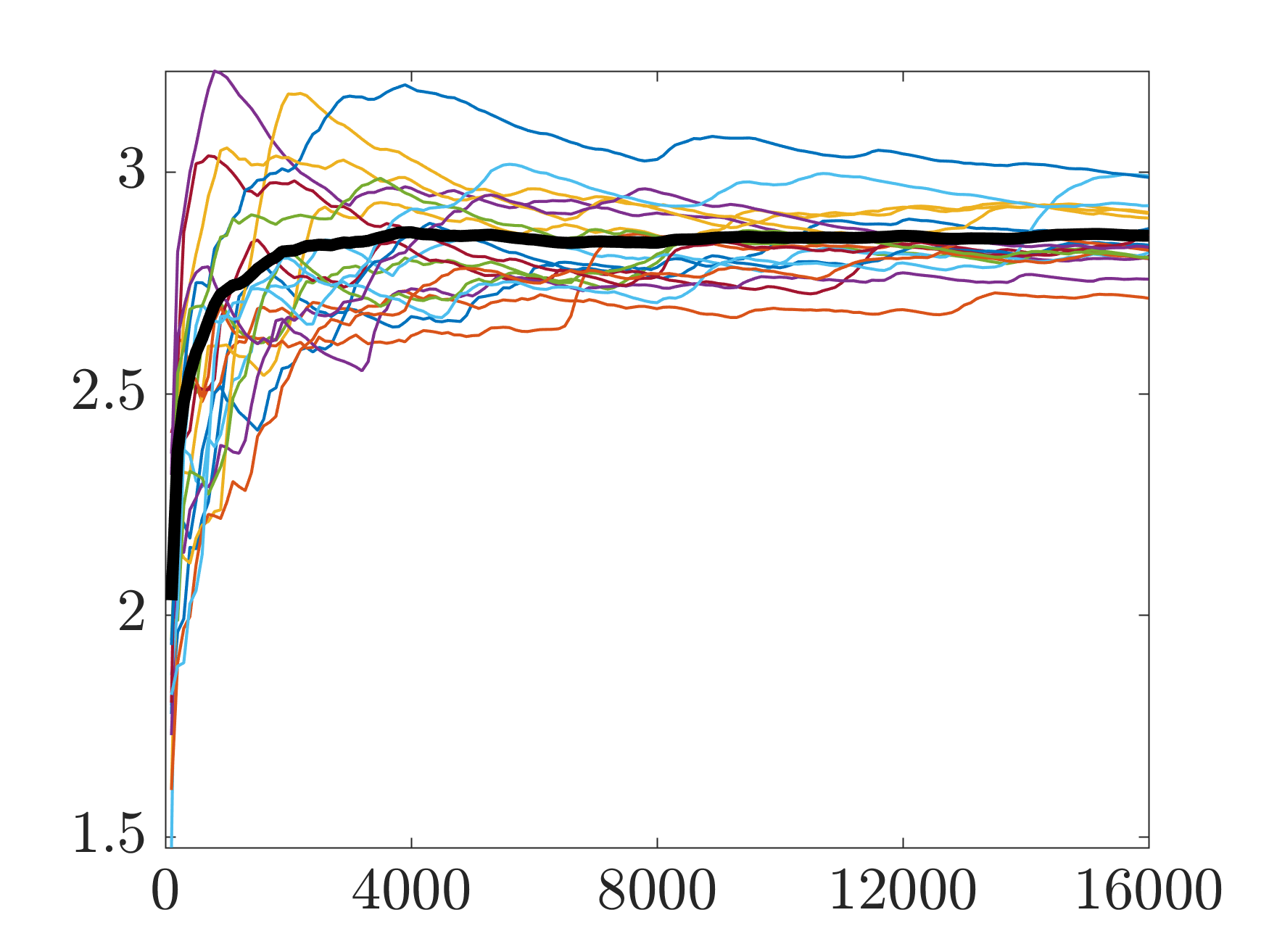}
		\\[.0cm]
		{\normalsize $t$}
		\end{tabular}
	\end{tabular}
	\caption{Time evolution of the {receptivity coefficient $C_R$} for twenty realizations of near-wall stochastic forcing to linearized dynamics~\eqref{eq.lnse1} with $(k_x,k_z)=(0.19,0.005)$ and $Re_0=232$. The {receptivity coefficient} averaged over all simulations is marked by the thick black line.}			
    \label{fig.simu}
	\end{centering}
\end{figure}

\begin{figure}
	\begin{centering}
	\vspace{.15cm}
	\begin{tabular}{cccc}
        \subfigure[]{\label{fig.parallelh2vzEng}}
        &&
        \subfigure[]{\label{fig.parallelh2vzrecepcoeff}}
        &
        \\[-.5cm]
        \hspace{.2cm}
        \begin{tabular}{c}
		\vspace{1cm}
		\rotatebox{90}{\normalsize {$E(k_z)$}}
		\end{tabular}
		&
		\begin{tabular}{c}
		\includegraphics[width=0.41\textwidth]{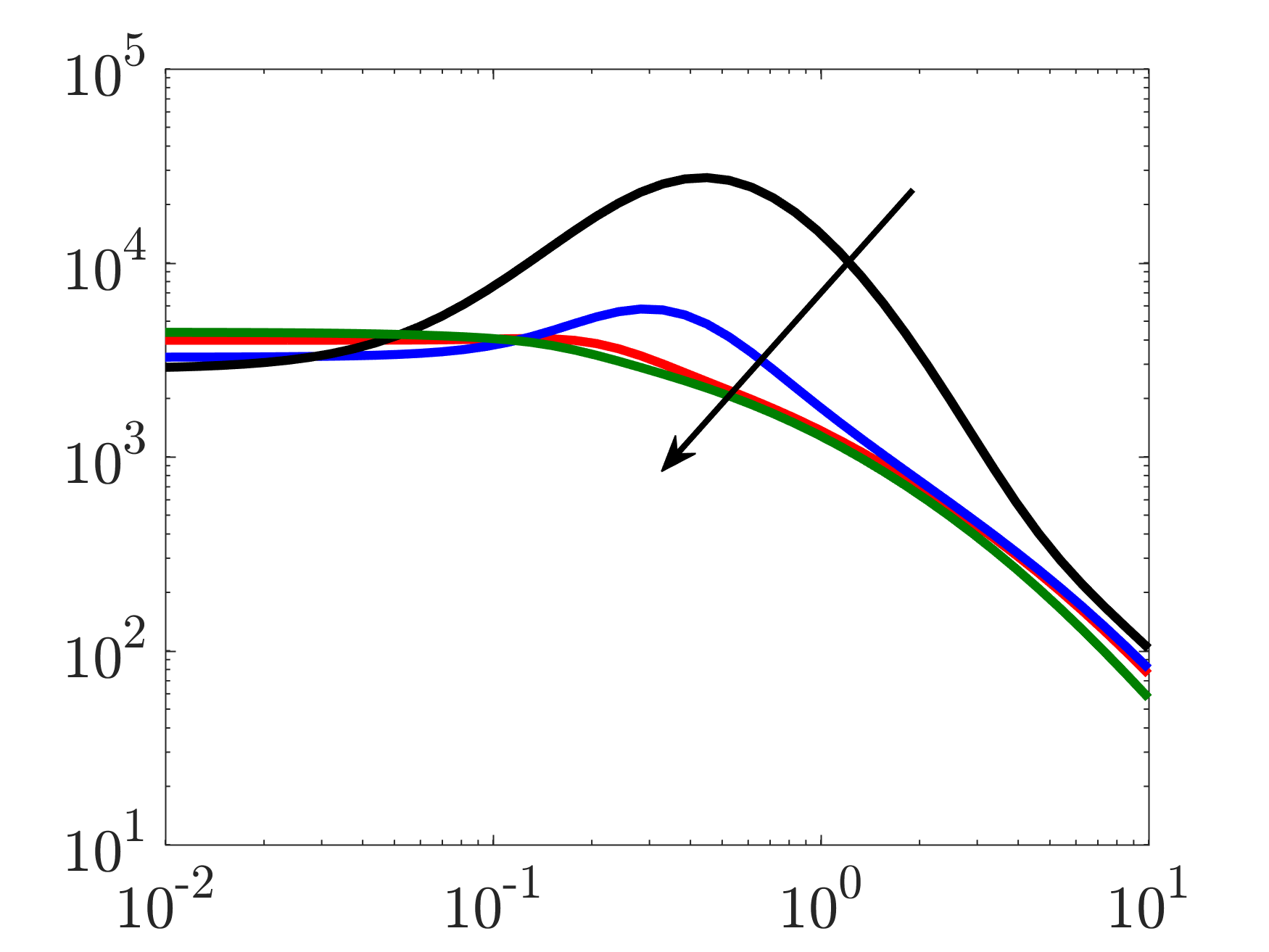}
		\\[.1cm]
		{\normalsize $k_z$}
		\end{tabular}
        &
        \hspace{.2cm}
		\begin{tabular}{c}
		\vspace{1cm}
		\rotatebox{90}{\normalsize {$C_R(k_z)$}}
		\end{tabular}
		&
		\begin{tabular}{c}
        \includegraphics[width=0.41\textwidth]{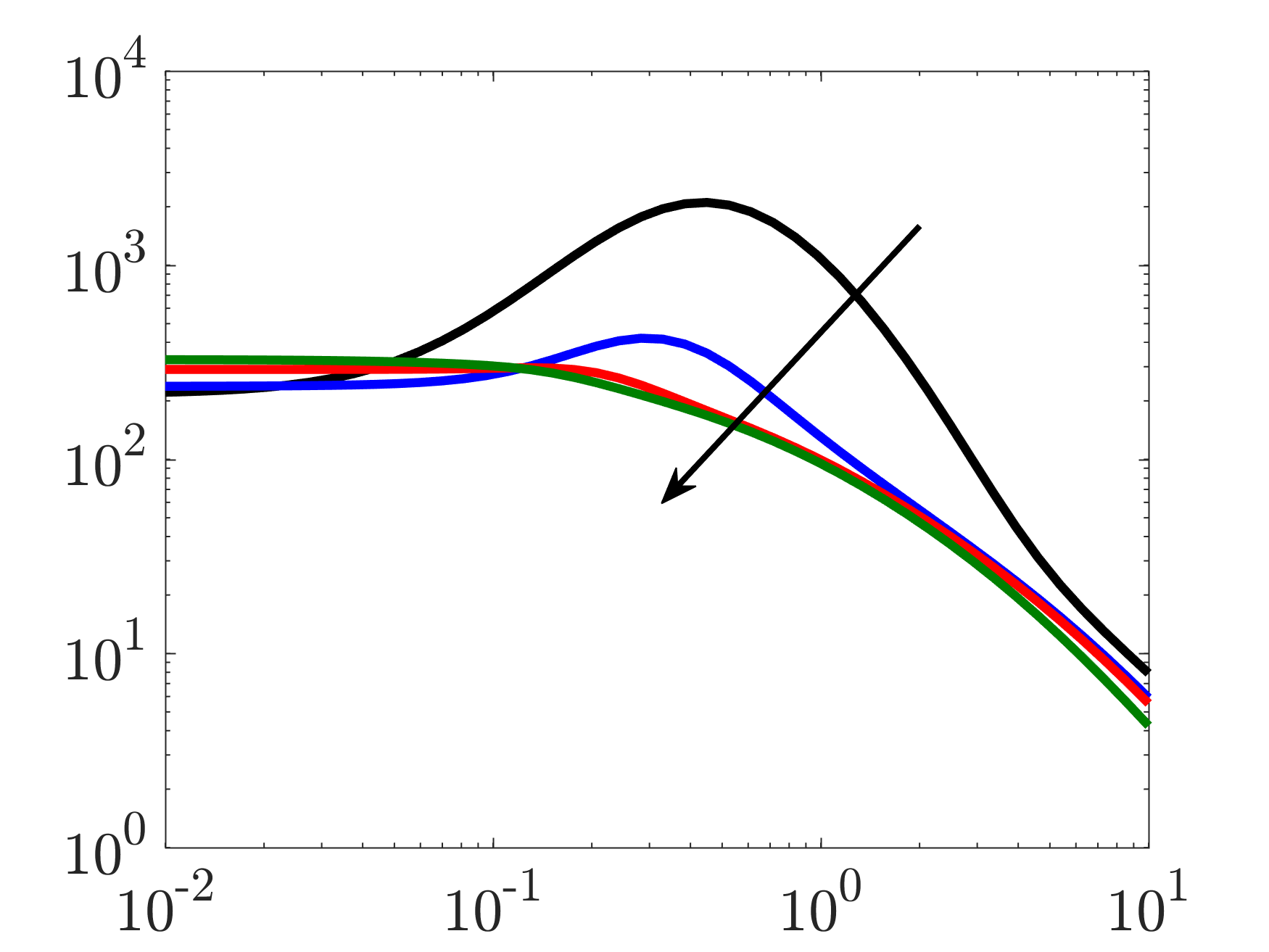}
		\\[.1cm]
		{\normalsize $k_z$}
		\end{tabular}
	\end{tabular}
	\caption{{(a) The one-dimensional energy spectrum, and (b) the receptivity coefficient for the parallel Blasius boundary layer flow with $Re_0=232$ subject to white stochastic excitation entering in the wall-normal regions covered in Table~\ref{tab.forcing-cases}; case 1 (black), case 2  (blue), case 3 (red), and case 4 (green). The forcing region moves away from the wall in the direction of the arrows.}}
    \label{fig.parallelh2vz}
	\end{centering}
\end{figure}

As noted in Section~\ref{sec.stats}, the solution $X$ to Lyapunov equation~\eqref{eq.standard_lyap} represents the steady-state (i.e., long-time average) covariance matrix of the state $\bpsi$ of stochastically forced linearized evolution model~\eqref{eq.lnse1}{, which can be used to compute the energy spectrum in Eq.~\eqref{eq.H2norm} or the receptivity coefficient in Eq.~\eqref{eq.recepcoeff} in a simulation-free manner}. To verify the {values of $C_R$} reported in Fig.~\ref{fig.receptivity-coeff}, we conduct stochastic simulations of the forced linearized flow equations at the wavenumber pair $(k_x,k_z) = (0.19,0.005)$, which is marked by the red dot in Fig.~\ref{fig.Eturb_pBBL_Re500_fy_0_5}. This wavenumber pair allows us to examine the amplification of TS {waves} identified in Fig.~\ref{fig.Eturb_pBBL_Re500_fy_0_5}. Since proper comparison with the result of the Lyapunov equation requires ensemble-averaging, rather than comparison at the level of individual stochastic simulations, we have conducted twenty simulations of system~\eqref{eq.lnse1}. The total simulation time was set to $1.6\times 10^4$ dimensionless time units. Figure~\ref{fig.simu} shows the time evolution of {$C_R$} for twenty realizations of white-in-time forcing $\bd$ to system~\eqref{eq.lnse1}. The {receptivity coefficient averaged} over all simulations is marked by the thick black line. The results indicate that the average of the sample sets asymptotically approaches the correct {steady-state} value of {$C_R$}.

The one-dimensional energy spectrum shown in {Fig.~\ref{fig.parallelh2vzEng} quantifies} the energy amplification $E$ over various spanwise wavenumbers when forcing enters at different distances {from} the wall. This quantity can be computed by integrating the energy spectrum {$E(k_x,k_z)$ (cf.\ Eq.~\eqref{eq.H2norm})} over streamwise wavenumbers. In Fig.~\ref{fig.parallelh2vz}, the locations at which the energy spectrum peaks correspond to the spanwise scale associated with streamwise elongated streaks. When the forcing region shifts away from the wall, the energy amplification decreases, indicating that the flow region in the immediate vicinity of the wall is more susceptible to external excitation. As mentioned earlier, we also observe that, when the forcing region shifts upward, the boundary layer streaks become wider in the spanwise direction. {Figure~\ref{fig.parallelh2vzrecepcoeff} shows similar trends in the receptivity coefficient as a function of spanwise wavenumber $k_z$, which is computed by integrating $C_R$ presented in Fig.~\ref{fig.receptivity-coeff} over streamwise wavenumbers.}

The eigenvalue decomposition of the {velocity covariance matrix $\Phi$} can be used to identify the energetically dominant flow structures resulting from stochastic excitation. In particular, symmetries in the wall-parallel directions can be used to express velocity components as
\be
	\label{eq.spatialstruct}
	\ba{rcl}
		u_j(x,z,t)
		&= &
		4\cos(k_z z)\, \mathrm{Re} \left( \tilde{u}_j(k_x,k_z)\, \mre^{\mri k_x x} \right),
		\\[.15cm]
		v_j(x,z,t)
		&=&
		4\cos(k_z z)\, \mathrm{Re} \left( \tilde{v}_j(k_x,k_z) \, \mre^{\mri k_x x} \right),
		\\[.15cm]
		w_j(x,z,t)
		&=&
		-4\sin(k_z z)\, \mathrm{Im} \left( \tilde{w}_j(k_x,k_z) \, \mre^{\mri k_x x} \right),
	\ea
\ee
Here, $\mathrm{Re}$ and $\mathrm{Im}$ denote real and imaginary parts, and $\tilde{u}_j$, $\tilde{v}_j$, and $\tilde{w}_j$ correspond to the streamwise, wall-normal, and spanwise components of the $j$th eigenvector of the matrix $\Phi$ in Eq.~\eqref{eq.outputcovariance}. {While all amplitudes have been normalized, the phase of these components have been modulated to ensure the compactness of $v_j(x,y,z)$ around $z=0$~\cite{moimos89}}; see~\cite[Appendix F]{moajovJFM12} for additional details.

\begin{figure}[!ht]
        \begin{tabular}{cccc}
        \subfigure[]{\label{fig.streak05eigspectrum}}
        &&
        \subfigure[]{\label{fig.streak510eigspectrum}}
        &
        \\[-.5cm]
        \hspace{.2cm}
        \begin{tabular}{c}
                \vspace{.4cm}
                \rotatebox{90}{\normalsize $\lambda_j/\sum_i \lambda_i$}
        \end{tabular}
        &
        \begin{tabular}{c}
                \includegraphics[width=.40\textwidth]{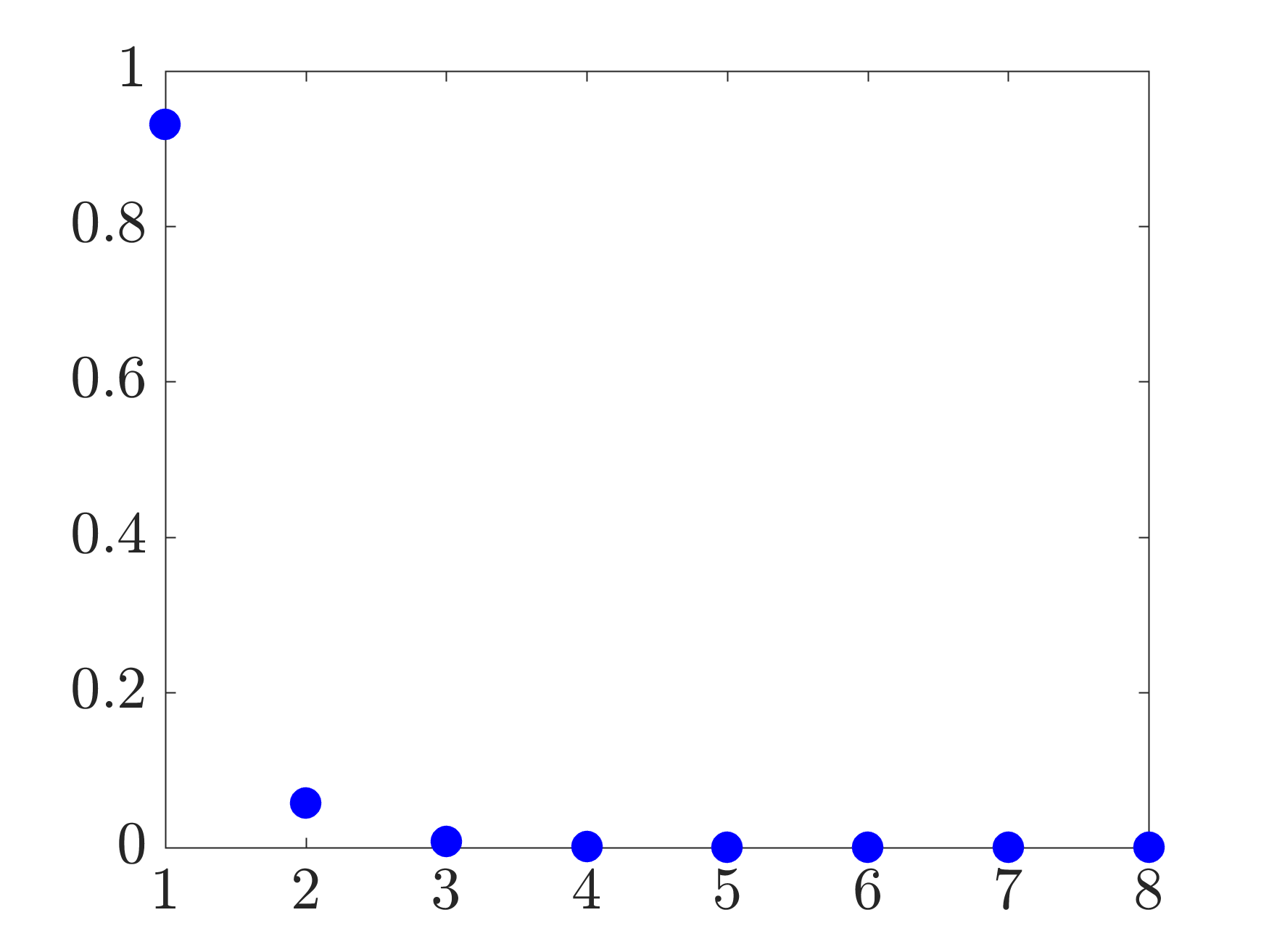}
        \end{tabular}
        \hspace{-.3cm}
        &
        \hspace{.2cm}
        \begin{tabular}{c}
                \vspace{.4cm}
                \rotatebox{90}{\normalsize $\lambda_j/\sum_i \lambda_i$}
        \end{tabular}
        &
        \begin{tabular}{c}
                \includegraphics[width=.40\textwidth]{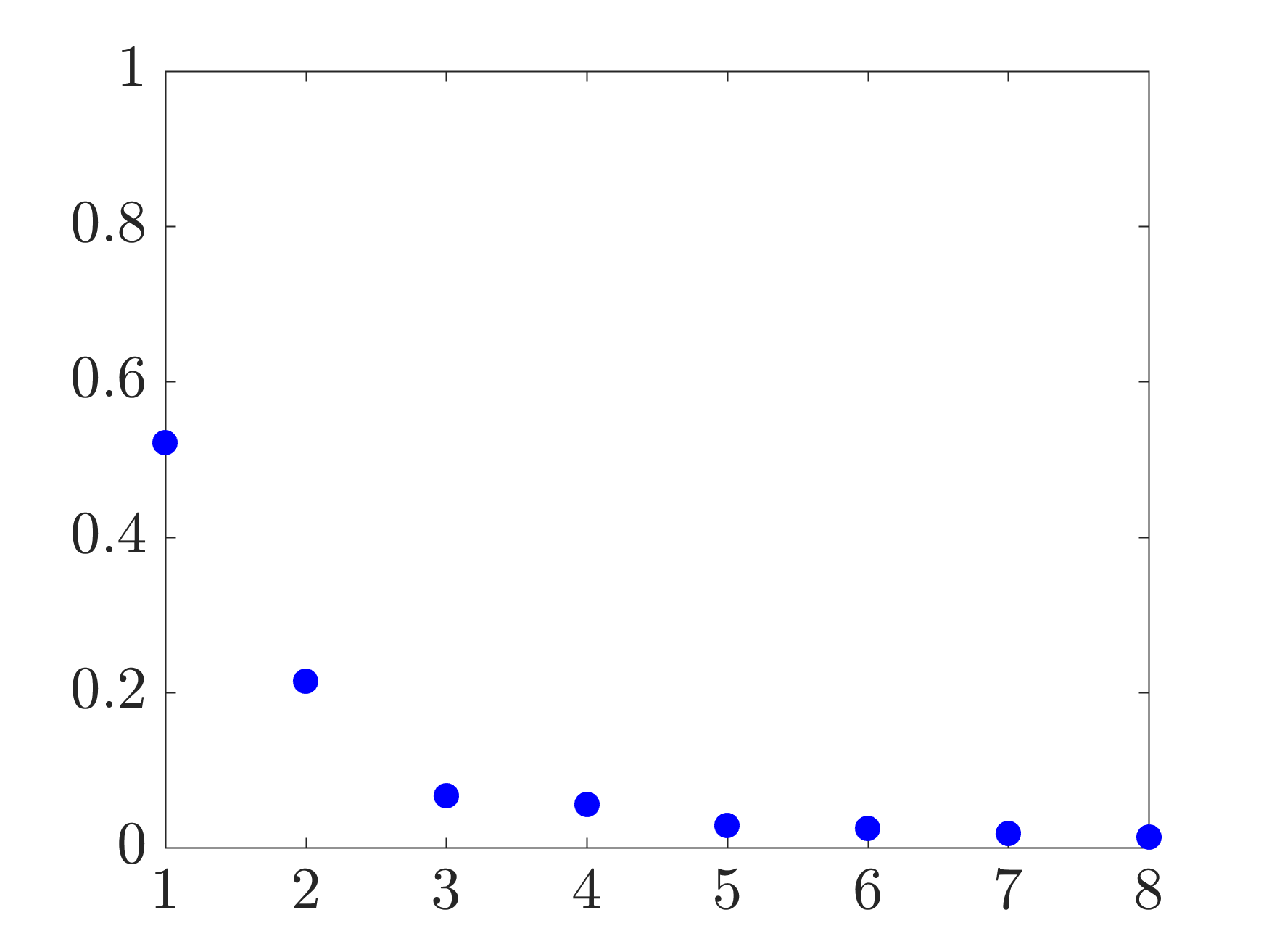}
        \end{tabular}
        \\[-.0cm]
        &
         {\normalsize $j$}
        &&
        {\normalsize $j$}
        \end{tabular}
        \caption{Contribution of the first $8$ eigenvalues of the velocity covariance matrix $\Phi$ of the Blasius boundary layer flow with $Re_0=232$ subject to {(a) near-wall, and (b) outer-layer white-in-time stochastic forcing}.}
        \label{fig.parallelstreakeigspectrum}
\end{figure}

\begin{figure}[!ht]
        \begin{tabular}{cccccc}
        \hspace{-0.4cm}
        \subfigure[]{\label{fig.parallelstreakxyz05}}
        &&
        \hspace{-.65cm}
        \subfigure[]{\label{fig.parallelstreakxy05}}
        &&
        \hspace{-.65cm}
        \subfigure[]{\label{fig.parallelstreakyz05}}
        &
        \\[-.2cm]
        \begin{tabular}{c}
		\vspace{1.2cm}
		\\{\normalsize $y$}\\\\\\\\\\\hspace{0.3cm}{\normalsize $z$}\\\\
	    \end{tabular}
        &
        \hspace{-.34cm}
        \begin{tabular}{c}
                \includegraphics[width=0.315\textwidth]{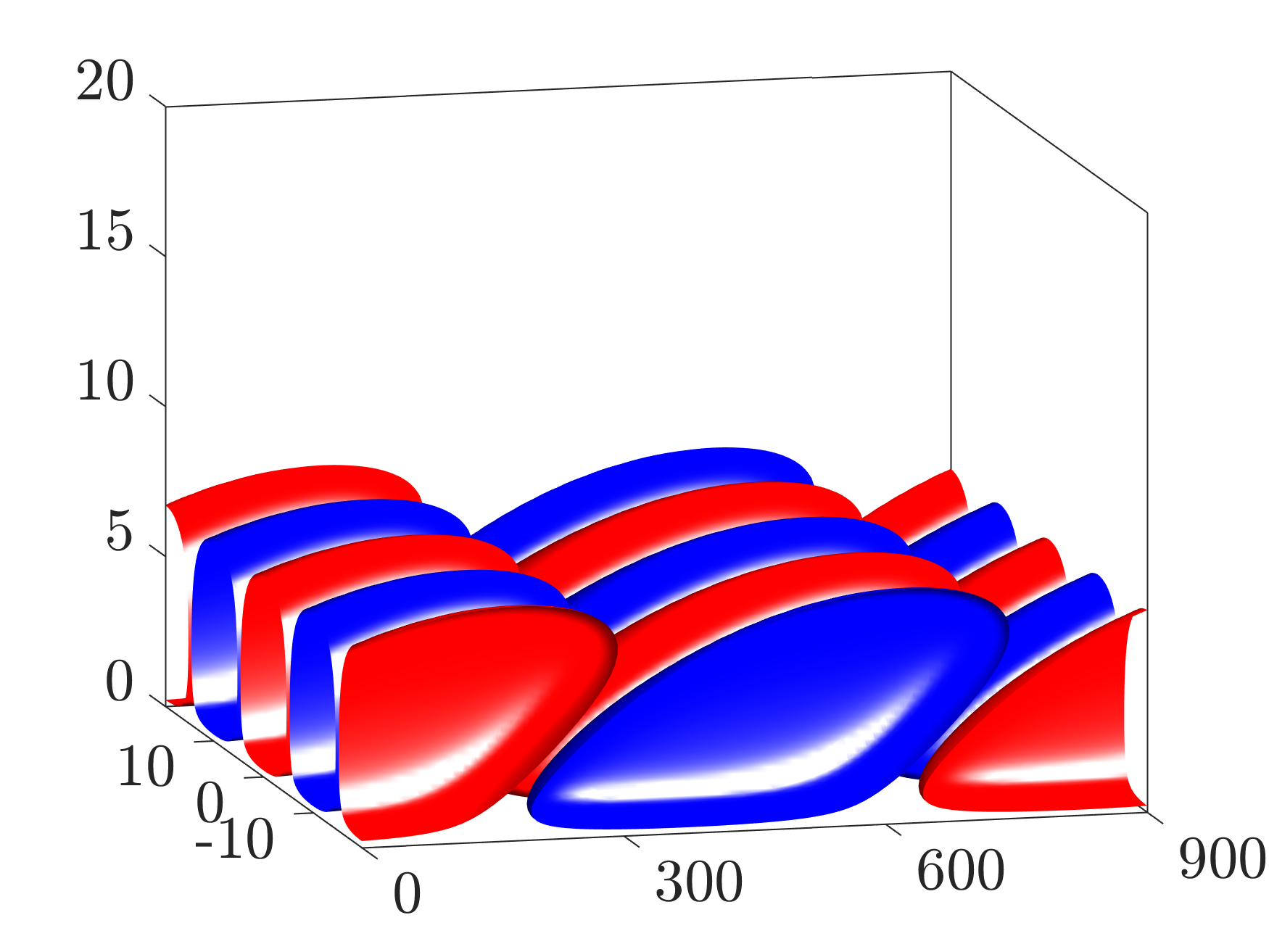}
        \end{tabular}
        &&
        \hspace{-.34cm}
        \begin{tabular}{c}
                \includegraphics[width=0.315\textwidth]{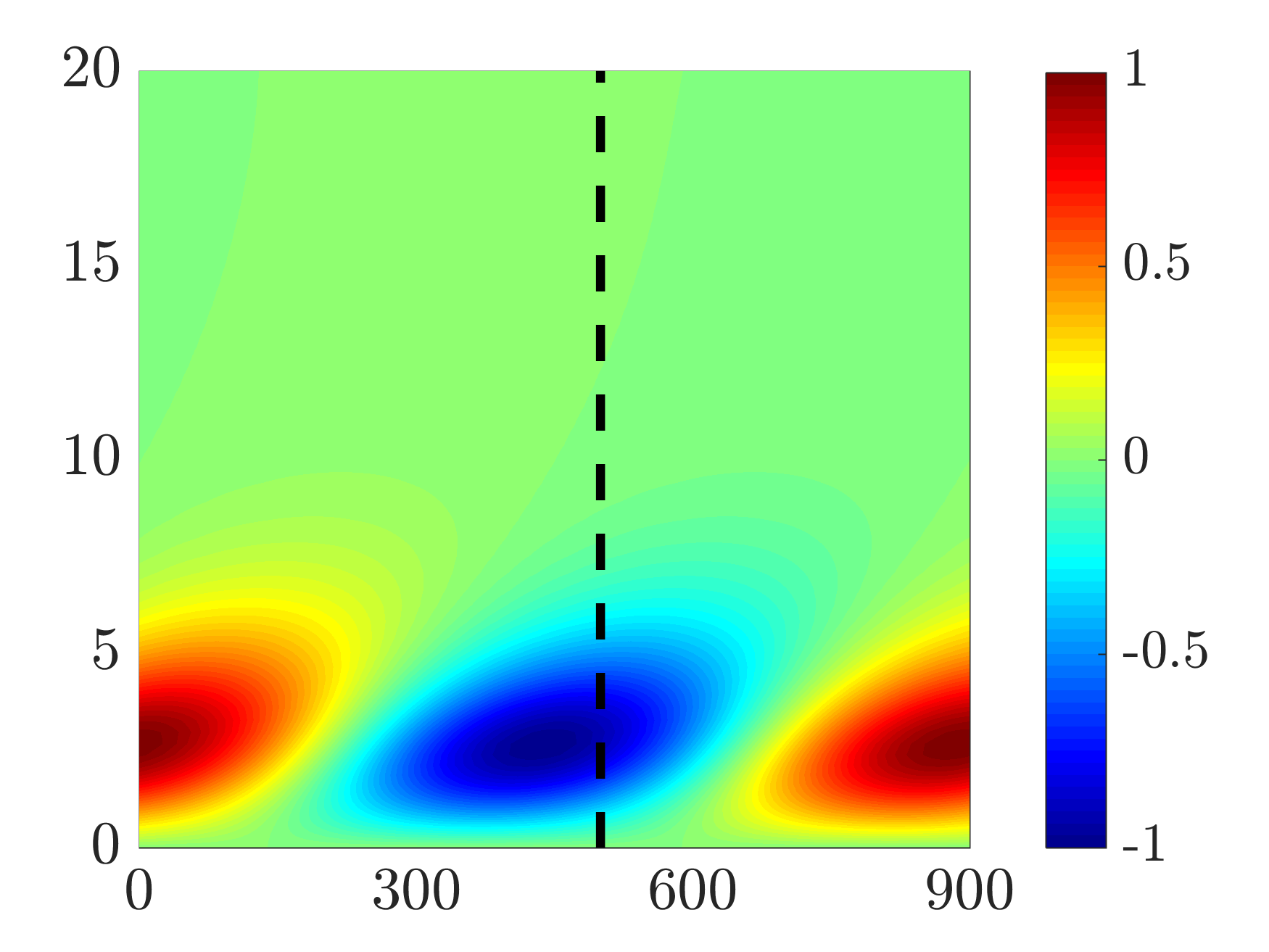}
        \end{tabular}
        &&
        \hspace{-.34cm}
        \begin{tabular}{c}
                \includegraphics[width=0.315\textwidth]{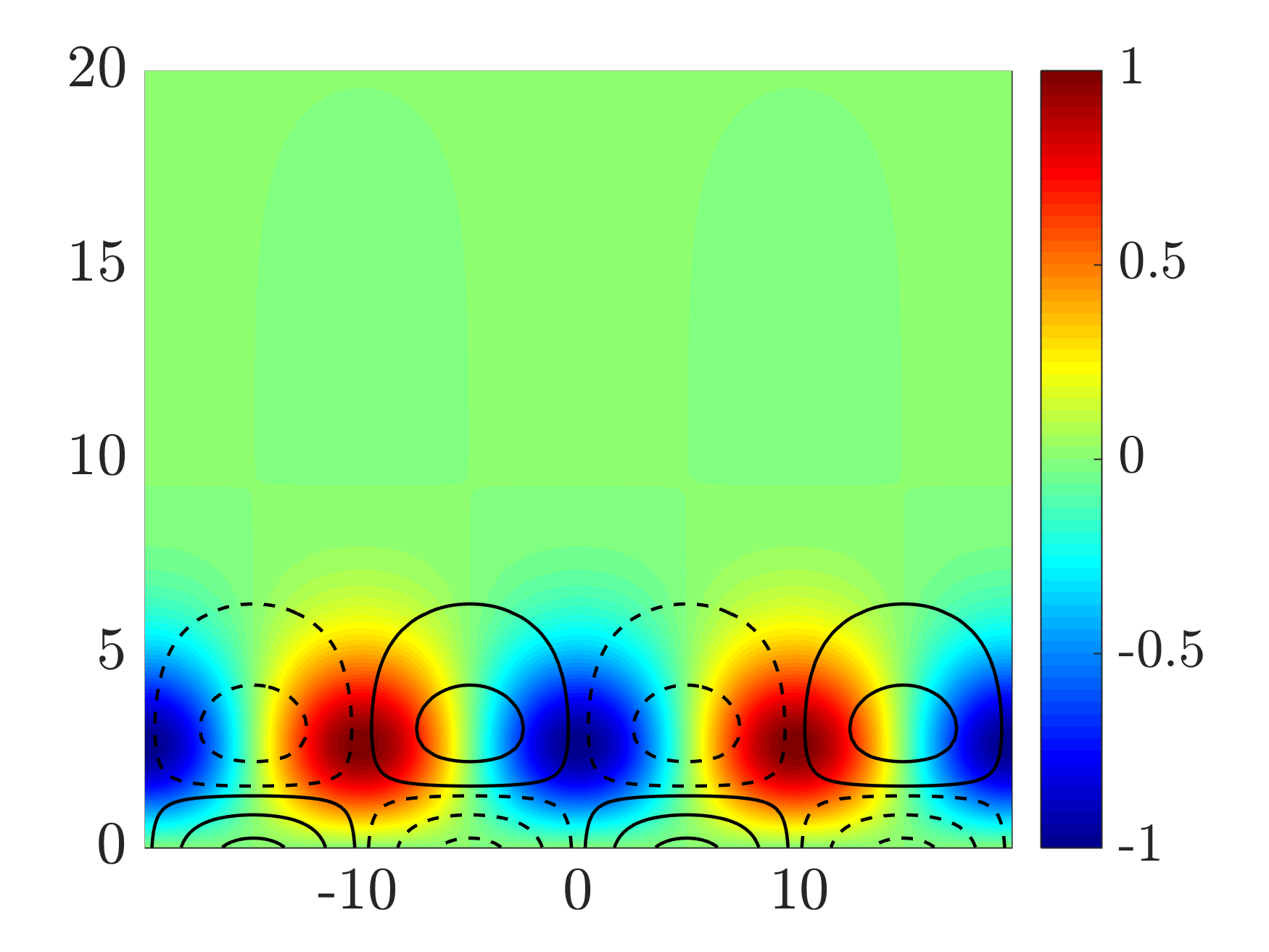}
        \end{tabular}
        \\[-.1cm]
        &
         {\normalsize $x$}
        &&
        \hspace{-.9cm}
        {\normalsize $x$}
        &&
        \hspace{-.8cm}
        {\normalsize $z$}
        \end{tabular}
        \caption{Principal modes with $(k_x,k_z)=(7\times10^{-3},0.32)$, resulting from excitation of the boundary layer flow with $Re_0=232$ in the vicinity of the wall. (a) Streamwise velocity component where red and blue colors denote regions of high and low velocity. (b) Streamwise velocity at $z=0$. (c) $y$-$z$ slice of streamwise velocity (color plots) and vorticity (contour lines) at $x=500$, which corresponds to the cross-plane slice indicated by the black dashed lines in (b).}
        \label{fig.parallelflowstructure-05}
\end{figure}

\begin{figure}[!ht]
        \begin{tabular}{cccccc}
        \hspace{-0.4cm}
        \subfigure[]{\label{fig.parallelstreakxyz1520}}
        &&
        \hspace{-.65cm}
        \subfigure[]{\label{fig.parallelstreakxy1520}}
        &&
        \hspace{-.65cm}
        \subfigure[]{\label{fig.parallelstreakyz1520}}
        &
        \\[-.2cm]
        \begin{tabular}{c}
		\vspace{1.2cm}
		\\{\normalsize $y$}\\\\\\\\\\\hspace{0.3cm}{\normalsize $z$}\\\\
	    \end{tabular}
        &
        \hspace{-.3cm}
        \begin{tabular}{c}
                \includegraphics[width=0.315\textwidth]{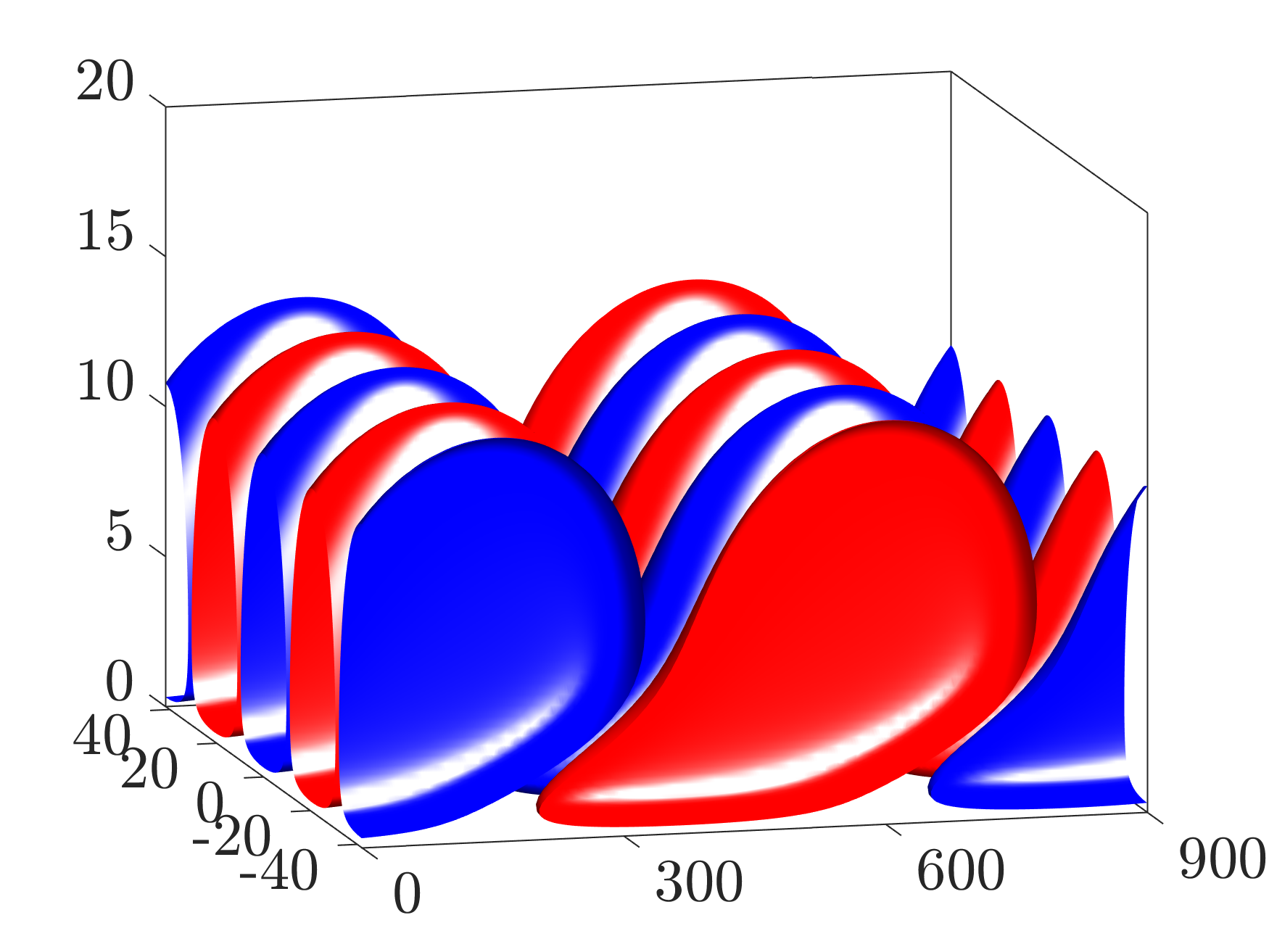}
        \end{tabular}
        &&
        \hspace{-.3cm}
        \begin{tabular}{c}
                \includegraphics[width=0.315\textwidth]{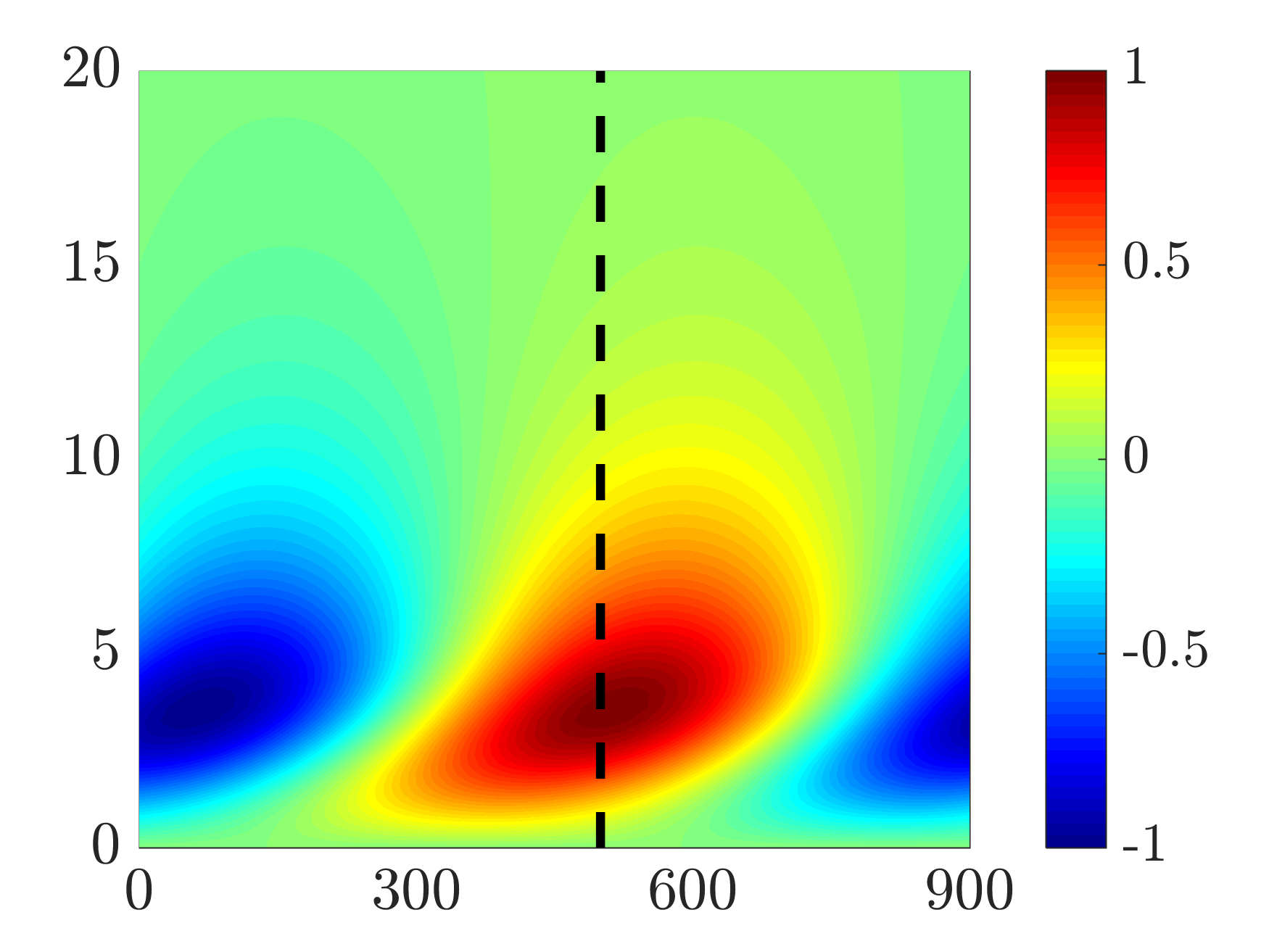}
        \end{tabular}
        &&
        \hspace{-.3cm}
        \begin{tabular}{c}
                \includegraphics[width=0.315\textwidth]{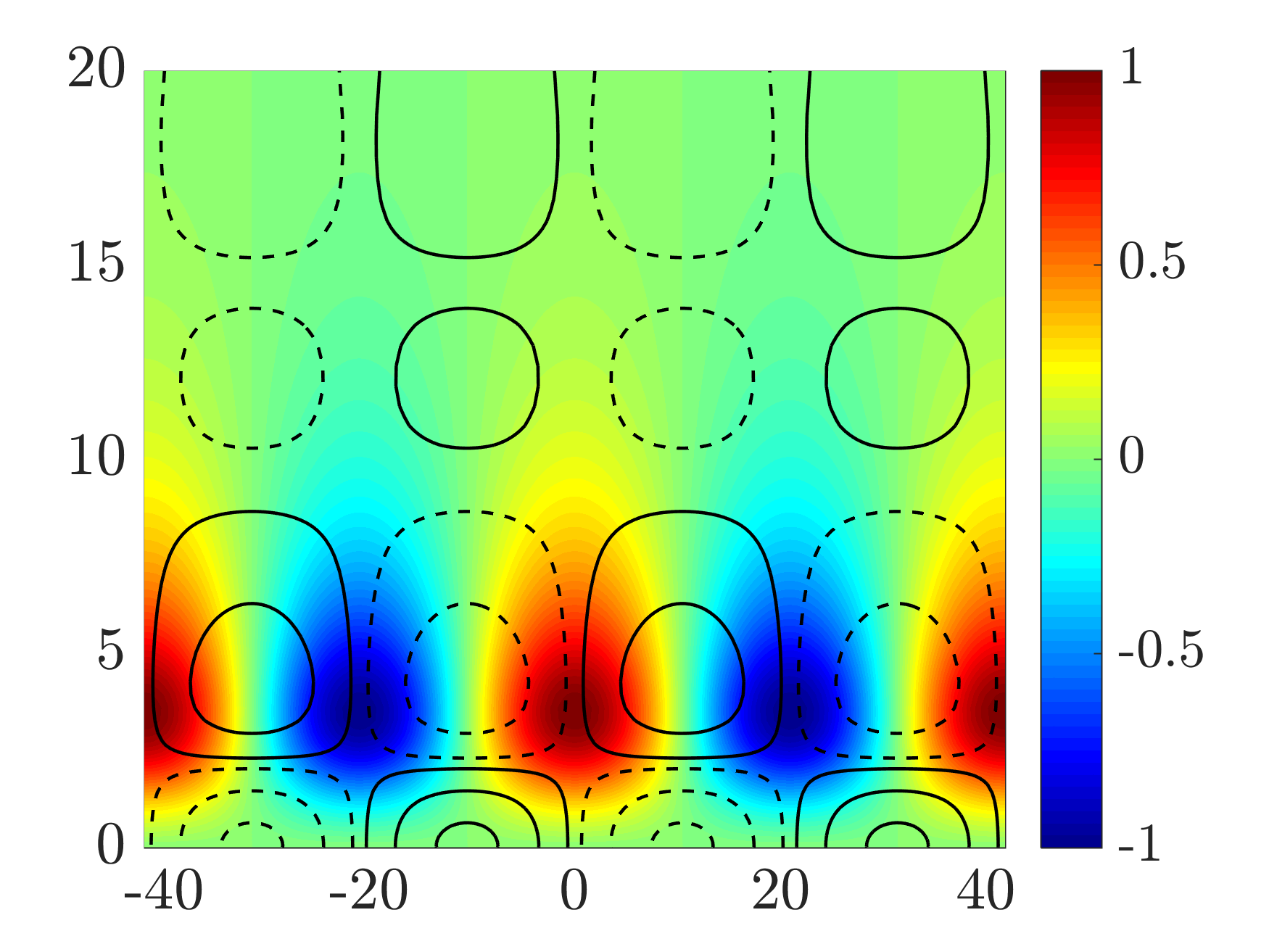}
        \end{tabular}
        \\[-.1cm]
        &
         {\normalsize $x$}
        &&
        \hspace{-.9cm}
        {\normalsize $x$}
        &&
        \hspace{-.8cm}
        {\normalsize $z$}
        \end{tabular}
        \caption{Principal modes with {$(k_x,k_z)=(7 \times 10^{-3},0.15)$}, resulting from {outer-layer excitation of the boundary layer flow} with $Re_0=232$. (a) Streamwise velocity component where red and blue colors denote regions of high and low velocity. (b) Streamwise velocity at $z=0$. (c) $y$-$z$ slice of streamwise velocity (color plots) and vorticity (contour lines) at $x=500$, which corresponds to the cross-plane slice indicated by the black dashed lines in (b).}
        \label{fig.parallelflowstructure-1520}
\end{figure}

While the sum of all eigenvalues of the matrix $\Phi$ determines the overall energy amplification reported in Fig.~\ref{fig.parallelh2vzEng}, it is also useful to examine the spatial structure of modes with dominant contribution to the energy of the flow. Figure~\ref{fig.parallelstreakeigspectrum} shows the contribution of the first $8$ eigenvalues of $\Phi$ to the energy amplification, $\lambda_j/\sum_i \lambda_i$ when the boundary layer flow is subject to stochastic forcing. For fluctuations with $(k_x,k_z)=(7\times10^{-3},0.32)$ and near-wall excitation (cross in Fig.~\ref{fig.Eturb_pBBL_Re500_fy_0_5}) the principal mode which corresponds to the largest eigenvalue, contains approximately $93\%$ of the total energy. {On the other hand, for fluctuations with $(k_x,k_z)=(7\times10^{-3},0.15)$} and {outer-layer} excitation (cross in Fig.~\ref{fig.Eturb_pBBL_Re232_fy_15_20}) the principal mode contains approximately {$52\%$} of the total energy. Figures~\ref{fig.parallelflowstructure-05} and~\ref{fig.parallelflowstructure-1520} show the flow structures associated with the streamwise component of these most significant modes. From Figs.~\ref{fig.parallelstreakxy05} and~\ref{fig.parallelstreakxy1520} it is evident that the core of streamwise elongated structures  moves away from the wall with the shift of the stochastically excited region. These streamwise elongated structures are situated between counter-rotating vortical motions in the cross-stream plane (cf.\ Figs.~\ref{fig.parallelstreakyz05} and~\ref{fig.parallelstreakyz1520}) and contain alternating regions of fast- and slow-moving fluid, which are slightly inclined (and detached) relative to the wall. Even though these structures do not capture the full complexity of transitional flow, as we show in Section~\ref{sec.global-BL}, they contain information about energetic streamwise elongated flow structures that are amplified by external excitation of the boundary layer flow. In particular, such alignment of counter-rotating vortices and streaks is closely related to the lift-up mechanism and the generation of streamwise elongated streaks~\cite{andberhen99,luc00,haczak15}.

\vspace*{-3ex}
\section{Global analysis of stochastically forced linearized NS equations}
\label{sec.global-BL}

The parallel flow assumption applied in Section~\ref{sec.parallel-BL} allows for the efficient parameterization of the governing equations over all wall-parallel wavenumbers $k_x$ and $k_z$. While this significantly reduces computational complexity, it excludes the effect of the spatially evolving base flow on the dynamics of velocity fluctuations. In global stability analysis, the NS equations are linearized around a spatially evolving Blasius boundary layer profile and the finite dimensional approximation is obtained by discretizing all inhomogeneous spatial directions. In this section, we employ global receptivity analysis to quantify the influence of stochastic excitation on the velocity fluctuations around the spatially evolving Blasius boundary layer base flow.

At any spanwise wavenumber $k_z$, the state $\bpsi = [\,v^T \; \eta^T\,]^T$ of linearized evolution model~\eqref{eq.lnse1} is a complex vector with ${2 N_x N_y}$ components, where $N_x$ and $N_y$ denote the number of collocation points used to discretize the differential operators in the streamwise and wall-normal directions, respectively. While this choice of state variables is not commonly used in conventional global stability analysis of boundary layer flows, in Appendix~\ref{sec.appendix-descriptor} we demonstrate that it yields consistent results with the descriptor form in which the state is determined by all velocity and pressure fluctuations. We consider a Reynolds number $Re_0=232$ and a computational domain with $L_x= 900$ and $L_y = 35$, where the differential operators are discretized using $N_x=101$ and $N_y=50$ Chebyshev collocation points in $x$ and $y$, respectively. {Similar to locally parallel} analysis, we {verify} convergence by doubling the number of grid points.

As in Section~\ref{sec.parallel-BL}, in the wall-normal direction we enforce homogenous Dirichlet boundary conditions on $\eta$ and homogeneous Dirichlet/Neumann boundary conditions on $v$. At the inflow, we impose homogeneous Dirichlet boundary conditions on $\eta$, i.e., $\eta(0, y) = 0$, and homogeneous Dirichlet/Neumann boundary conditions on $v$, i.e., $v(0, y) = v_y(0, y) = 0$. {At the outflow, we apply linear extrapolation conditions on both state variables ($v, \eta$) and the streamwise derivative of the wall-normal component ($v_x$)~\cite{theo03},
\begin{align*}
\centering
        \ba{rclcl}
        v(x(N_x), y)
        &=&
        \alpha\,v(x(N_x-1),y)
        &+&
        \beta\, v(x(N_x-2),y),
        \\[0.35cm]
        \eta(x(N_x), y)
        &=&
        \alpha\,\eta(x(N_x-1),y)
        &+&
        \beta\,\eta(x(N_x-2),y),
        \\[0.35cm]
        v_y(x(N_x), y)
        &=&
        \alpha\,v_y(x(N_x-1),y)
        &+&
        \beta\, v_y(x(N_x-2),y),
        \ea
\end{align*}
\vspace{-.45cm}
\begin{align*}
        \alpha
        \;=\;
        \dfrac{x(N_x)\,-\,x(N_x-2)}{x(N_x-1)\,-\,x(N_x-2)},
        ~
        \beta
        \;=\;
        \dfrac{x(N_x-1)\,-\,x(N_x)}{x(N_x-1)\,-\,x(N_x-2)}.
\end{align*}}
\noindent We also introduce sponge layers at the inflow and outflow to mitigate the influence of boundary conditions on the fluctuation dynamics within the computational domain~\cite{niclel11,man12}; see~\cite{ranzarnicjovAIAA17} for an in-depth study on the effect of sponge layer strength in the global stability analysis of boundary layer flow. {The results presented in this section are obtained after adjusting the} sponge layer parameters to match the energy amplification obtained via the descriptor form of the linearized dynamics; see Appendix~\ref{sec.appendix-descriptor} for~details.

\begin{figure}
	\begin{centering}
	\vspace{.15cm}
	\begin{tabular}{cccc}
        \subfigure[]{\label{fig.globalh2vzEng}}
        &&
        \subfigure[]{\label{fig.globalh2vzrecepcoeff}}
        &
        \\[-.5cm]
        	\hspace{.2cm}
		\begin{tabular}{c}
        \vspace{1cm}
        \rotatebox{90}{\normalsize $E(k_z)$}
		\end{tabular}
		&
		\begin{tabular}{c}
		\includegraphics[width=0.41\textwidth]{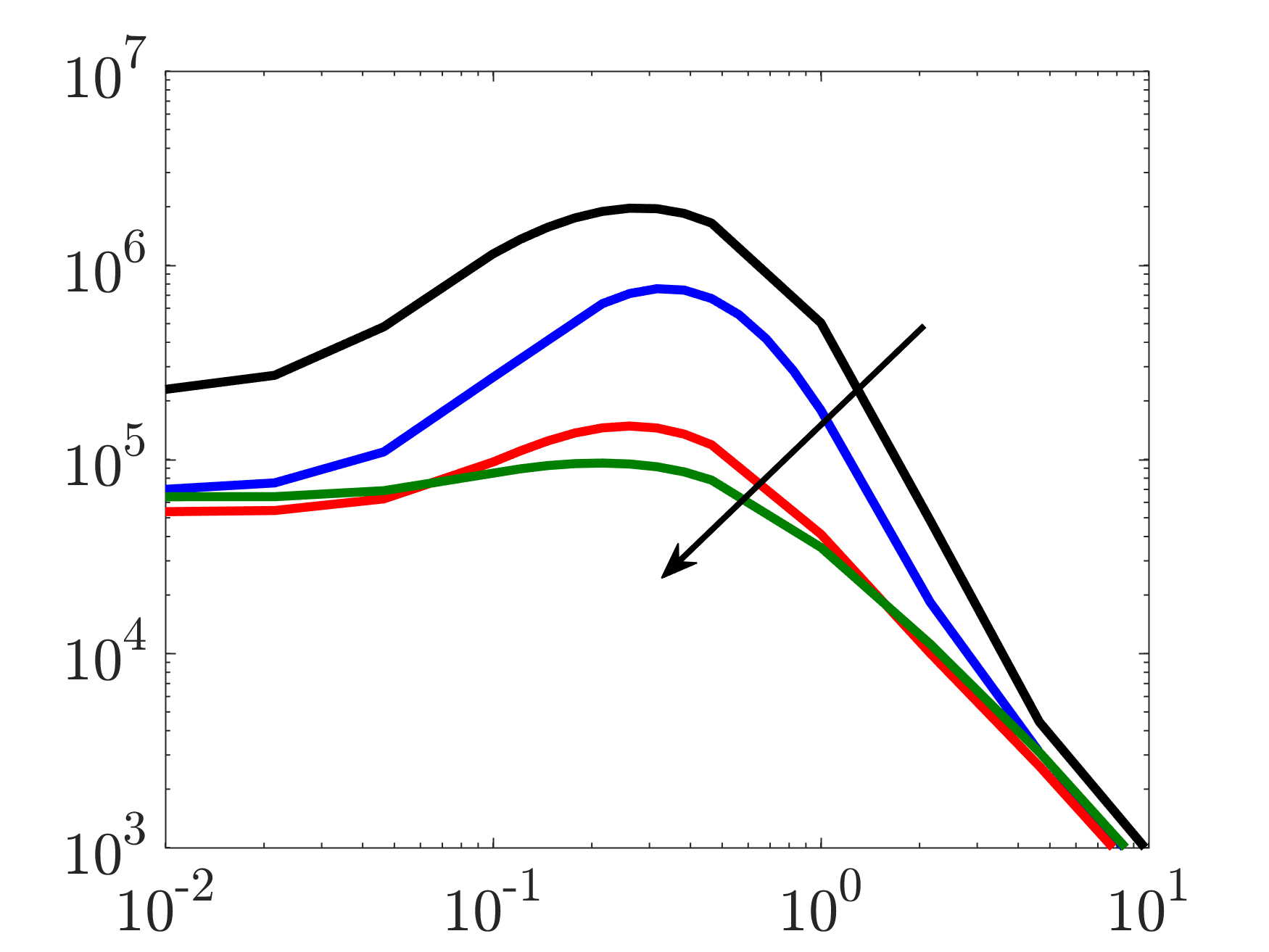}
		\\[.1cm]
		{\normalsize $k_z$}
		\end{tabular}
        &
        	\hspace{.2cm}
        \begin{tabular}{c}
        \vspace{1cm}
		\rotatebox{90}{\normalsize {$C_R(k_z)$}}
		\end{tabular}
		&
		\begin{tabular}{c}
        \includegraphics[width=0.41\textwidth]{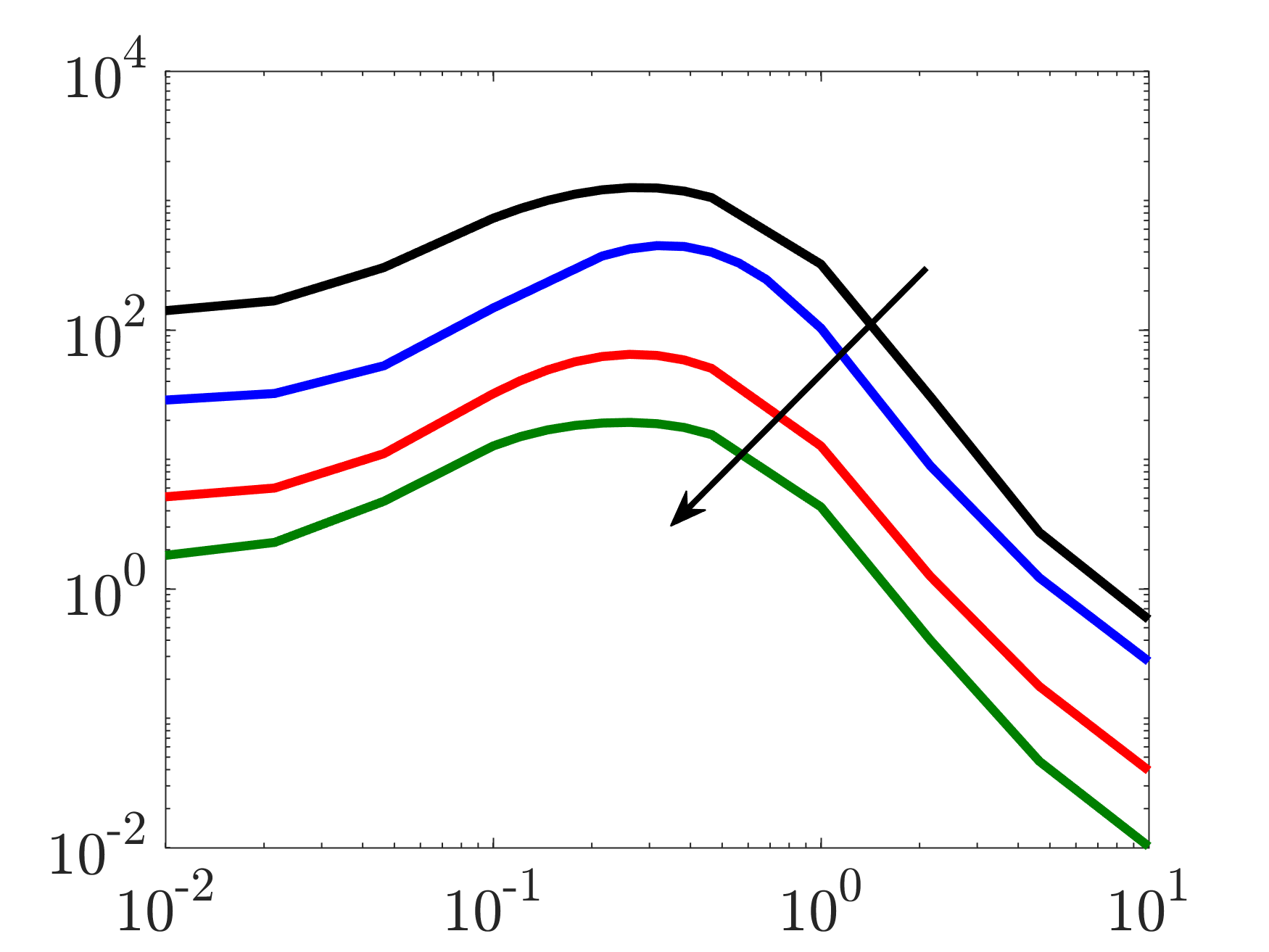}
		\\[.1cm]
		{\normalsize $k_z$}
		\end{tabular}
	\end{tabular}
	\caption{{ (a) Energy amplification and (b) receptivity coefficient resulting from stochastic excitation of the linearized NS equations around a spatially varying Blasius profile with $Re_0=232$. Stochastic forcing enters at the wall-normal regions covered in Table~\ref{tab.forcing-cases}; case 1 (black), case 2  (blue), case 3 (red), and case 4 (green). The forcing region moves away from the wall in the direction of the arrows.}}
	\label{fig.globalh2vz}
	\end{centering}
\end{figure}

{For boundary layer flows, the global operator in Eqs.~\eqref{eq.lnse1} has no exponentially growing eigenmodes~\cite{huemon90}; see Remark~\ref{remark1}.} Thus, the steady-state covariance of the fluctuating velocity field can be obtained from the solution to Lyapunov equation~\eqref{eq.standard_lyap} and the energy amplification can be computed using~Eq.~\eqref{eq.H2norm}. As in Section~\ref{sec.parallel-BL}, we examine the influence of {streamwise-invariant ($h(x)=1$)} white-in-time stochastic forcing with covariance $W=I$ which enters at various wall-normal regions; this is achieved by filtering the forcing using the function $f(y)$ in~\eqref{eq.fy}. Figure~\ref{fig.globalh2vz} shows the $k_z$-dependence of energy amplification {and receptivity coefficient} for stochastic excitation entering at various wall-normal regions. {Our computations show that the energy amplification increases as the region of influence for the external forcing approaches the wall, which qualitatively matches the result of the locally parallel analysis in Section~\ref{sec.parallel-BL}. In particular, for $Re_0=232$, the energy amplification reduces from $2.0\times10^6$ (for stochastic excitation that enters in the vicinity of the wall (case 1 in Table~\ref{tab.forcing-cases}) with $k_z=0.32$) to $9.6\times10^4$ (for stochastic excitation that enters away from the wall (case 4 in Table~\ref{tab.forcing-cases}) with $k_z=0.21$). Moreover, the structures that correspond to the largest energy amplification {or receptivity coefficient} become slightly wider in the spanwise direction, but this shift to smaller values of $k_z$ is not as pronounced as in parallel flows (cf.\ Fig.~\ref{fig.parallelh2vz}). The largest energy amplification and receptivity are observed for structures with $k_z \in [ 0.21, 0.32 ]$, which is in close agreement with previous experimental~\cite{matalf01} and theoretical studies~\cite{andberhen99,luc00}.}

\begin{figure}[!ht]
        \begin{tabular}{cccc}
        \subfigure[]{\label{fig.globalstreak05eigspectrumbeta6}}
        &&
        \subfigure[]{\label{fig.globalstreak510eigspectrumbeta6}}
        &
        \\[-.5cm]
        \hspace{.2cm}
        \begin{tabular}{c}
                \vspace{.4cm}
                \rotatebox{90}{\normalsize $\lambda_j/\sum_i \lambda_i$}
        \end{tabular}
        &
        \begin{tabular}{c}
                \includegraphics[width=0.41\textwidth]{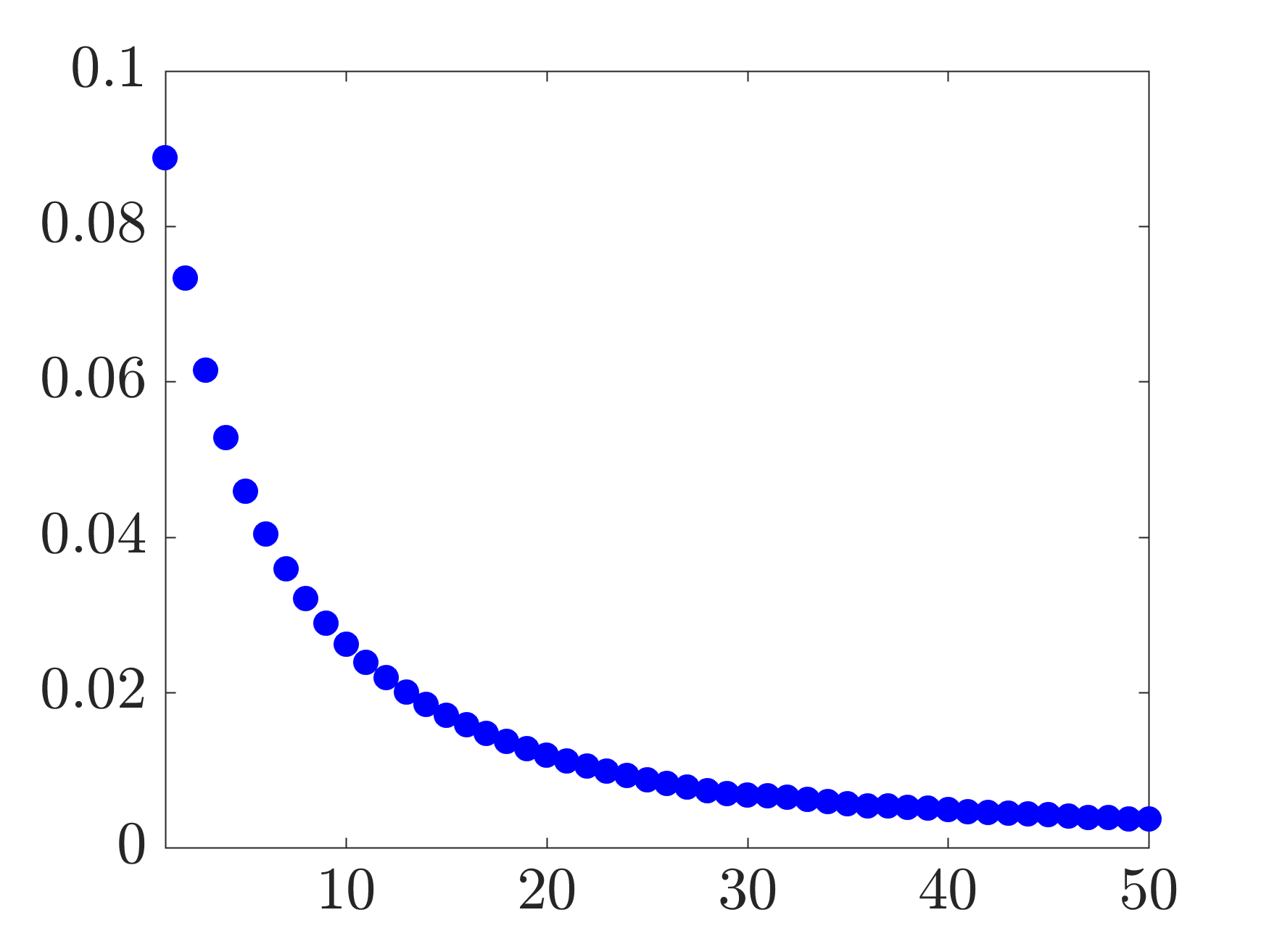}
        \end{tabular}
        \hspace{-.3cm}
        &
        \hspace{.2cm}
        \begin{tabular}{c}
                \vspace{.4cm}
                \rotatebox{90}{\normalsize $\lambda_j/\sum_i \lambda_i$}
        \end{tabular}
        &
        \begin{tabular}{c}
                \includegraphics[width=0.41\textwidth]{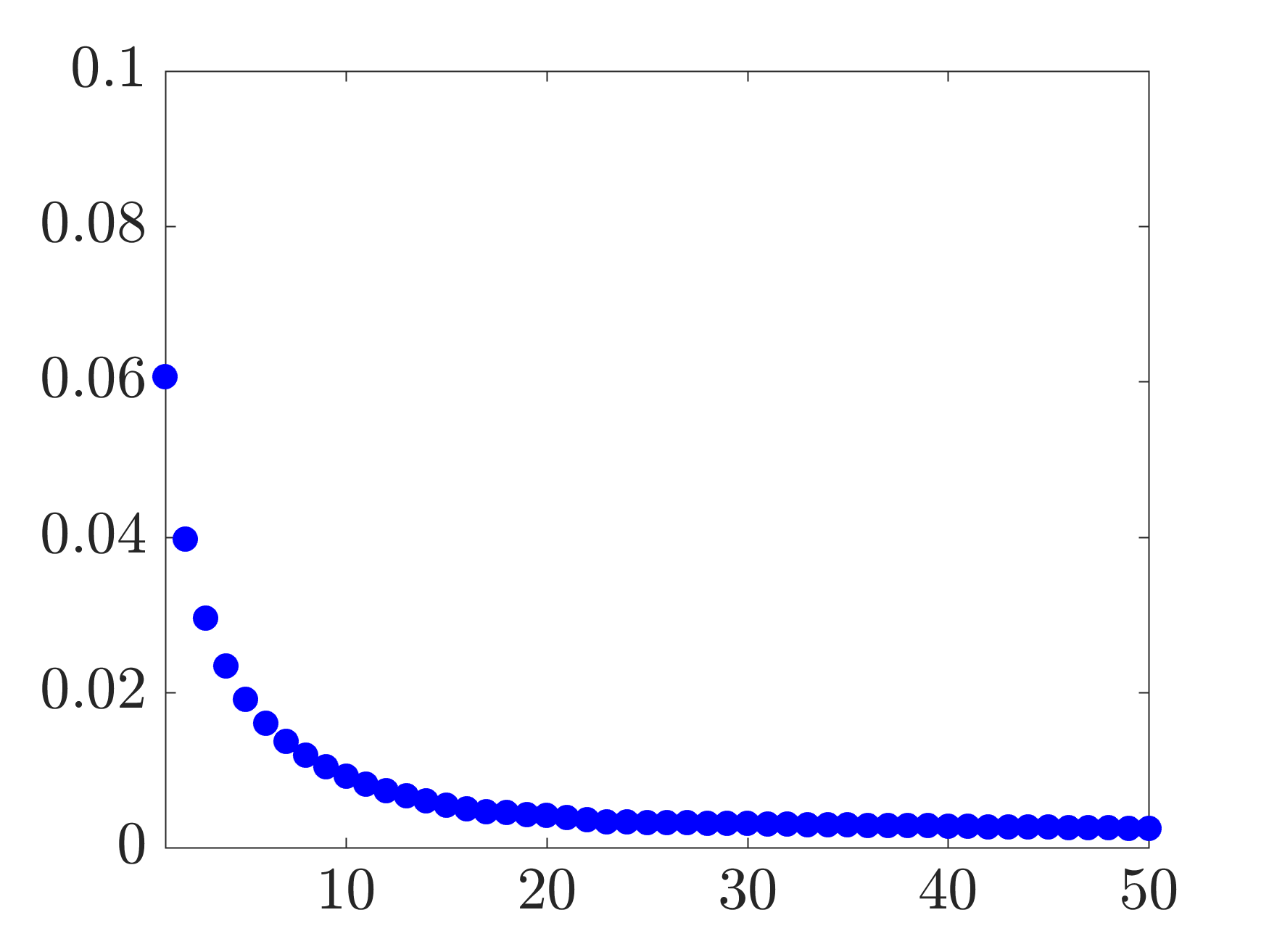}
        \end{tabular}
        \\[-.1cm]
        &
        {\normalsize $j$}
        &&
        {\normalsize $j$}
        \end{tabular}
        \caption{{ Contribution of the first $50$ eigenvalues of the velocity covariance matrix $\Phi$ of the Blasius boundary layer flow with $Re_0=232$ subject to white-in-time stochastic forcing (a) in the vicinity of the wall with spanwise wavenumber $k_z=0.32$; and (b) away from the wall with spanwise wavenumber $k_z=0.21$.}}
        \label{fig.globalstreakeigspectrum}
\end{figure}

\begin{figure}[!ht]
        \begin{tabular}{cccccc}
        \hspace{-.4cm}
        \subfigure[]{\label{fig.streamwisevelocity-3D-0-5}}
        &&
        \hspace{-.65cm}
        \subfigure[]{\label{fig.streamwisevelocity-2D-0-5}}
        &&
        \hspace{-.65cm}
        \subfigure[]{\label{fig.streamwisevelocity-vorticity-0-5}}
        &
        \\[-.2cm]
        \begin{tabular}{c}
		\vspace{1.2cm}
		\\{\normalsize $y$}\\\\\\\\\\\hspace{0.3cm}{\normalsize $z$}\\\\
	    \end{tabular}
        &
        \hspace{-.34cm}
        \begin{tabular}{c}
                \includegraphics[width=0.315\textwidth]{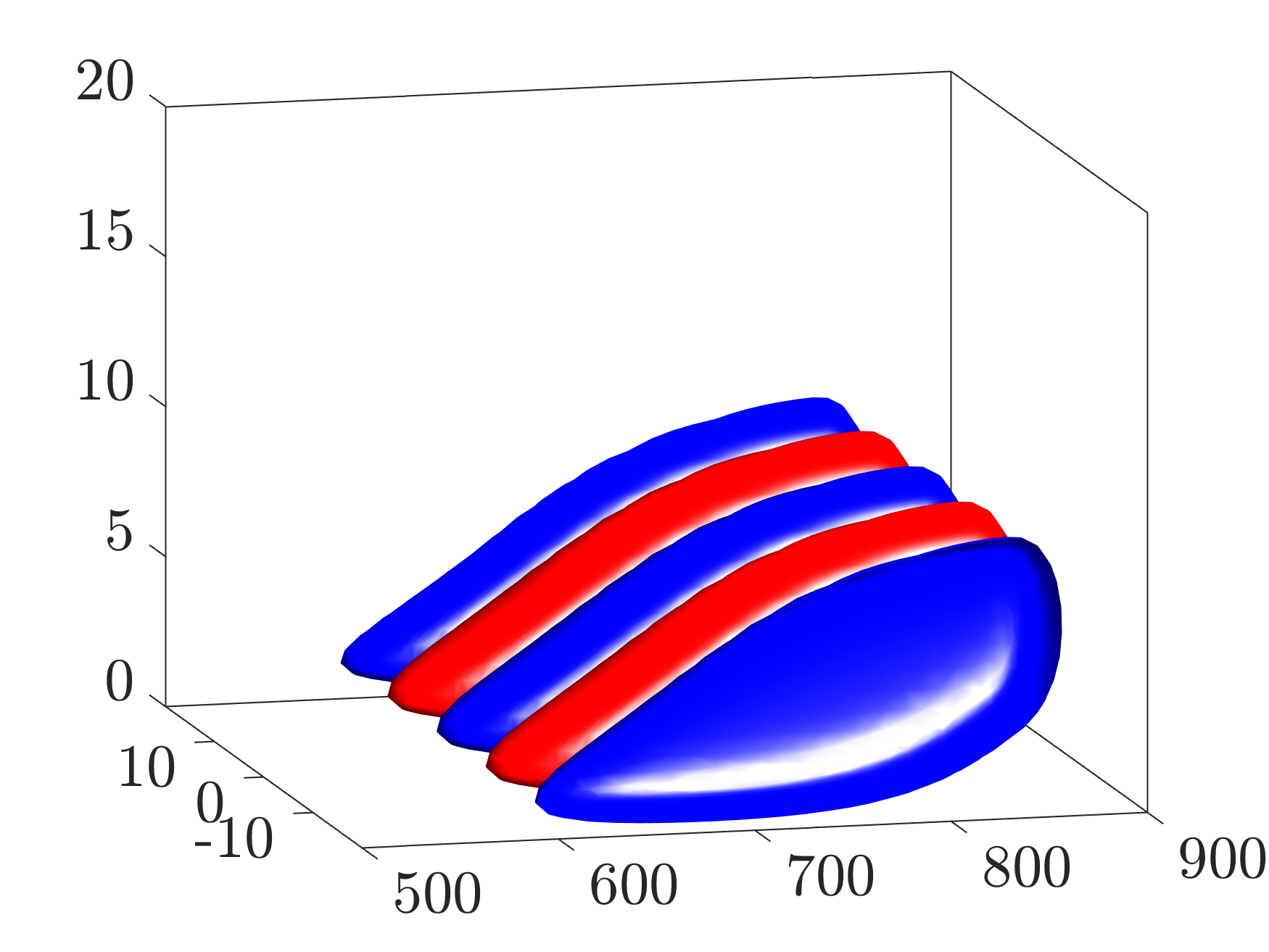}
        \end{tabular}
        &&
        \hspace{-.34cm}
        \begin{tabular}{c}
                \includegraphics[width=0.315\textwidth]{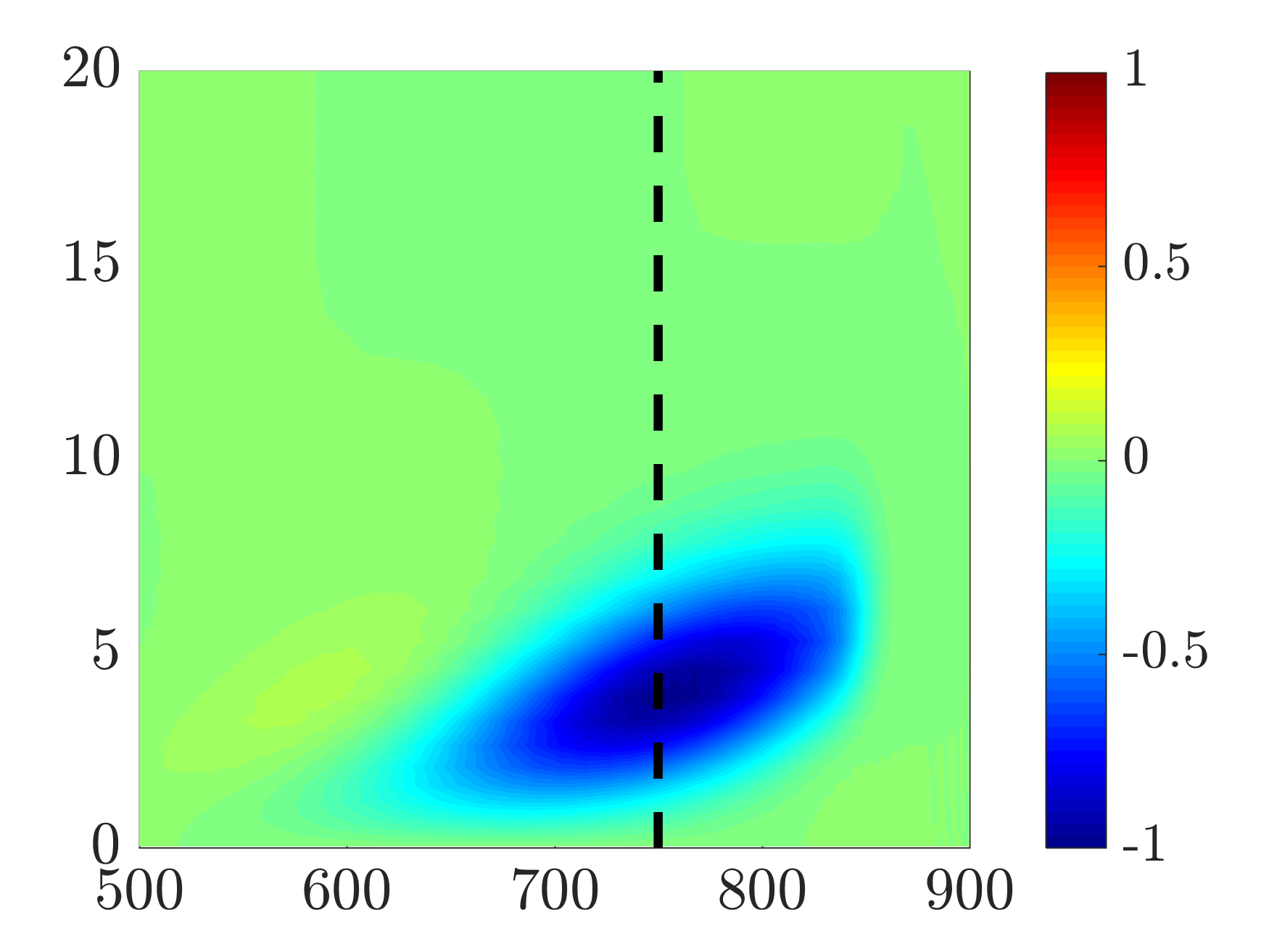}
        \end{tabular}
        &&
        \hspace{-.34cm}
        \begin{tabular}{c}
                \includegraphics[width=0.315\textwidth]{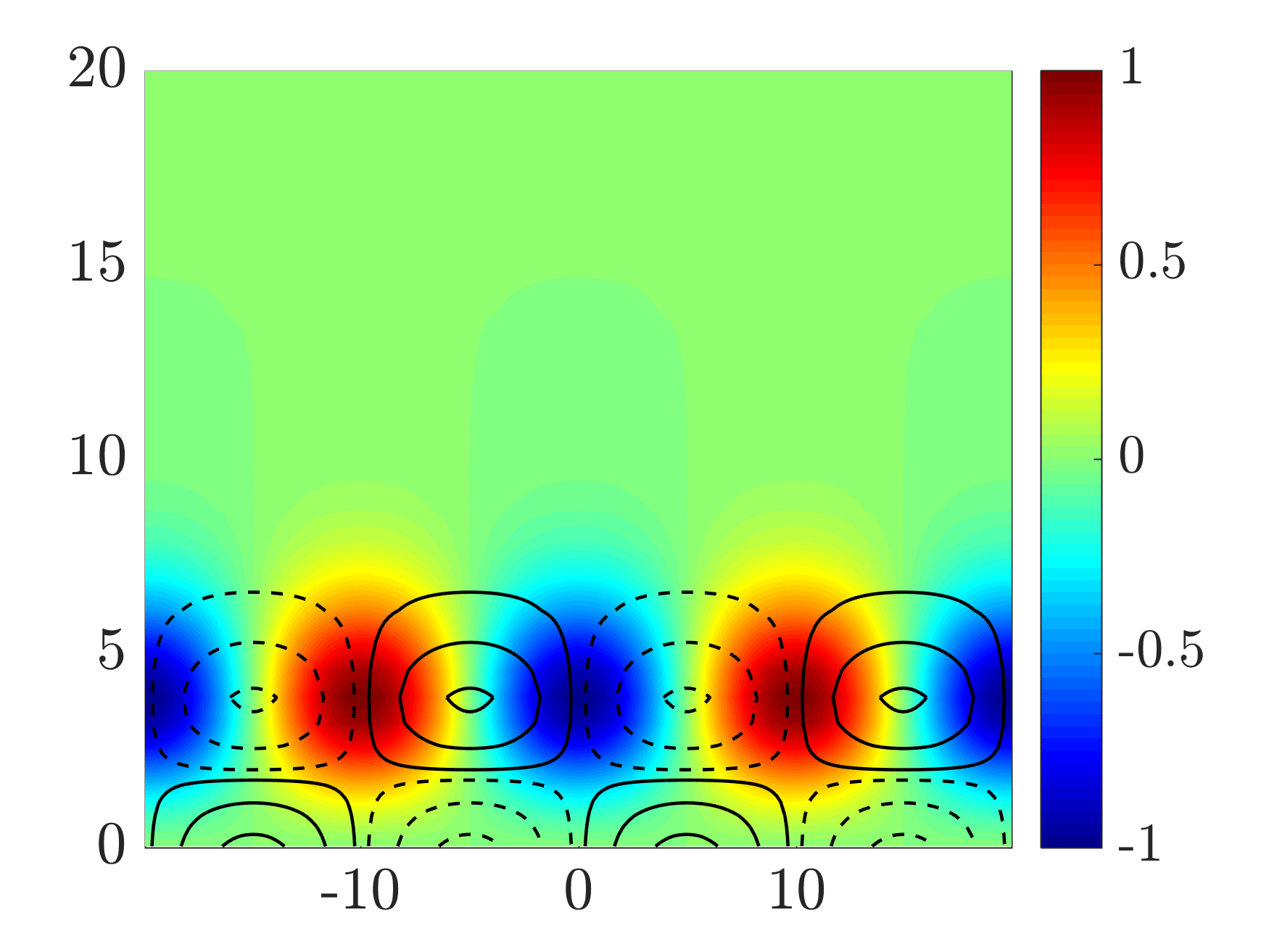}
        \end{tabular}
        \\[-.1cm]
        &
         {\normalsize $x$}
        &&
        \hspace{-.9cm}
        {\normalsize $x$}
        &&
        \hspace{-.6cm}
        {\normalsize $z$}
        \end{tabular}
        \caption{Principal modes with $k_z=0.32$, resulting from {near-wall excitation of the boundary layer flow (case 1 in Table~\ref{tab.forcing-cases}) with $Re_0=232$}. (a) Streamwise velocity component where red and blue colors denote regions of high and low velocity. (b) Streamwise velocity at $z=0$. (c) $y$-$z$ slice of streamwise velocity (color plots) and vorticity (contour lines) at $x=750$, which corresponds to the cross-plane slice indicated by the black dashed lines in (b).}
        \label{fig.streamwisevelocity-vorticity-05}
\end{figure}

\begin{figure}[!ht]
        \begin{tabular}{cccccc}
        \hspace{-.4cm}
        \subfigure[]{\label{fig.streamwisevelocity-3D-15-20}}
        &&
        \hspace{-.65cm}
        \subfigure[]{\label{fig.streamwisevelocity-2D-15-20}}
        &&
        \hspace{-.65cm}
        \subfigure[]{\label{fig.streamwisevelocity-vorticity-15-20}}
        &
        \\[-.2cm]
        \begin{tabular}{c}
		\vspace{1.2cm}
		\\{\normalsize $y$}\\\\\\\\\\\hspace{0.3cm}{\normalsize $z$}\\\\
	    \end{tabular}
        &
        \hspace{-.34cm}
        \begin{tabular}{c}
                \includegraphics[width=0.315\textwidth]{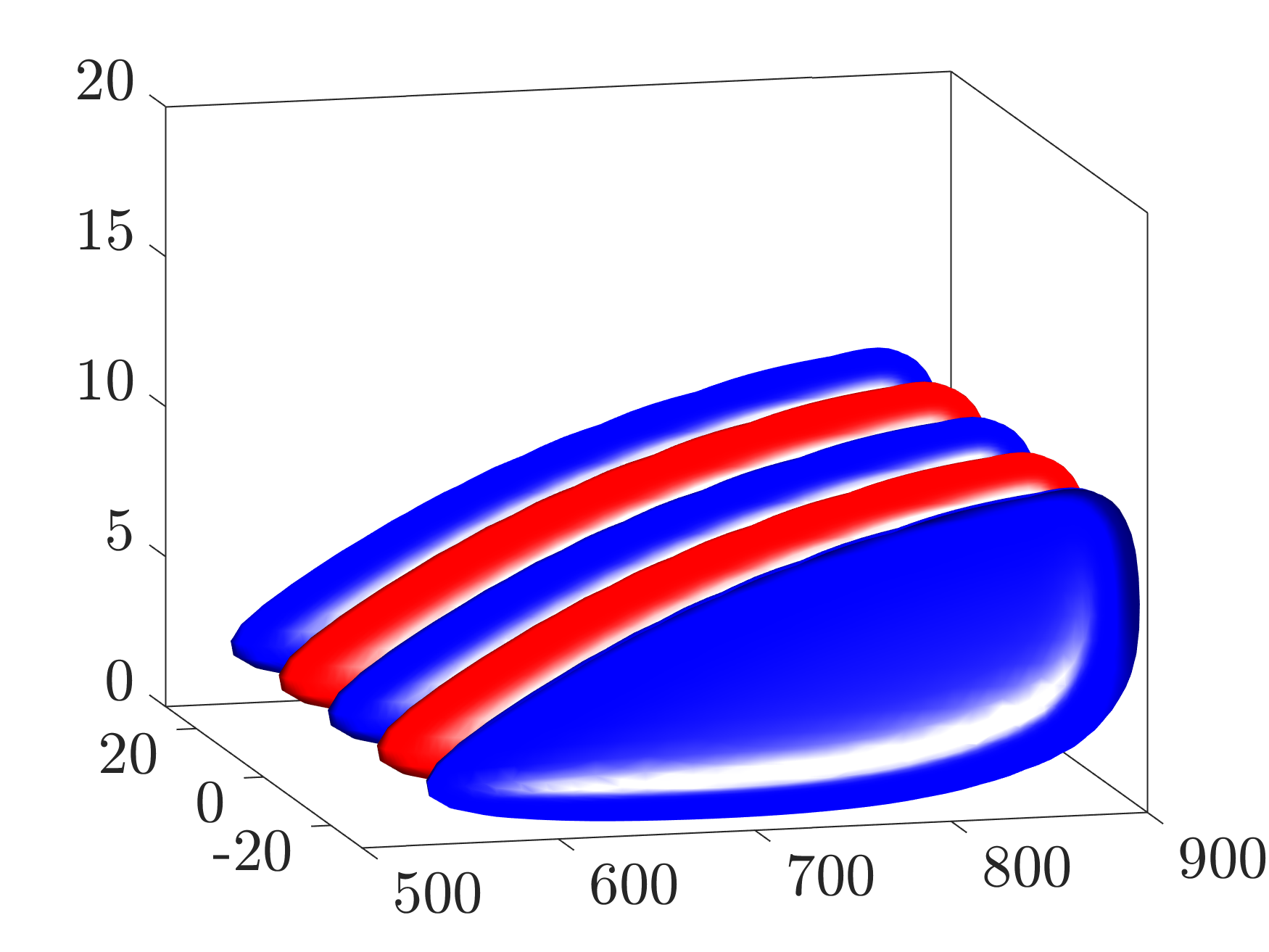}
        \end{tabular}
        &&
        \hspace{-.34cm}
        \begin{tabular}{c}
                \includegraphics[width=0.315\textwidth]{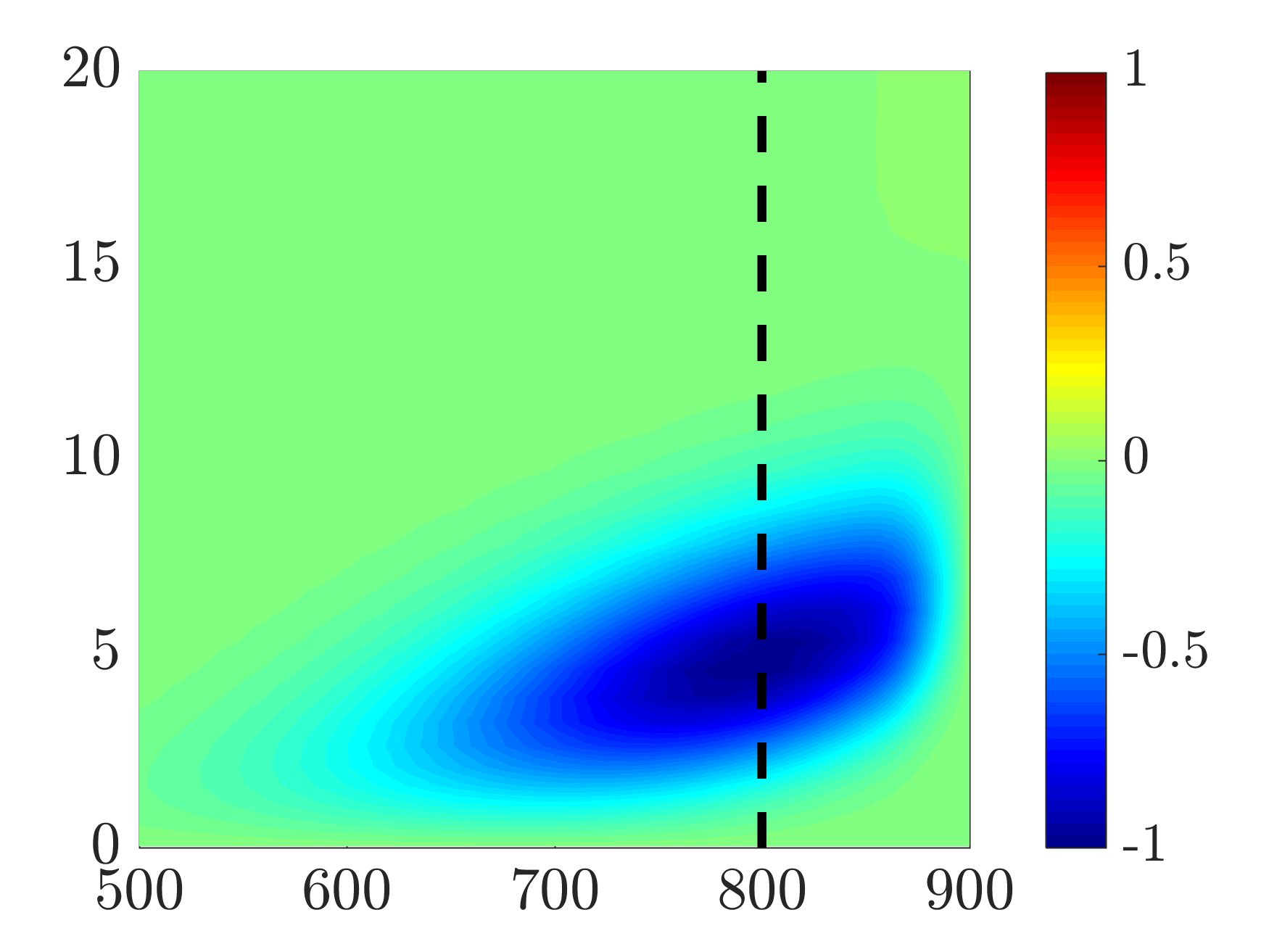}
        \end{tabular}
        &&\hspace{-.34cm}
        \begin{tabular}{c}
                \includegraphics[width=0.315\textwidth]{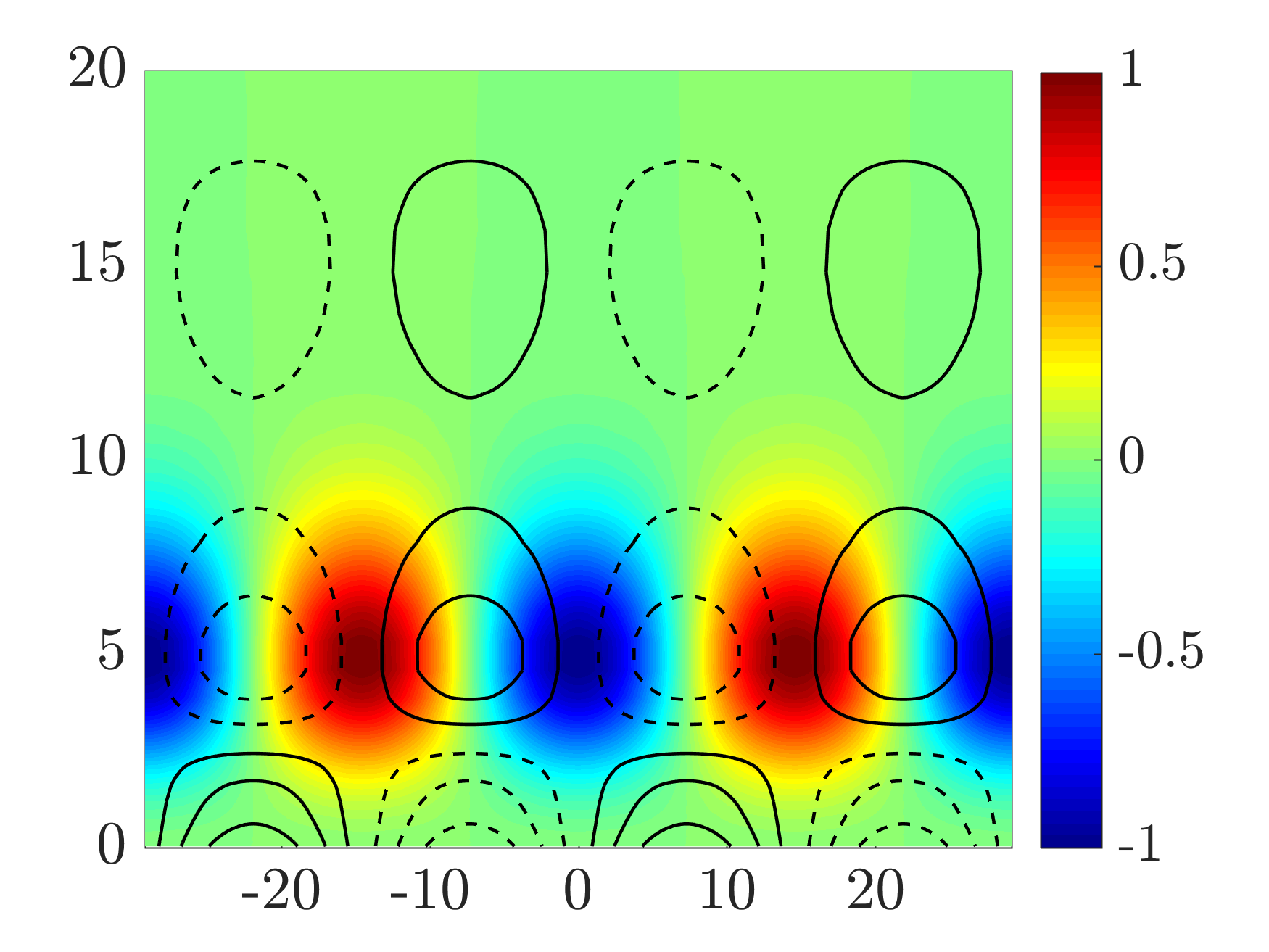}
        \end{tabular}
        \\[-.1cm]
        &
         {\normalsize $x$}
        &&
        \hspace{-.9cm}
        {\normalsize $x$}
        &&
        \hspace{-.6cm}
        {\normalsize $z$}
        \end{tabular}
        \caption{Principal modes with {$k_z=0.21$}, resulting from {outer-layer excitation of the boundary layer flow {(case 4 in Table~\ref{tab.forcing-cases})} with $Re_0=232$}. (a) Streamwise velocity component where red and blue colors denote regions of high and low velocity. (b) Streamwise velocity at $z=0$. (c) $y$-$z$ slice of streamwise velocity (color plots) and vorticity (contour lines) at $x=800$, which corresponds to the cross-plane slice indicated by the black dashed lines in (b).}
        \label{fig.streamwisevelocity-vorticity-1520}
\end{figure}

For $k_z = 0.32$, Fig.~\ref{fig.globalstreakeigspectrum} shows the contribution of the first $50$ eigenvalues of the velocity covariance matrix $\Phi$ {resulting from near-wall and outer-layer stochastic excitation}. In contrast to locally parallel analysis (cf.~Fig.~\ref{fig.parallelstreakeigspectrum}), we observe that other eigenvalues play a more prominent role. {The implication is} that in global analysis the principal eigenmode of $\Phi$ cannot capture the full complexity of the spatially evolving flow. Nevertheless, we examine the shape of such flow structures to gain insight into the effect of stochastic excitation on the eigenmodes {of the covariance matrix $\Phi$} that comprise the fluctuation field. Figures~\ref{fig.streamwisevelocity-3D-0-5} and~\ref{fig.streamwisevelocity-3D-15-20} show the spatial structure of the streamwise component of the principal response to white-in-time stochastic forcing that enters in the vicinity of the wall and {in the outer-layer}, respectively. The streamwise growth of the streaks can be observed. Figures~\ref{fig.streamwisevelocity-2D-0-5} and~\ref{fig.streamwisevelocity-2D-15-20} display the cross-section of these streamwise elongated structures at $z=0$. As the forcing region gets detached from the wall, the cores of the streaky structures also move away from it. As shown in Figs.~\ref{fig.streamwisevelocity-vorticity-0-5} and~\ref{fig.streamwisevelocity-vorticity-15-20}, these streaky structures are situated between counter-rotating vortical motions in the cross-stream plane and they contain alternating regions of fast- and slow-moving fluid that are slightly inclined to the~wall.

\begin{figure}[!ht]
        \begin{tabular}{cccc}
        \hspace{-.4cm}
        \subfigure[]{\label{fig.globalstreak05mode1}}
        &&
        \hspace{-.4cm}
        \subfigure[]{\label{fig.globalstreak05mode2}}
        &
        \\[-.1cm]
        \hspace{-.2cm}
        \begin{tabular}{c}
                \vspace{.4cm}
                \rotatebox{90}{\normalsize $y$}
        \end{tabular}
        &
        \begin{tabular}{c}
                \includegraphics[width=.45\textwidth]{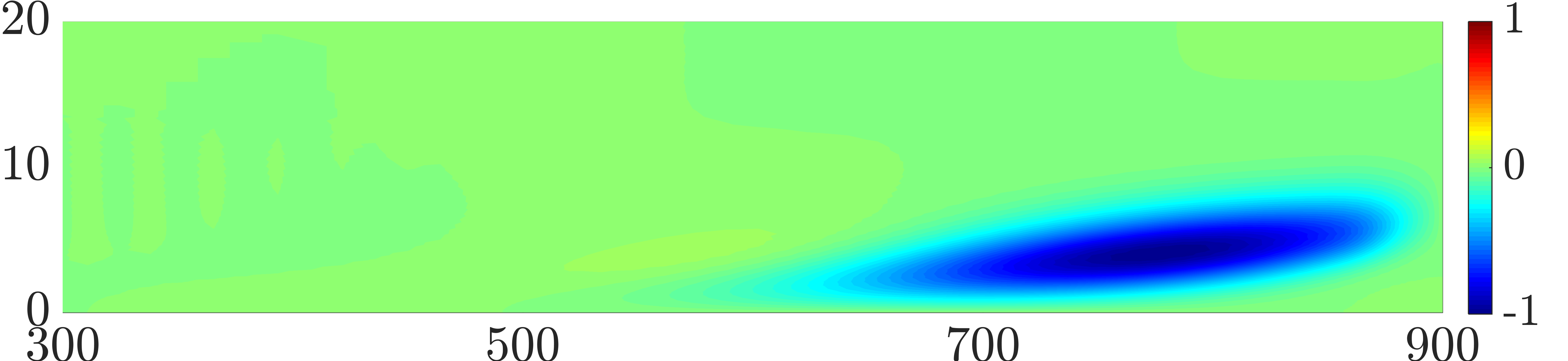}
        \end{tabular}
        &&
        \hspace{.15cm}
        \begin{tabular}{c}
                \includegraphics[width=.45\textwidth]{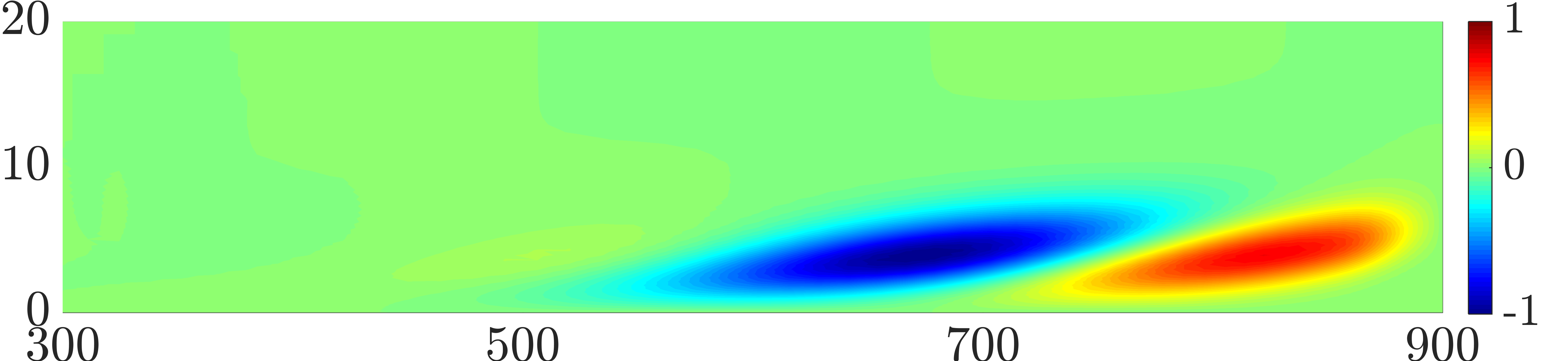}
        \end{tabular}
        \\
        \hspace{-.4cm}
        \subfigure[]{\label{fig.globalstreak05mode3}}
        &&
        \hspace{-.4cm}
        \subfigure[]{\label{fig.globalstreak05mode4}}
        &
        \\[-.1cm]
        \hspace{-.2cm}
        \begin{tabular}{c}
                \vspace{.4cm}
                \rotatebox{90}{\normalsize $y$}
        \end{tabular}
        &
        \begin{tabular}{c}
                \includegraphics[width=.45\textwidth]{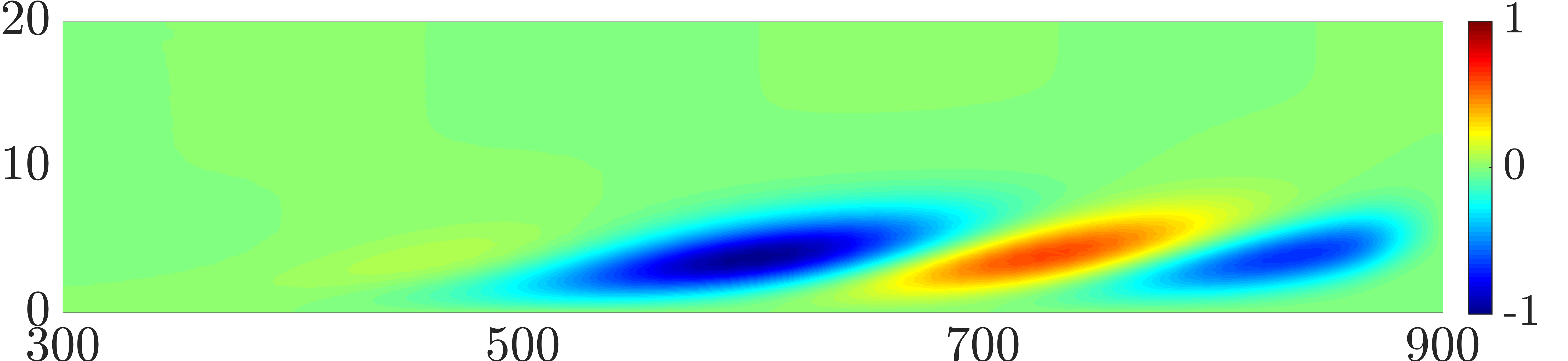}
        \end{tabular}
        &&
        \hspace{.15cm}
        \begin{tabular}{c}
                \includegraphics[width=.45\textwidth]{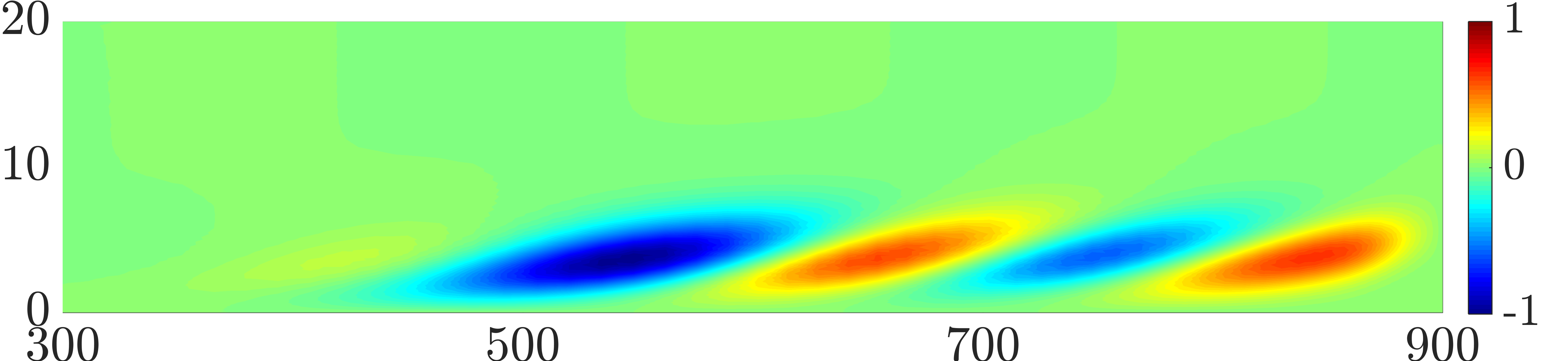}
        \end{tabular}
        \\
        \hspace{-.4cm}
        \subfigure[]{\label{fig.globalstreak05mode5}}
        &&
        \subfigure[]{\label{fig.globalstreak05mode6}}
        &
        \\[-.1cm]
        \hspace{-.2cm}
        \begin{tabular}{c}
                \vspace{.4cm}
                \rotatebox{90}{\normalsize $y$}
        \end{tabular}
        &
        \begin{tabular}{c}
                \includegraphics[width=.46\textwidth]{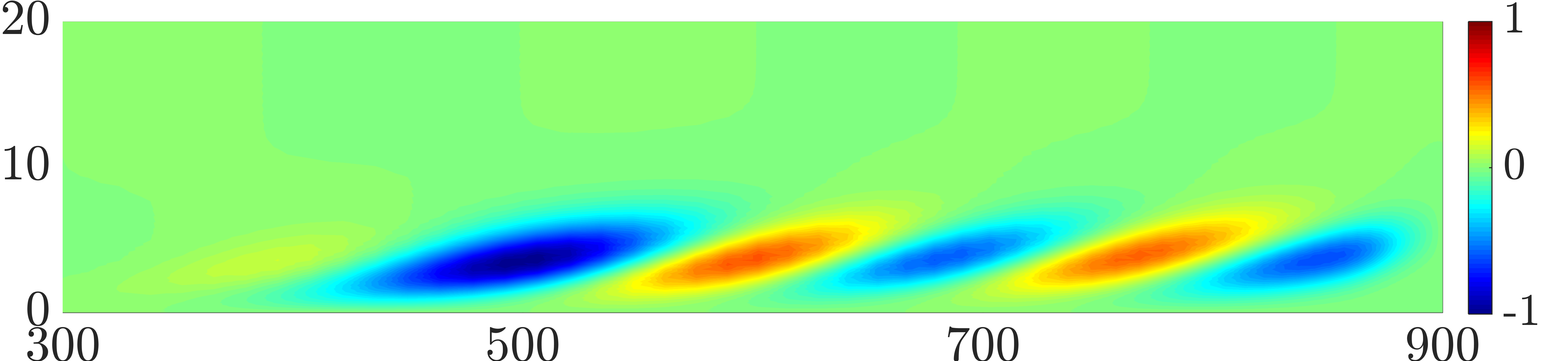}
        \end{tabular}
        &&
        \hspace{.15cm}
        \begin{tabular}{c}
                \includegraphics[width=.46\textwidth]{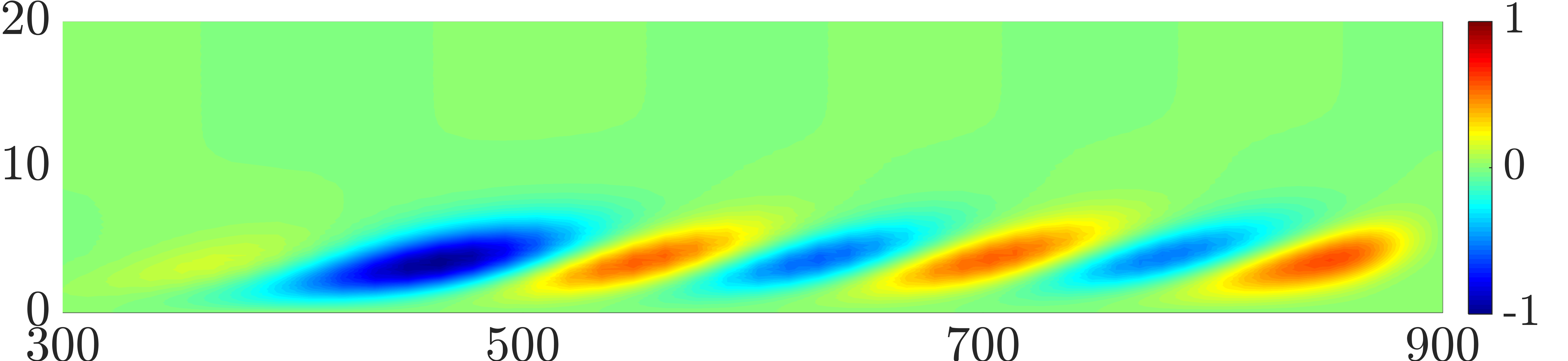}
        \end{tabular}
        \\
        &
        {\normalsize $x$}
        &&
        {\normalsize $x$}
        \end{tabular}
        \caption{Streamwise velocity at $z=0$ corresponding to the first six eigenmodes of the steady-state covariance matrix $\Phi$ resulting from near-wall excitation of the boundary layer flow with $Re_0=232$ and at $k_z=0.32$; (a) $j=1$, (b) $j=2$, (c) $j=3$, (d) $j=4$, (e) $j=5$, and (f) $j=6$ where $j$ corresponds to ordering in Fig.~\ref{fig.globalstreak05eigspectrumbeta6}.}
        \label{fig.globalstreak05modes}
\end{figure}

We next examine the spatial structure of less energetic eigenmodes of $\Phi$. As illustrated in Fig.~\ref{fig.globalstreak05eigspectrumbeta6}, for near-wall stochastic forcing the first six eigenmodes respectively contribute $8.9\%$,  $7.3\%$, $6.1\%$, $5.3\%$, $4.6\%$, and $4.0\%$ to the total energy amplification. We again use the streamwise velocity component to study the spatial structure of the corresponding eigenmodes. As shown in Fig.~\ref{fig.globalstreak05mode2}, while the principal mode consists of a single streamwise-elongated streak, the second mode is comprised of two shorter high- and low-speed streaks. Similarly, the third and fourth modes respectively contain three and four streaks. These streaks become shorter in the streamwise direction and {their energy content reduces}; see Figs.~\ref{fig.globalstreak05mode3} and~\ref{fig.globalstreak05mode4}. As the mode number increases, the streamwise extent of these structures further reduces, they appear at an earlier streamwise location, and their peak value moves closer to the leading edge. This breakup into shorter streaks for higher modes {can be} related to the dominant modes identified in {locally parallel analysis} for increasingly larger streamwise wavenumbers and at various streamwise locations (or Reynolds numbers).

\begin{figure}[!ht]
        \begin{tabular}{cccccc}
        \hspace{-.4cm}
        \subfigure[]{\label{fig.TSsample2D}}
        &&
        \hspace{-.65cm}
        \subfigure[]{\label{fig.TSwave-FFT}}
        &&
        \hspace{-.65cm}
        \subfigure[]{\label{fig.globalh2}}
        &
        \\[-.2cm]
        \begin{tabular}{c}
		\vspace{.6cm}
		\rotatebox{90}{\normalsize $y$}
	    \end{tabular}
	    \hspace{-.25cm}
        &
        \hspace{-.2cm}
        \begin{tabular}{c}
                \includegraphics[width=.315\textwidth]{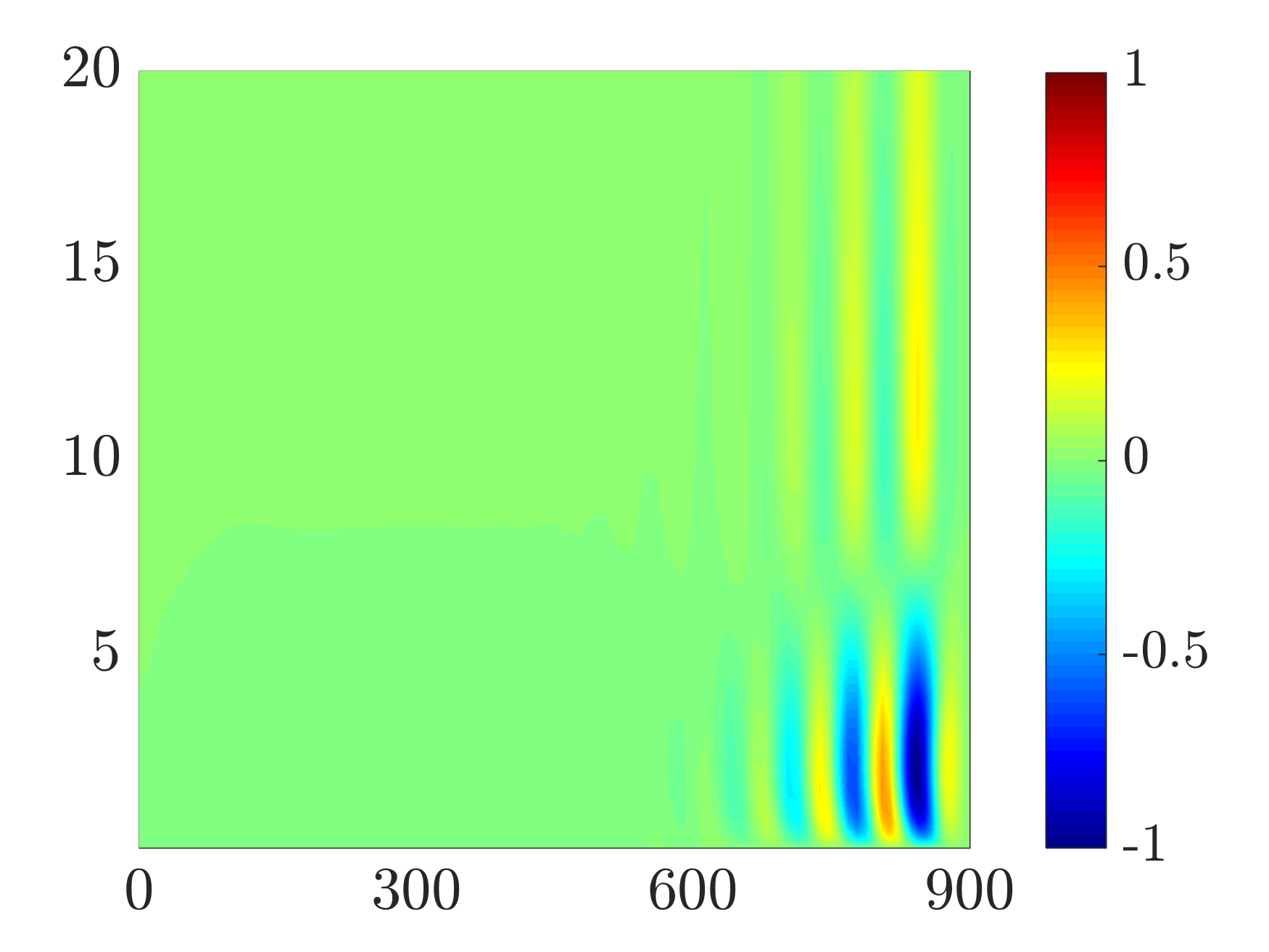}
        \end{tabular}
        &
        \hspace{-.45cm}
        \begin{tabular}{c}
		\vspace{.6cm}
		\rotatebox{90}{\normalsize $y$}
	    \end{tabular}
        \hspace{-0.3cm}
        &
        \hspace{-.2cm}
        \begin{tabular}{c}
                \includegraphics[width=.315\textwidth]{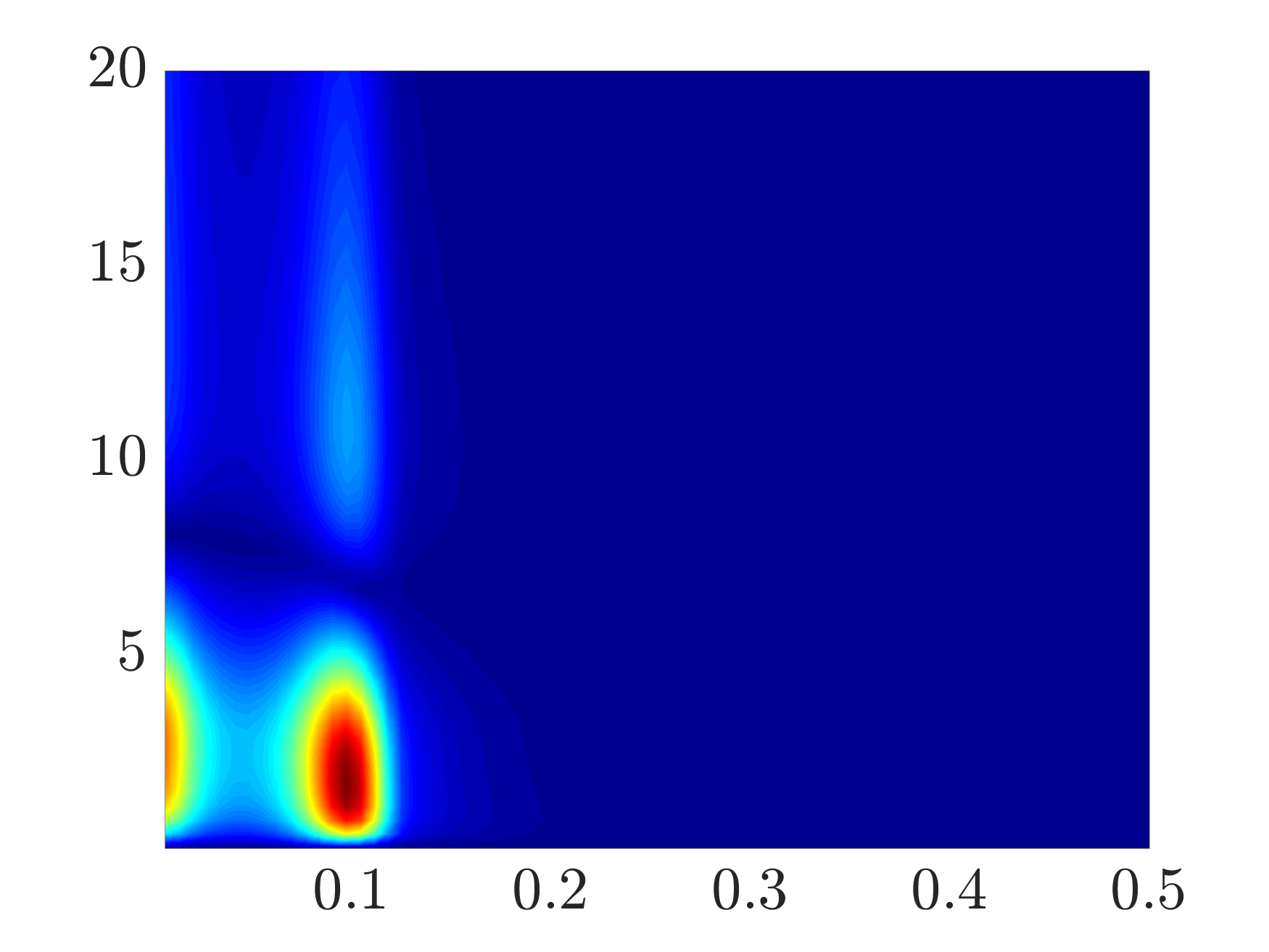}
        \end{tabular}
        &
        \hspace{-.45cm}
        \begin{tabular}{c}
		\vspace{.6cm}
		\rotatebox{90}{\normalsize $k_x$}
	    \end{tabular}
        \hspace{-0.3cm}
        &
        \begin{tabular}{c}
                \includegraphics[width=.315\textwidth]{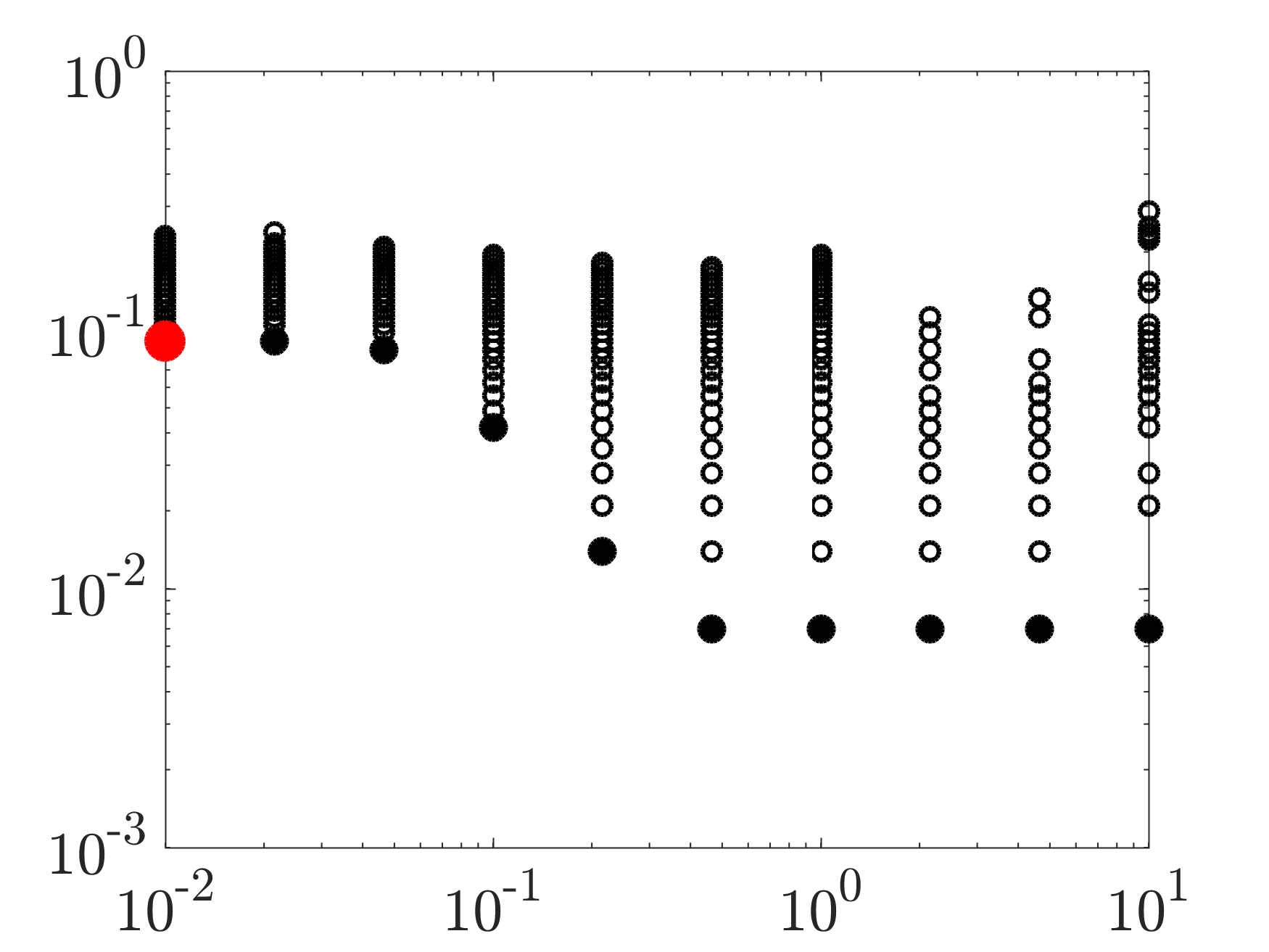}
        \end{tabular}
        \\[-.1cm]
        &
        \hspace{-.9cm}
         {\normalsize $x$}
        &&
        {\normalsize $k_x$}
        &&
        {\normalsize $k_z$}
        \end{tabular}
        \caption{Blasius boundary layer flow with initial Reynolds number $Re_0=232$ subject to white-in-time stochastic excitation of the near-wall region {(case 1 in Table~\ref{tab.forcing-cases})}. (a) The TS wave-like spatial structure of the streamwise velocity component of the principal eigenmode of the matrix $\Phi$ at $k_z=0.01$; {(b) Fourier transform in streamwise dimension}; and (c) the distribution of streamwise length-scales obtained from various eigenmodes of the covariance matrix $\Phi$ for various values of the spanwise wavenumber $k_z$. {Filled} dots represent the {dominant} streamwise wavenumber associated with the principal eigenmode of $\Phi$, {the red dot corresponds to the fundamental wavenumber extracted from (b),} and circles are the streamwise wavenumbers resulting from less significant eigenmodes.}
        \label{fig.h2norm}
\end{figure}

As shown in Fig.~\ref{fig.globalstreak05modes}, spatial visualization of various eigenmodes of $\Phi$ resulting from global receptivity analysis uncovers approximately periodic flow structures in the streamwise direction. The fundamental spatial frequency extracted from the streamwise variation of the principal eigenmode of $\Phi$ provides information about the streamwise length-scales associated with the dominant flow structures. Figure~\ref{fig.TSsample2D} shows the dominant TS wave-like spatial structure that results from near-wall stochastic excitation of the boundary layer flow with $Re_0=232$ and $k_z=0.01$. The Fourier transform in the streamwise direction can be used to extract the fundamental value of $k_x$ associated with this spatial structure. As illustrated in Fig.~\ref{fig.TSwave-FFT}, the Fourier coefficient peaks at $k_x\approx0.1$, which corresponds to the most significant streamwise flow structures (cf. Fig.~\ref{fig.TSsample2D}). The identified fundamental wavenumber is representative of the streamwise variation of this flow structure and it provides a good approximation of the dominant value of $k_x$ that is excited by the near-wall forcing. For different values of $k_z$, the filled black dots in Fig.~\ref{fig.globalh2} denote the streamwise wavenumbers extracted from the principal eigenmodes of the covariance matrix $\Phi$, which contribute most to the energy amplification. The circles represent the tail of streamwise wavenumbers extracted from other eigenmodes of the matrix $\Phi$. As shown in Fig.~\ref{fig.globalstreak05modes}, for any $k_z$, less significant eigenmodes are associated with flow structures that are shorter in the streamwise direction. The observed trends are in close agreement with the results obtained using locally parallel analysis (cf.\ Fig.~\ref{fig.Eturb_pBBL_Re500_fy_0_5}). In particular, streamwise elongated structures are most amplified for $k_z \approx 0.3$. On the other hand, for low spanwise wavenumbers, the TS wave-like structures are most amplified for $k_x \gtrsim 0.1$ (cf.\ $k_x \approx 0.19$ from locally parallel analysis).

	\vspace*{-2ex}
\subsection{Modeling the effect of homogeneous isotropic turbulence}
\label{sec.HIT}
	
So far, we have studied the energy amplification of the boundary layer flow subject to persistent white-in-time stochastic excitation with a trivial covariance matrix ($W=I$). It is also of interest to model the effect of free-stream turbulence on the boundary layer flow using Homogeneous Isotropic Turbulence (HIT)~\cite{braschhen04}. {The spectrum of HIT has been previously used as an initial condition to study transient growth in boundary layer flows based on the temporal evolution of the solution to the differential Lyapunov equation~\cite{hoebra08}. Herein, we consider the persistent stochastic forcing $\bd$ in system~\eqref{eq.lnse1} to be of the form defined in Eq.~\eqref{eq.forcing}. The filter function $h(x) \DefinedAs 10^{-r x/L_x}$ is used to model the streamwise decay of turbulence intensity (cf.~\cite[Fig.\ 2]{braschhen04}) and the spatial covariance matrix $W$ of the forcing term $\bd_s$ is selected to match the spectrum of HIT; see Appendix~\ref{sec.appendix-HIT} for additional details. We utilize such forcing model as well as the input matrix $B$ in the infinite-horizon Lyapunov equation~\eqref{eq.standard_lyap} to compute the steady-state covariance matrix $X$ and determine the corresponding energy spectrum via~Eq.~\eqref{eq.H2norm}.}

\begin{figure}[!ht]
        \begin{tabular}{cccc}
        \subfigure[]{\label{fig.globalHIT}}
        &&
        \subfigure[]{\label{fig.globalHITwhite}}
        &
        \\[-.5cm]
        \hspace{.2cm}
        \begin{tabular}{c}
		\vspace{.6cm}
		\rotatebox{90}{{\normalsize $C_R(k_z)$}}
	    \end{tabular}
        &
        \begin{tabular}{c}
                \includegraphics[width=0.41\textwidth]{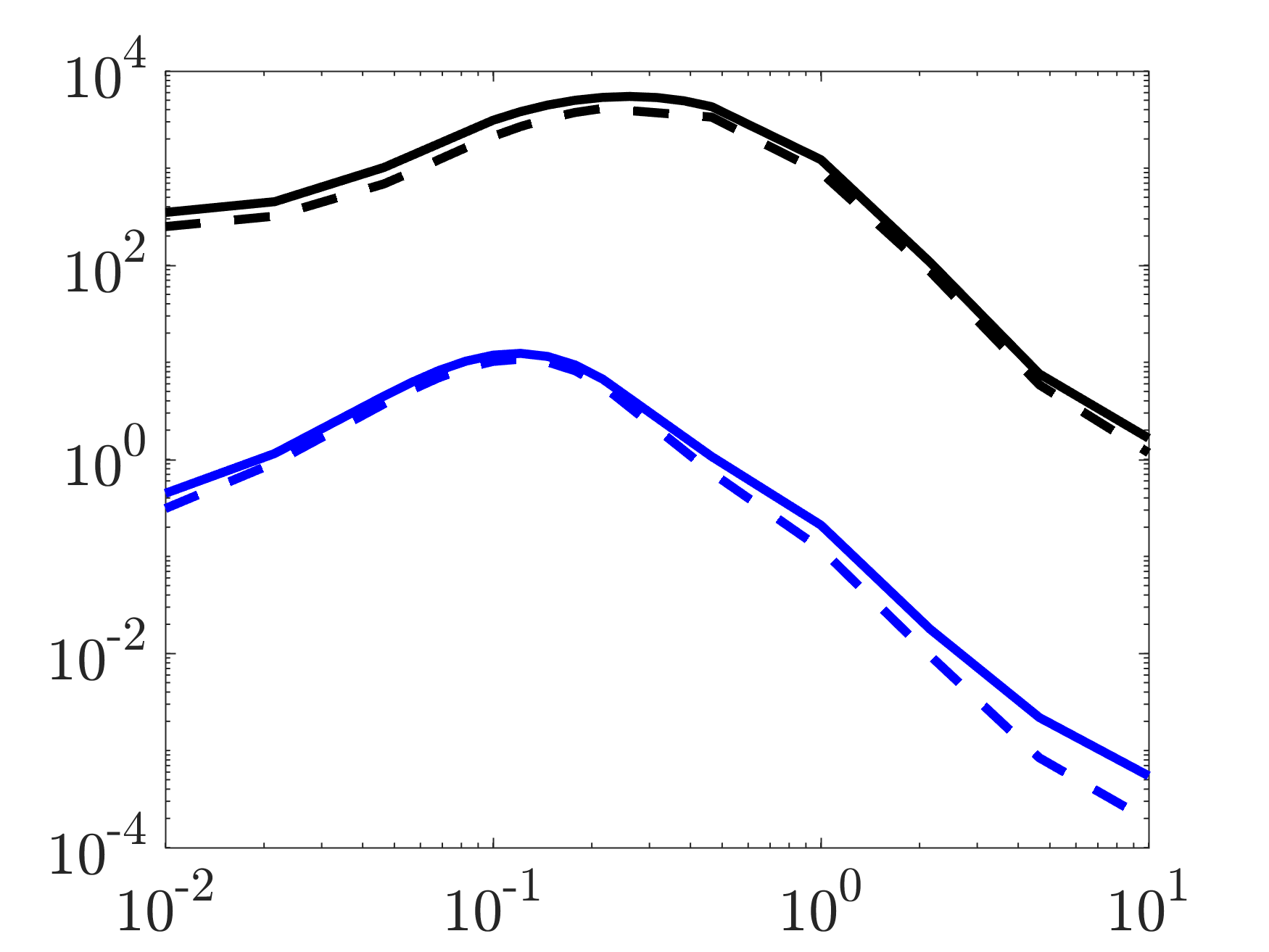}
        \end{tabular}
        &
        \hspace{.2cm}
        \begin{tabular}{c}
                \vspace{.4cm}
                \rotatebox{90}{{\normalsize $C_R(k_z)$}}
        \end{tabular}
        &
        \begin{tabular}{c}
                \includegraphics[width=0.41\textwidth]{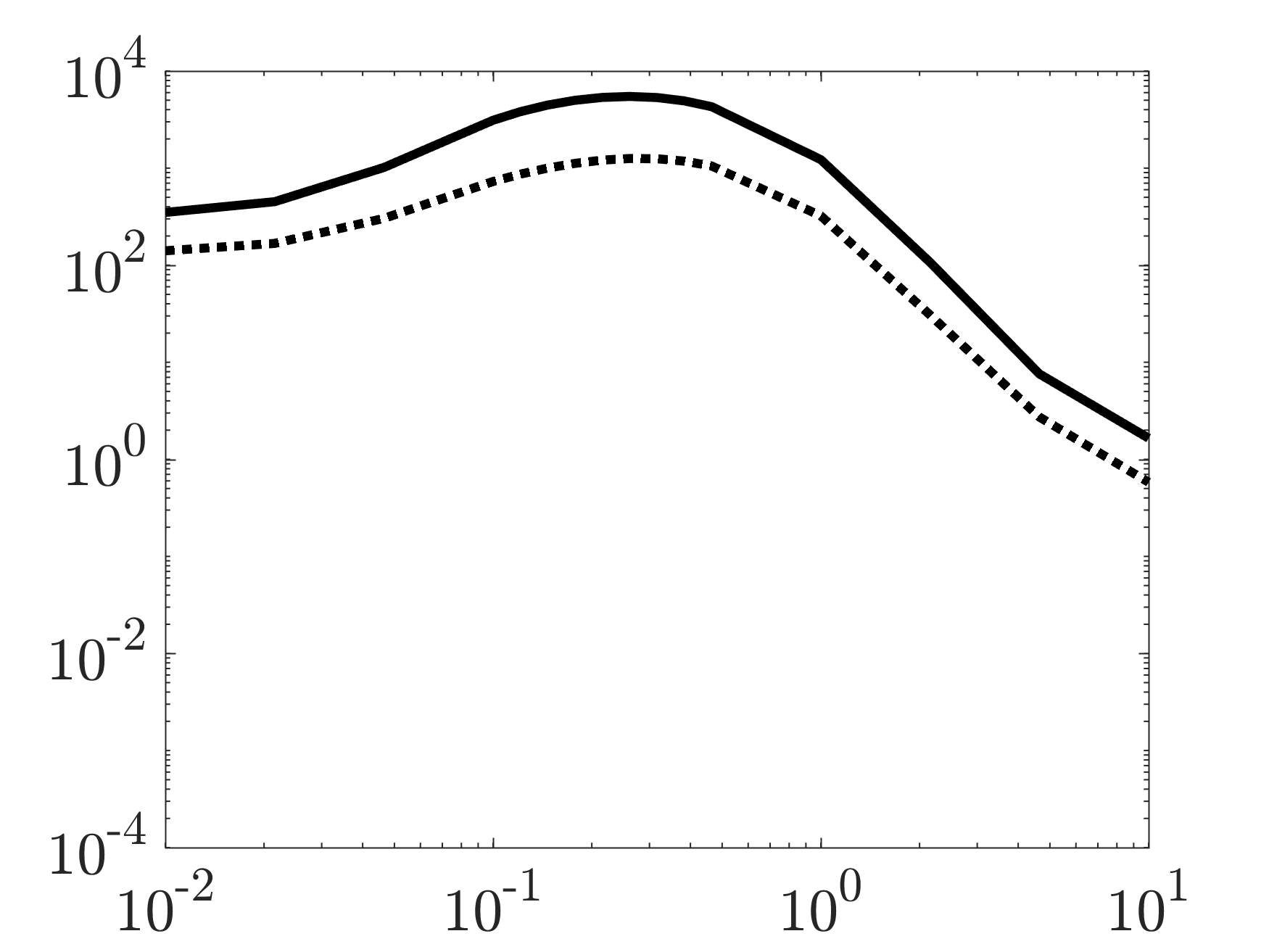}
        \end{tabular}
        \\[-.0cm]
        &
        {\normalsize $k_z$}
        &&
        {\normalsize $k_z$}
        \end{tabular}
        \caption{(a) {Receptivity coefficient resulting from HIT-based stochastic excitation of system~\eqref{eq.lnse1} with $Re_0=232$ in the near-wall (black) and outer-layer (blue) regions. The solid lines correspond to streamwise-invariant} forcing ($r=0$ in $h(x)$) and the dashed lines correspond to streamwise decaying forcing with a decay rate of $r=1.5$ in $h(x)$. (b) {Receptivity coefficient corresponding to the streamwise-invariant ($r=0$ in $h(x)$) HIT-based forcing (solid) and white-in-time forcing with covariance $W=I$ (dotted) entering in the near-wall region.}}
        \label{fig.globalHITwhiteh2vz}
\end{figure}

{We first study the receptivity of the linearized NS equations to HIT-based stochastic forcing. The receptivity coefficient as a function of spanwise wavenumber $k_z$ is shown in Fig.~\ref{fig.globalHIT}. As shown in this figure, the streamwise decay of forcing using the filter function $h(x) = 10^{-r x/L_x}$ has a minimal damping effect on the receptivity coefficient.} Figure~\ref{fig.globalHITwhite} illustrates a {similar trend in the receptivity coefficient} obtained from both types of white-in-time stochastic forcing, which suggests that stochastic forcing with covariance $W=I$ provides a reasonable approximation of the effect of HIT. {However, it is clear that the boundary layer flow is more receptive to the scale-dependent distribution of energy (von K\'{a}rm\'{a}n spectrum) realized by the HIT-based forcing.}

\begin{figure}[!ht]
        \begin{tabular}{cccc}
        \subfigure[]{\label{fig.HITstreaky05xy}}
        &&
        \subfigure[]{\label{fig.DecayHITstreaky05xy}}
        &
        \\[-.2cm]
        \begin{tabular}{c}
                \vspace{.4cm}
                \rotatebox{90}{\normalsize $y$}
        \end{tabular}
        &
        \begin{tabular}{c}
                \includegraphics[width=.46\textwidth]{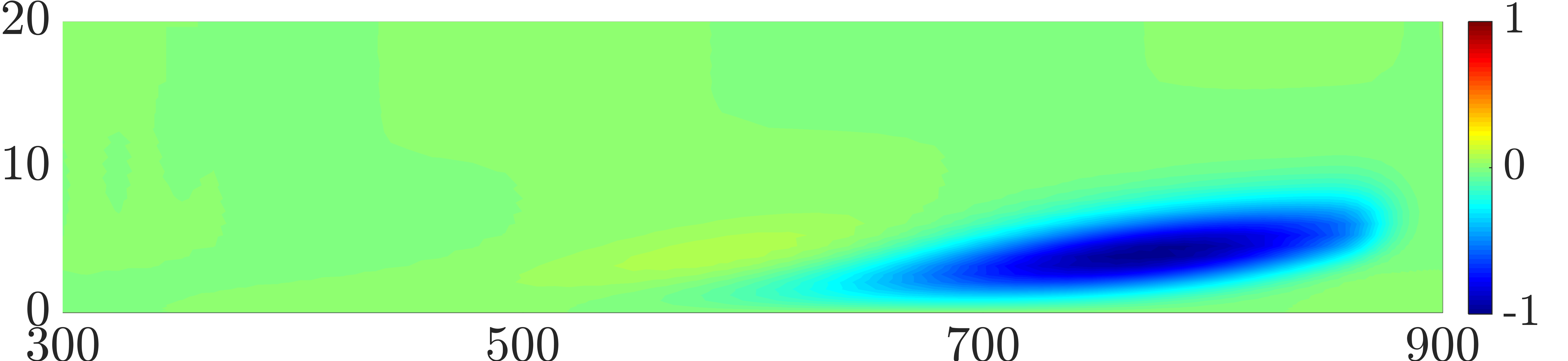}
        \end{tabular}
        &&
        \hspace{.15cm}
        \begin{tabular}{c}
                \includegraphics[width=.46\textwidth]{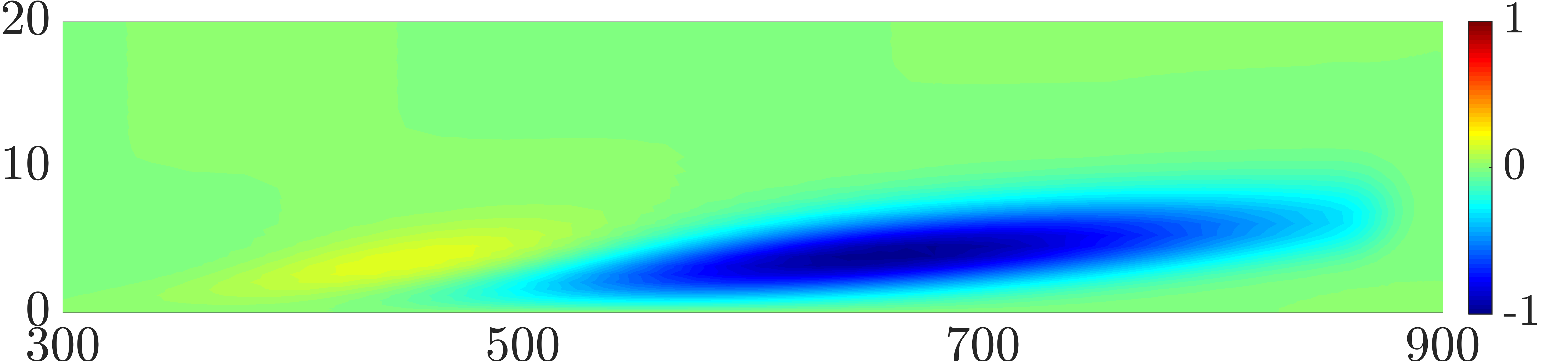}
        \end{tabular}
        \\[.0cm]
        &
        {\normalsize $x$}
        &&
        {\normalsize $x$}
        \end{tabular}
        \caption{The $x$-$y$ slice of the streamwise component of the principal eigenmode from the covariance matrix $\Phi$ at $k_z=0.26$ resulting from near-wall HIT-based excitation of the boundary layer flow with $Re_0=232$. (a) Streamwise-invariant forcing ($r=0$ in $h(x)$); and (b) streamwise-decaying forcing ($r=1.5$ in $h(x)$).}
        \label{fig.HITstreaky05xymodes}
\end{figure}

\begin{figure}[!ht]
	\begin{tabular}{rc}
		\begin{tabular}{c}
		\vspace{.8cm}
		\rotatebox{90}{\normalsize $y$}
		\end{tabular}
		&
		\begin{tabular}{c}
		\includegraphics[width=.46\textwidth]{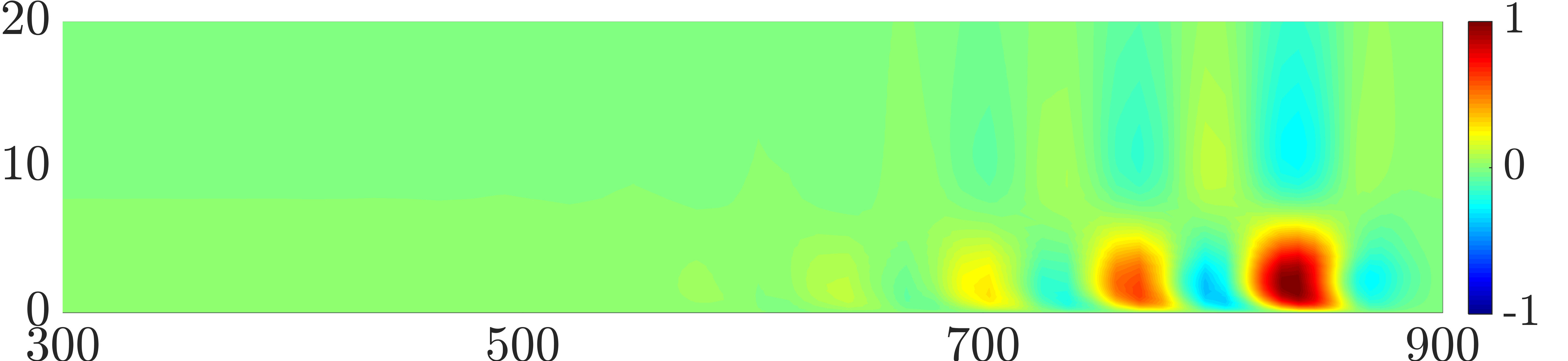}
		\\
		\hspace{-.5cm} {\normalsize  $x$}
		\end{tabular}
	\end{tabular}
	\caption{The TS wave-like spatial structure of the streamwise component of the principal eigenmode of matrix $\Phi$ at $k_z=0.01$ {resulting from global analysis of the boundary layer flow subject to near-wall streamwise-invariant ($r=0$ in $h(x)$) HIT-based stochastic forcing.}}
	\label{fig.TS-HIT}
\end{figure}

Figure~\ref{fig.HITstreaky05xymodes} shows the streamwise component of the {principal} eigenmodes of the velocity covariance matrix $\Phi$ resulting from near-wall HIT-based excitation of the boundary layer flow with $k_z=0.26$. The flow structures closely resemble the streamwise elongated streaks presented in Fig.~\ref{fig.streamwisevelocity-2D-0-5}. From Fig.~\ref{fig.DecayHITstreaky05xy} we conclude that an exponentially decaying excitation further elongates the streaks in the streamwise direction. We note that the amplification of streaks and their prominence in the downstream regions persists, even if the streamwise-decaying forcing completely vanishes towards the end of the domain. Figure~\ref{fig.TS-HIT} shows the dominant flow structure that results from near-wall HIT-based forcing of the boundary layer flow with $k_z = 0.01$. This figure demonstrates that our stochastic analysis is able to predict {the amplification of TS wave-like structures arising from persistent excitation that matches the spectrum of HIT, which is in agreement with the global stability analysis of~\cite{alirob07}. In contrast, similar stochastic analysis of the parallel flow dynamics fails to capture such structures;} see~\cite{linjovACC08} for the predictions resulting from locally parallel analysis.

\begin{figure}
	\begin{centering}
	\vspace{.15cm}
	\begin{tabular}{rc}
		\begin{tabular}{c}
		\vspace{1cm}
		\rotatebox{90}{\normalsize  $u_{\mathrm{rms}}$}
		\end{tabular}
		&
		\hspace{-.3cm}
		\begin{tabular}{c}
		\includegraphics[width=.40\textwidth]{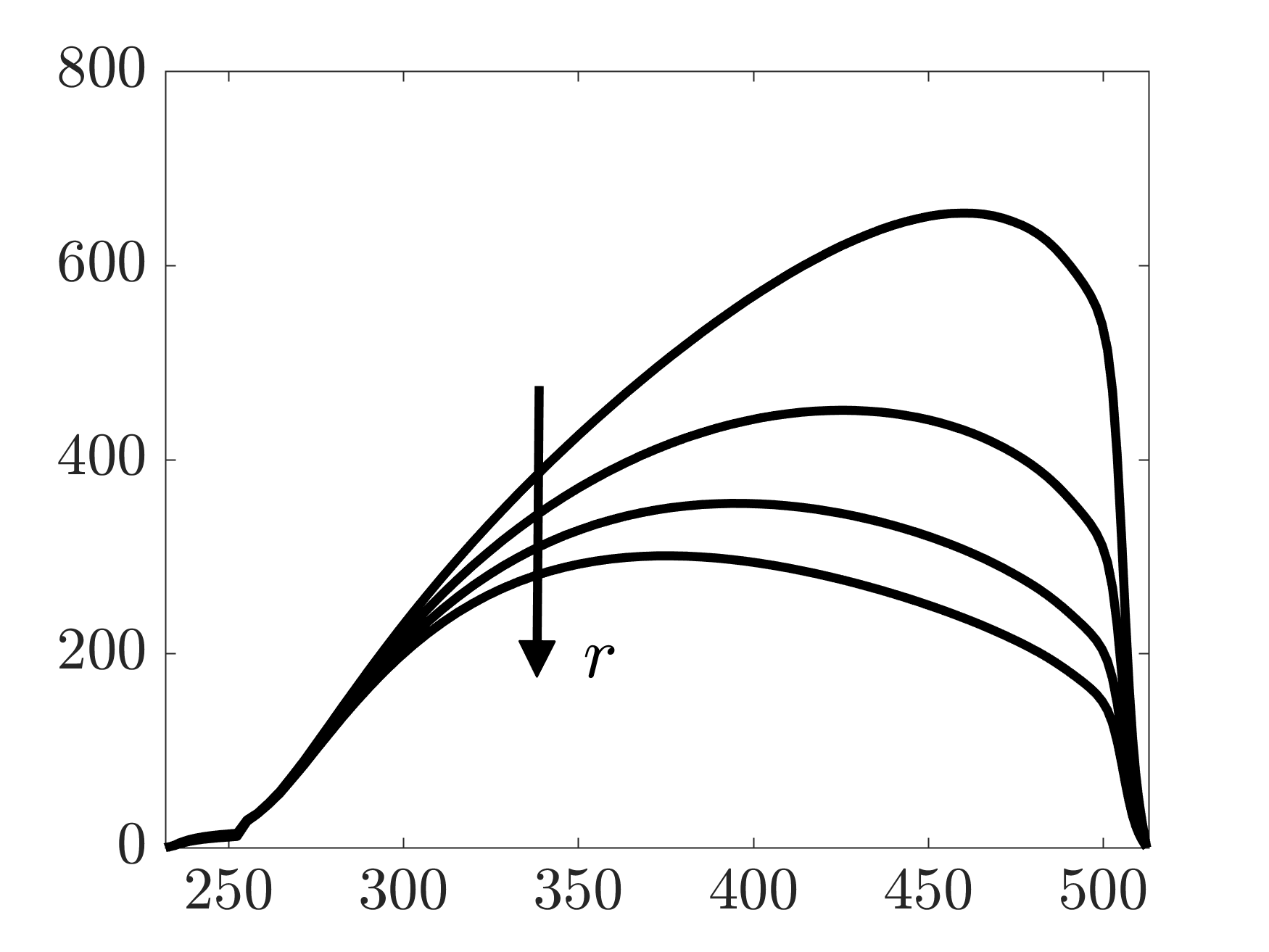}
		\\
		{\normalsize $Re$}
		\end{tabular}
	\end{tabular}
	\caption{The rms amplitude of the streamwise velocity resulting from stochastic excitation that corresponds to the spectrum of HIT entering in the near-wall region3 {(case 1 in Table~\ref{tab.forcing-cases})}. The decay rate for the intensity of stochastic forcing, $r$, increases in the direction of the arrow as $r = 0$, $0.5$, $1$, and $1.5$.}
	\label{fig.RmsvsRex}
	\end{centering}
\end{figure}

Figure~\ref{fig.RmsvsRex} illustrates the growth of the root-mean-square (rms) amplitude of the streamwise velocity resulting from HIT-based stochastic forcing with various streamwise decay rates; $r = 0$, $0.5$, $1$, and $1.5$. This figure is obtained by integrating the steady-state response ($\diag (\Phi)$) over $50$ logarithmically spaced spanwise wavenumbers with $0.01 < k_z < 10$. When the forcing is not damped ($r=0$), the growth is linear and proportional to the Reynolds number {for $Re<400$}, which is in agreement with previous studies based on linear stability theory~\cite{andberhen99,wungol01}. We observe that this linear trend is no longer present {for stochastic forcing with large streamwise decay rates $r$.}

	\vspace*{-2ex}
{
\section{Discussion}
\label{sec.discussion}

In this section, we provide connections between the spatial flow structures obtained via locally parallel and global analyses and examine frequency responses of the boundary layer flow subject to near-wall stochastic excitation.

	\vspace*{-2ex}
\subsection{Relations between locally parallel and global analyses}
\label{sec.comparison-local-global}

The eigenmodes resulting from locally parallel and global stability analysis are closely related~\cite{huemon90,alirob07}. As shown in the previous sections, both locally parallel and global receptivity analyses predict largest amplification of streamwise elongated structures and the appearance of TS waves. However, the size of flow structures and their wall-normal extent can vary with the streamwise location (Reynolds number). For a proper comparison between the streamwise/wall-normal extent of flow structures, herein, we adjust the Reynolds number used in locally parallel analysis to capture the dominant flow structures toward the end of the global streamwise domain. Moreover, a shorter global domain length $L_x$ should be considered to accommodate subcritical Reynolds numbers ($Re \lesssim 360$) beyond which the local dynamics are unstable. To ensure stability of the global dynamics, we extend the streamwise domain in the upstream direction to $Re_0=133$, but for consistency, display results for $Re \geq 232$ after appropriate scaling based on the Blasius length-scale at $Re=232$.

\begin{figure}
        \begin{tabular}{cccc}
        \subfigure[]{\label{fig.Re300y05kz0p32kx0p11}}
        &&
        \subfigure[]{\label{fig.streakmode6strechRe300}}
        &
        \\[-.5cm]
        \hspace{.2cm}
        \begin{tabular}{c}
                \vspace{.4cm}
                \rotatebox{90}{\normalsize $y$}
        \end{tabular}
        &
        \begin{tabular}{c}
                \includegraphics[width=.40\textwidth]{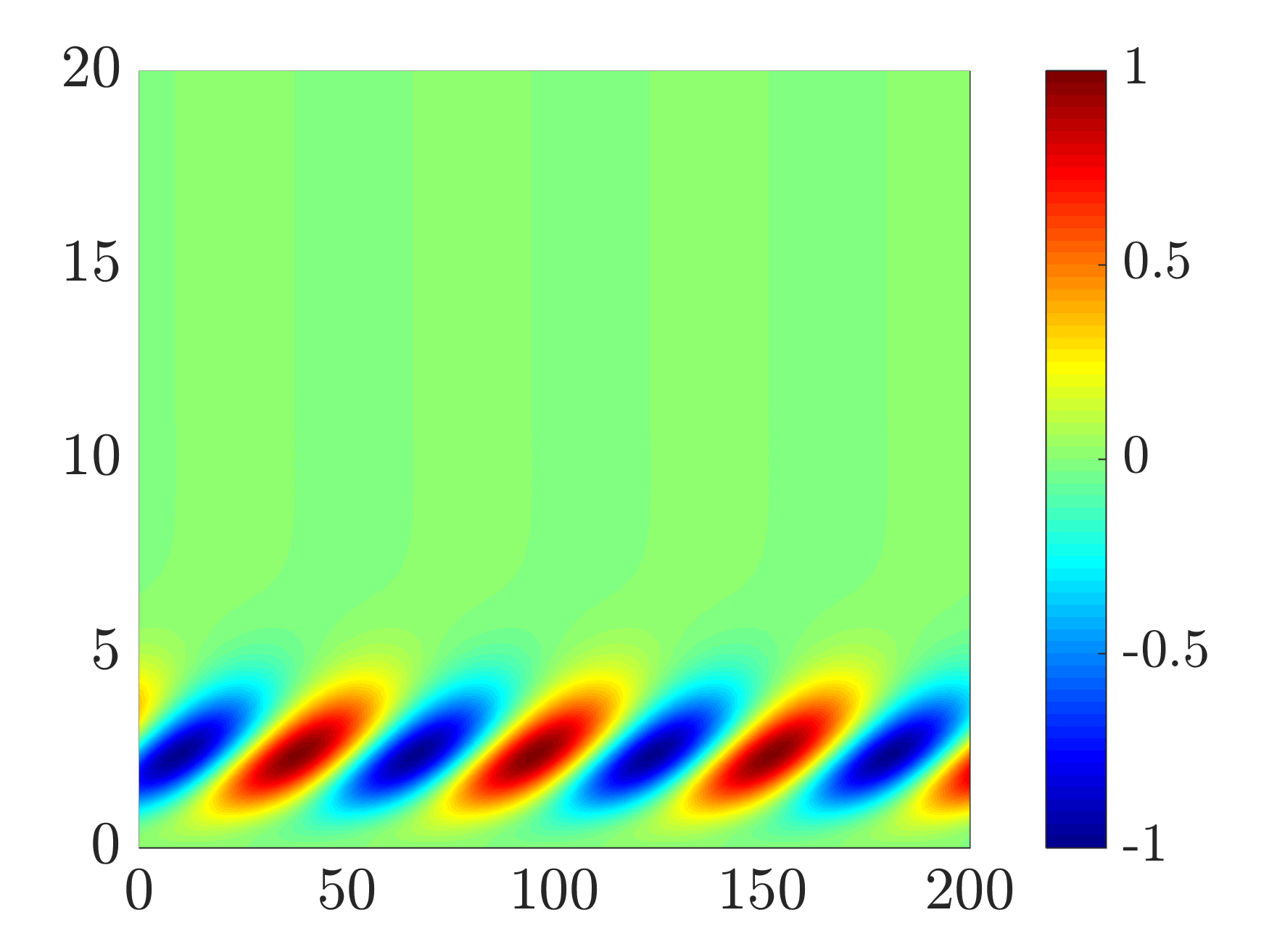}
        \end{tabular}
        &
        \hspace{.2cm}
        \begin{tabular}{c}
                \vspace{.4cm}
                \rotatebox{90}{\normalsize $y$}
        \end{tabular}
        &
        \begin{tabular}{c}
                \includegraphics[width=.40\textwidth]{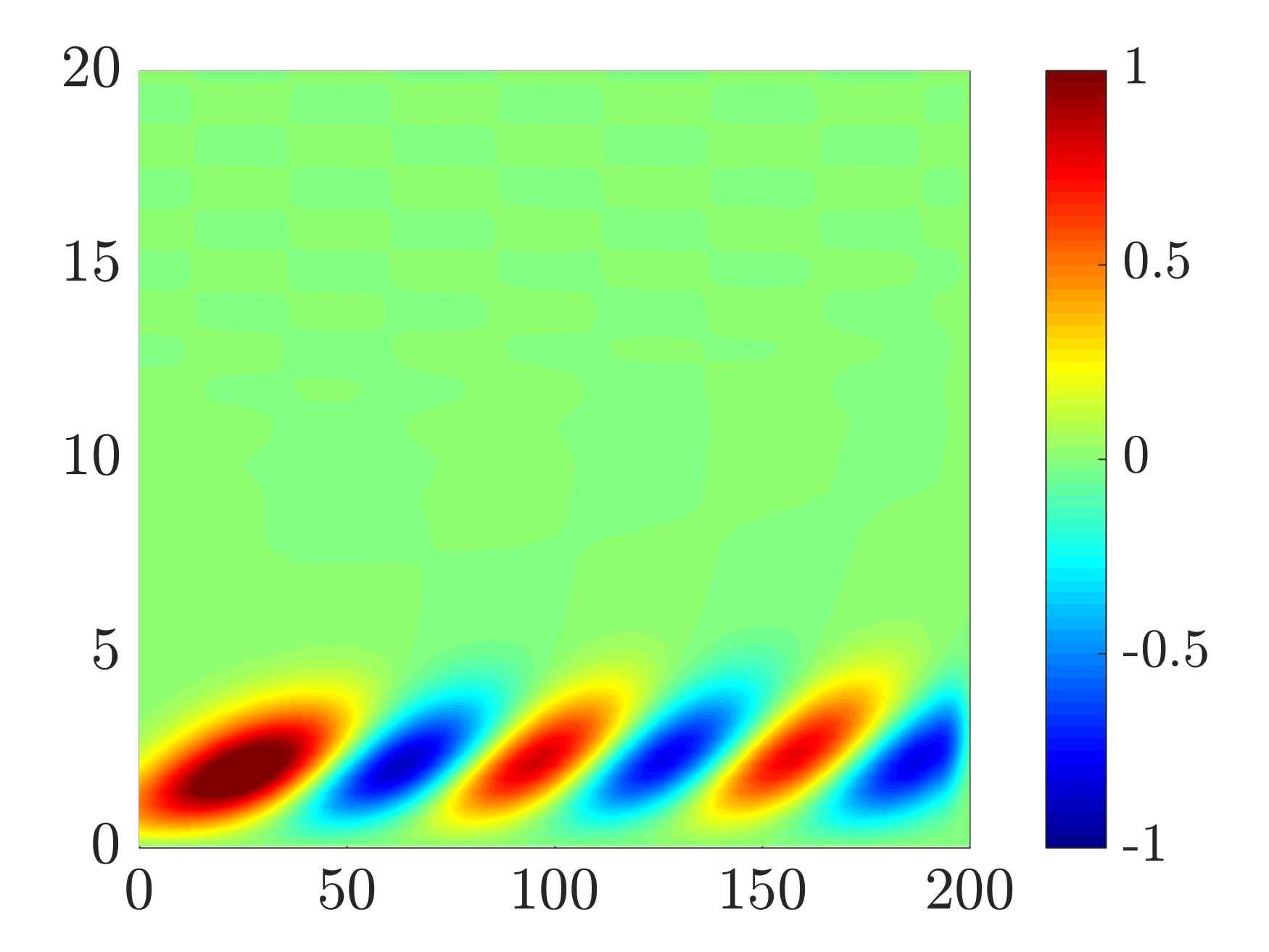}
        \end{tabular}
        \\[-.1cm]
        &
        \hspace{-.9cm}
        {\normalsize $x$}
        &&
        \hspace{-.9cm}
        {\normalsize $x$}
        \end{tabular}
        \caption{{Streamwise velocity fluctuations resulting from near-wall stochastic excitation of the boundary layer flow. (a) Principal eigenmode of $\Phi$ obtained in locally parallel analysis with $Re=300$ and $(k_x,k_z)=(0.11, 0.32)$; and (b) $6$th eigenmode of $\Phi$ resulting from global analysis with $k_z=0.32$. In the global computations $L_x=200$ and the dominant flow structures appear at $Re\approx300$.}}
        \label{fig.streakcompare}
\end{figure}

For near-wall stochastic excitation (case 1 in Table~\ref{tab.forcing-cases}), both locally parallel and global receptivity analyses predict the dominant amplification of streamwise elongated structures with $k_z\approx 0.3$; see Figs.~\ref{fig.parallelh2vz} and~\ref{fig.globalh2vz}. For near-wall excitations with $k_z=0.32$, Fig.~\ref{fig.streakcompare} shows that locally parallel analysis of the flow with $Re=300$ subject to near-wall excitation yieds similar flow structures (with $k_x=0.11$) to those appearing at $Re\approx300$ in the $6$th eigenmode of the covariance matrix $\Phi$ resulting from global analysis. Here, $k_x=0.11$ is the wavenumber extracted from spatial Fourier transform of the $6$th eigenmode of $\Phi$. Moreover, for long spanwise wavelengths, both models predict the amplification of similar TS wave-like structures in the presence of near-wall excitation (see Fig.~\ref{fig.LyapTSmodes}).

\begin{figure}[!ht]
        \begin{tabular}{cccc}
        \subfigure[]{\label{fig.LyapTSur1-parallel}}
        &&
        \subfigure[]{\label{fig.LyapTSur1-global}}
        &
        \\[-.5cm]
        \hspace{.2cm}
        \begin{tabular}{c}
                \vspace{.4cm}
                \rotatebox{90}{\normalsize $y$}
        \end{tabular}
        &
        \begin{tabular}{c}
                \includegraphics[width=.40\textwidth]{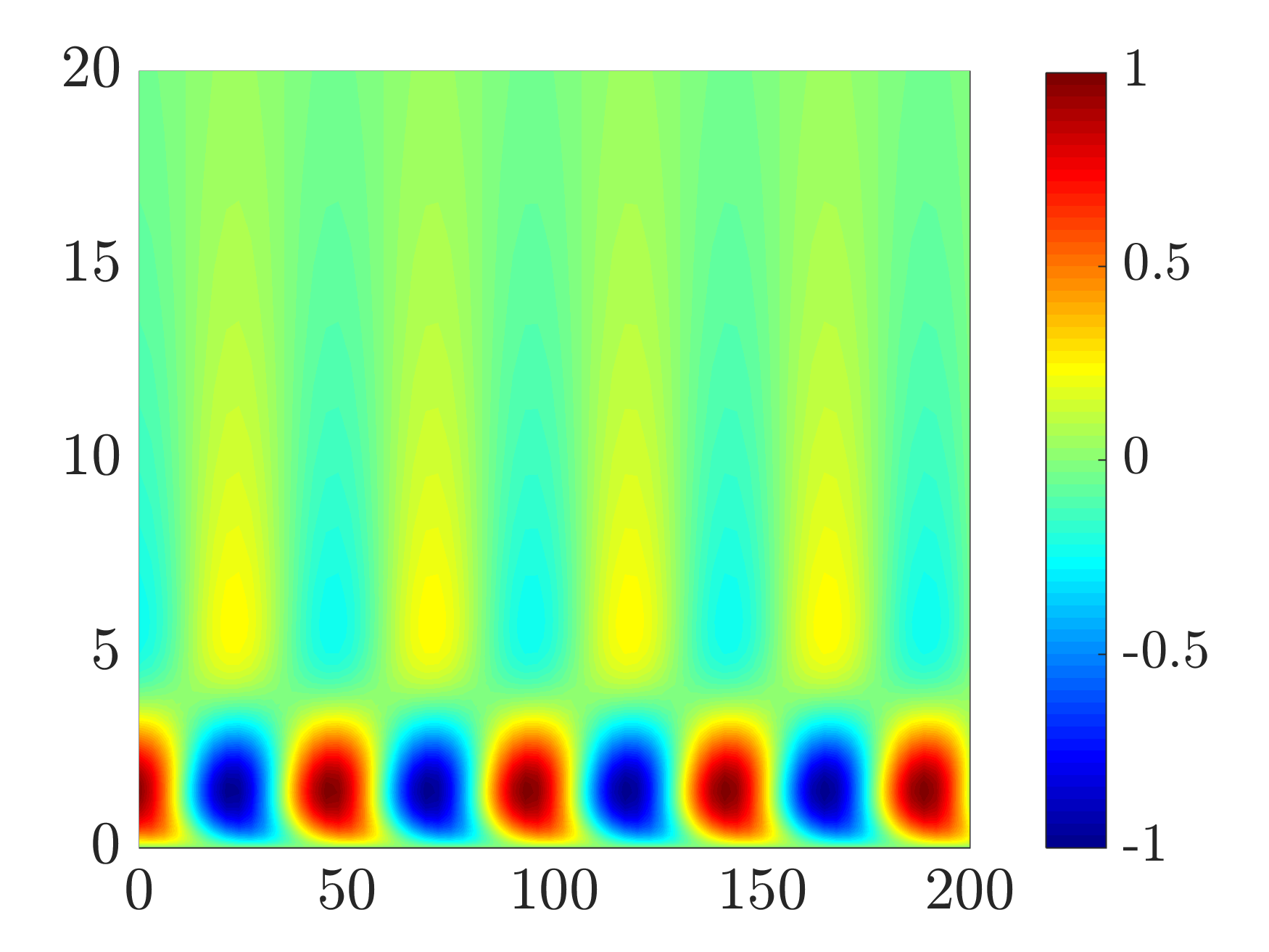}
        \end{tabular}
        &
        \hspace{.2cm}
        \begin{tabular}{c}
                \vspace{.4cm}
                \rotatebox{90}{\normalsize $y$}
        \end{tabular}
        &
        \begin{tabular}{c}
                \includegraphics[width=.40\textwidth]{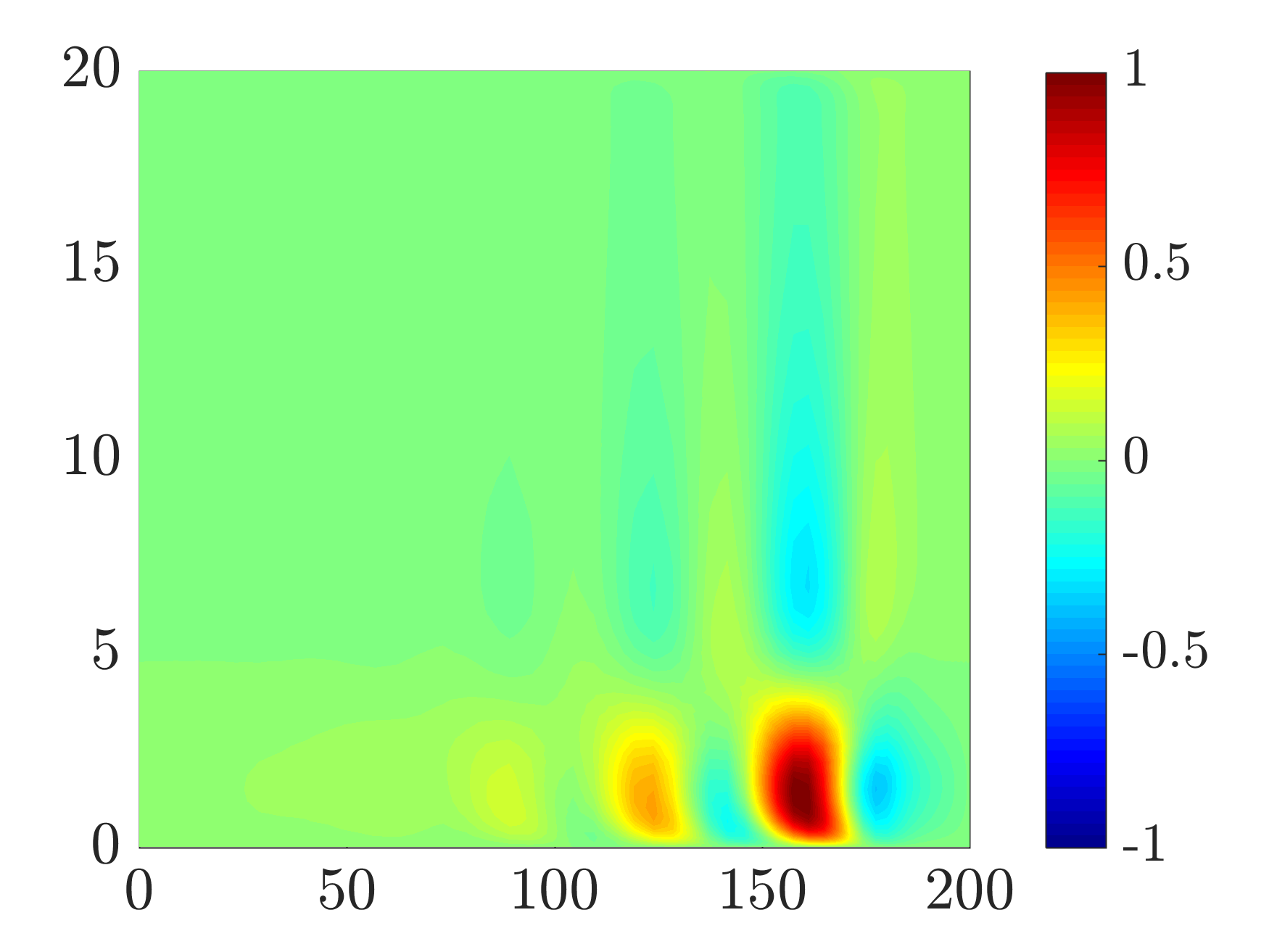}
        \end{tabular}
        \\[-.1cm]
        &
        \hspace{-.9cm}
        {\normalsize $x$}
        &&
        \hspace{-.9cm}
        {\normalsize $x$}
        \end{tabular}
        \caption{The TS wave-like spatial structure of the streamwise velocity component of the principal eigenmode of the matrix $\Phi$ resulting from near-wall stochastic excitation of the boundary layer flow. (a) {Locally parallel} analysis with $Re=300$ and $(k_x,k_z)=(0.13, 0.01)$; and (b) global flow analysis with $k_z=0.01$. The wavenumber pair for the {locally parallel} analysis corresponds to the TS wave branch in the energy spectrum of velocity fluctuations. In the global computations $L_x=200$ and the dominant flow structures appear at $Re\approx300$.}
        \label{fig.LyapTSmodes}
\end{figure}

In certain scenarios, locally parallel analysis can extract information about streamwise scales that may be hidden in global analysis. This feature of locally parallel analysis can be attributed to the parameterization of the velocity field over streamwise wavenumbers, which enables the separate study of various streamwise length-scales. For example, for wavenumbers at which the global receptivity analysis of the flow subject to outer-layer excitation is dominated by near-wall streaks, locally parallel analysis can uncover the trace of weakly growing outer-layer oscillations at TS frequencies. This is in agreement with experiments~\cite{ken98} which observe outer-layer oscillations of comparable length to width ($k_x\approx k_z$) that travel at the phase speed of free-stream velocity with similar temporal frequency as TS waves.

To further investigate this observation, we re-examine the flow structures that can be extracted from locally parallel and global flow analyses of the boundary layer flow at $Re\approx300$ subject to stochastic excitation covering the entire free stream region. In particular, the parameters in Eq.~\eqref{eq.fy} are set to $y_1=7$, $y_2 = 33$, and $a=10$ for locally parallel analysis, and $y_1=\delta_{0.99}+2$, $y_2 = 33$, and $a=10$ for global flow analysis. Note that $\delta_{0.99}$ in the global analysis is a function of $x$. By comparing the phase speed of the outer-layer oscillations to that of TS waves ($c \approx 0.4\, U_\infty$ obtained from local temporal stability analysis with $Re_0=232$ and $k_x\approx0.19$) we obtain $\omega\approx0.076$. Finally, Taylor's hypothesis ($c \approx U_\infty$) can be used to obtain $k_x \approx 0.076$ for outer-layer oscillations.

Figures~\ref{fig.Re300outlayerdXuparallel} and~\ref{fig.Re300outlayerdXuglobal} show the streamwise component of the steady-state response of the boundary layer flow with $Re=300$ and $k_z = 0.076$ resulting from locally parallel and global flow analyses, respectively. As aforementioned, locally parallel analysis considers $k_x= k_z=0.076$, which is in concert with the experimentally observed outer-layer oscillations. These flow structures represent the aggregate contribution of all eigenmodes of $\Phi$ and they have been obtained from $\diag \, ( C_u X C_u^*)$, where $C_u$ is the streamwise component of the output matrix $C$. Note that the spatial structure shown in Fig.~\ref{fig.Re300outlayerdXuparallel} is obtained by enforcing streamwise periodicity with $k_x=0.076$. While locally parallel analysis of the stochastically forced flow predicts the amplification of structures that reside in the outer-layer, the response obtained in global analysis is dominated by inner-layer streaks and a weaker amplification of outer-layer fluctuations is observed in the presence of stochastic forcing. As shown in Fig.~\ref{fig.Re300outlayerglobalmodes}, such weak outer-layer oscillations can be observed in the $7$th mode of the covariance matrix $\Phi$ resulting from global analysis. Figure~\ref{fig.Re300outlayeroscg7maxy} shows the streamwise variation of these flow structures at $y=20$, which corresponds to the wall-normal location where the largest amplitude occurs. The streamwise wavelength of this signal is approximately the same as the parallel flow estimate ($\lambda_x = 81$ vs $\lambda_x = 82.7$). Such flow structures may be dominated by higher amplitude streaks as their contribution to the total energy amplification is much smaller than the contribution of the principal mode ($0.15\%$ vs $1.9\%$). Nonetheless, similar to the cascade shown in Fig.~\ref{fig.globalstreak05mode6}, their presence in the eigenmodes of the covariance matrix points to the physical relevance of flow structures that are identified via locally parallel analysis.

\begin{figure}
        \begin{tabular}{cccc}
        \subfigure[]{\label{fig.Re300outlayerdXuparallel}}
        &&
        \subfigure[]{\label{fig.Re300outlayerdXuglobal}}
        &
        \\[-.5cm]
        \begin{tabular}{c}
		\vspace{.4cm}
		\rotatebox{90}{\normalsize $y$}
	    \end{tabular}
        \hspace{-.3cm}
        &
        \begin{tabular}{c}
                \includegraphics[width=.40\textwidth]{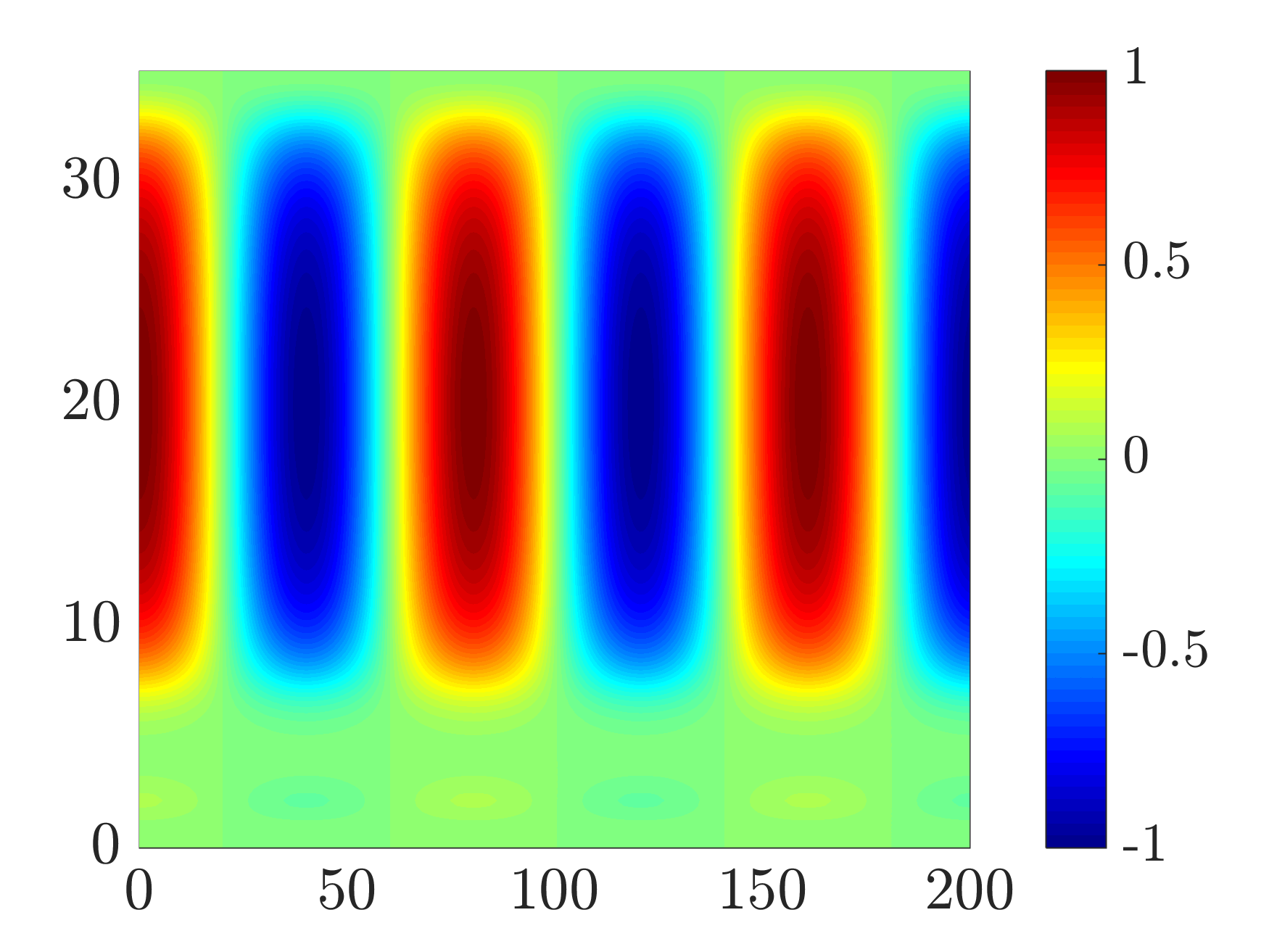}
        \end{tabular}
        &
        \hspace{.1cm}
        \begin{tabular}{c}
		\vspace{.4cm}
		\rotatebox{90}{\normalsize $y$}
	    \end{tabular}
        \hspace{-.3cm}
        &
        \begin{tabular}{c}
                \includegraphics[width=.40\textwidth]{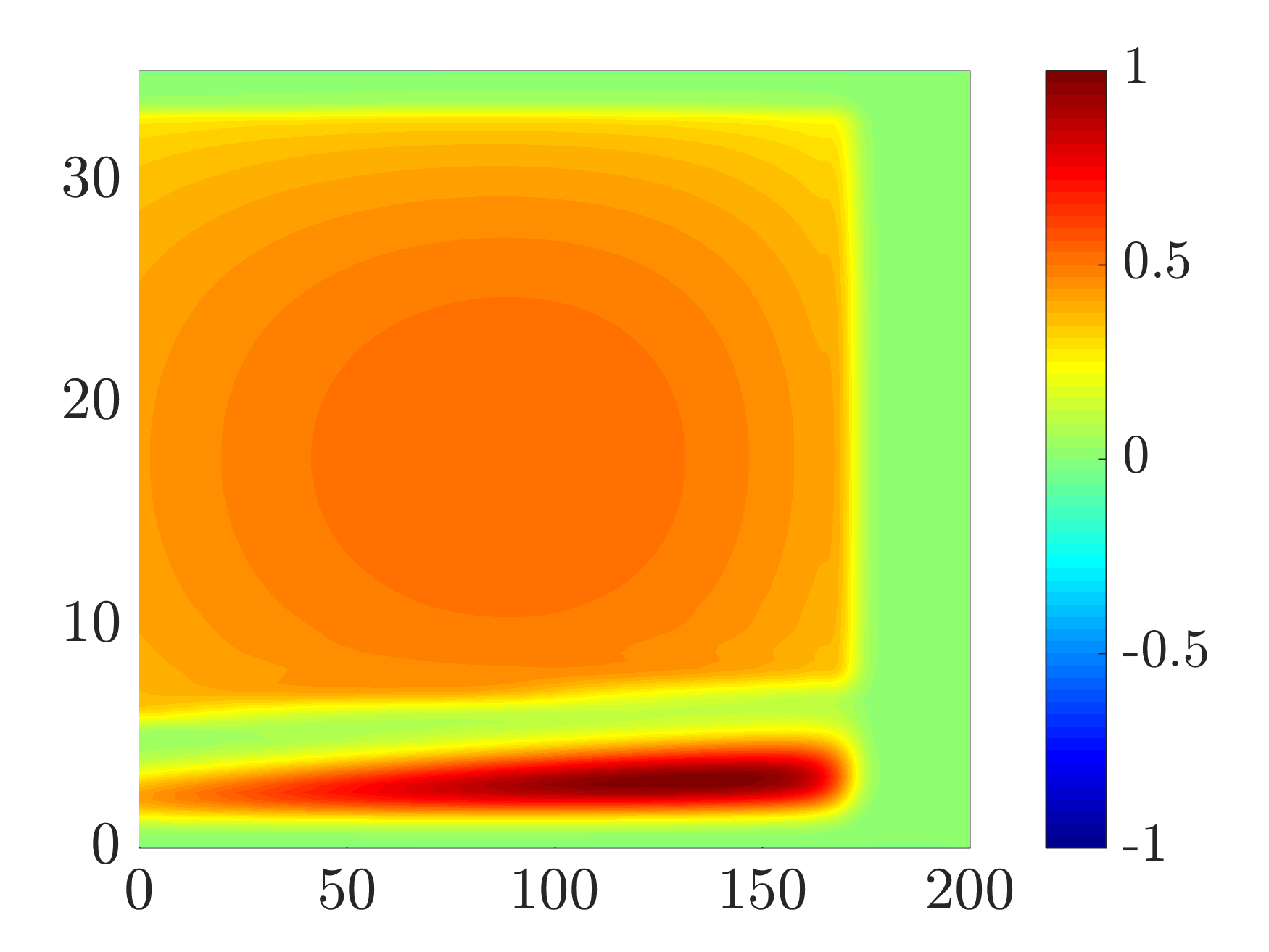}
        \end{tabular}
        \\
        &
        \hspace{-.9cm}
        {\normalsize $x$}
        &&
        \hspace{-.9cm}
        {\normalsize $x$}
        \end{tabular}
        \\[-.2cm]
        \begin{tabular}{rlrl}
        \subfigure[]{\label{fig.Re300outlayerglobalmodes}}
        &&
        \subfigure[]{\label{fig.Re300outlayeroscg7maxy}}
        &
        \\[-.2cm]
        \hspace{-.3cm}
        &
        \begin{tabular}{c}
                \includegraphics[width=.59\textwidth]{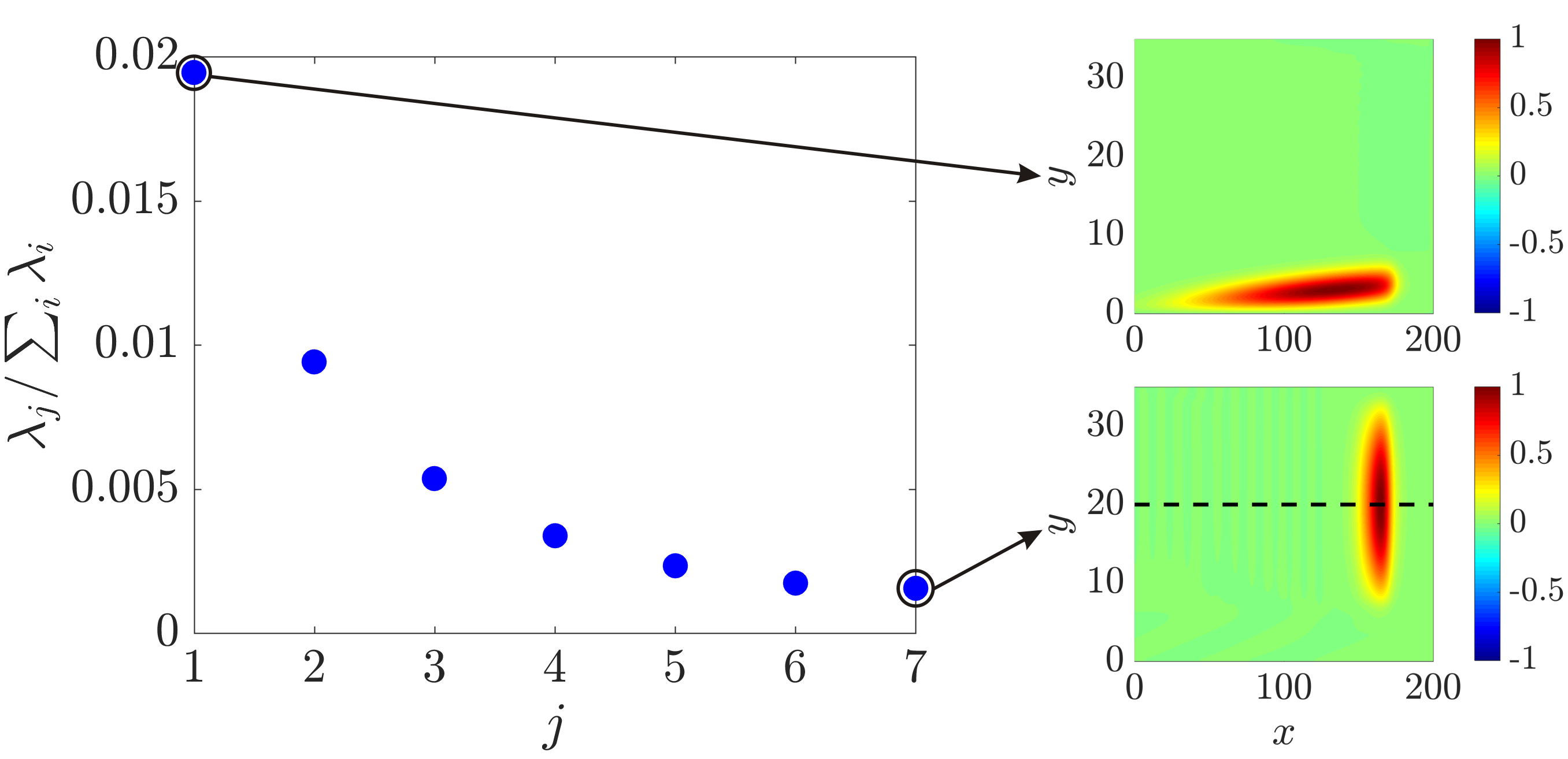}
        \end{tabular}
        &&
        \hspace{-.3cm}
        \begin{tabular}{c}
                \includegraphics[width=.39\textwidth]{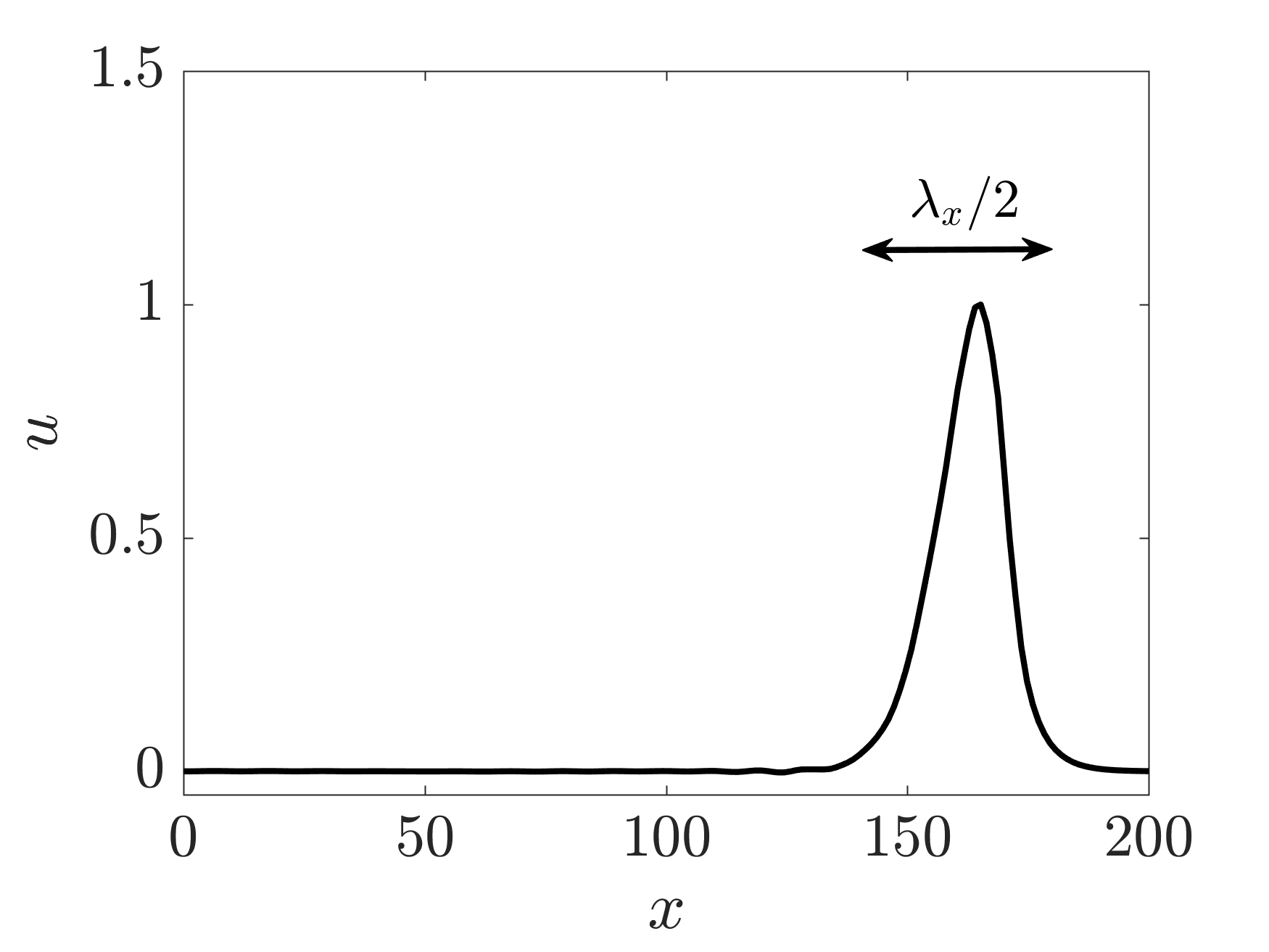}
        \end{tabular}
        \end{tabular}
        \caption{{The rms amplitude of the streamwise velocity component of the response ($\diag \, ( C_u X C_u^*)$) obtained from receptivity analysis  of boundary layer flow with $Re=300$ subject to full outer-layer stochastic excitation: (a) locally parallel analysis with $k_x= k_z=0.076$; (b) global flow analysis with $k_z = 0.076$. (c) Contribution of the first $7$ eigenvalues of the velocity covariance matrix $\Phi$ obtained via global receptivity analysis and the flow structures corresponding to the first and $7$th eigenmodes. (d) The streamwise velocity profile at $y=20$ from the $7$th eigenmode of $\Phi$ illustrated in (c). In the global computations, we set $L_x=200$ and the outer-layer oscillating structures appear at $Re\approx300$.}}
        \label{fig.Re300outlayerosc}
\end{figure}

	\vspace*{-2ex}
\subsection{Frequency response analysis}
\label{sec.resolvent}

The receptivity analysis conducted in this paper quantifies the energy amplification of stochastically-forced linearized NS equations and identifies the dominant flow structures in statistical steady-state. We utilize the solution $X$ to the algebraic Lyapunov equation~\eqref{eq.standard_lyap} to avoid the need for performing either costly stochastic simulations or integration over all temporal frequencies. This approach facilitates efficient computations by aggregating the impact of different frequencies on energy amplification. In what follows, we illustrate how additional insight into temporal aspects of the linearized dynamics can be obtained by examining the spectral density associated with velocity fluctuations~\eqref{eq.psd}.

Application of the temporal Fourier transform on system~\eqref{eq.lnse1} in combination with a coordinate transformation
	\be
	\bd(t)
	\; = \;
	W^{1/2}
	\,
	\tilde{\bd} (t),
	\non
	\ee
where $\bd(t)$ and $\tilde{\bd} (t)$ are white-in-time forcings with the spatial covariance matrices $W$ and $I$, respectively, yields
\begin{align}
	\label{eq.temporesponse}
	\bv(\bk,\omega)
	\;=\;
	T_{\bv \bd}(\bk, \omega) \,\bd(\bk,\omega)
	\;=\;
	T_{\bv \tilde{\bd}}(\bk, \omega) \, \tilde{\bd} (\bk,\omega).
\end{align}
Here, $\bk$ denotes the spatial wavenumbers, $\omega$ is the temporal frequency, $T_{\bv \bd}(\bk, \omega)$ is the frequency response of system~\eqref{eq.lnse1} given in Eq.~\eqref{eq.resolvent}, and
\begin{align}
\label{eq.resolvent}
	T_{\bv \tilde{\bd}}(\bk, \omega)
	\; \DefinedAs \;
	T_{\bv \bd} (\bk, \omega) \, W^{1/2}
	\;=\;
	C \left( \mri \omega I \,-\, A \right)^{-1} \! B \, W^{1/2}.
\end{align}
Singular value decomposition of $T_{\bv \tilde{\bd}}(\bk, \omega)$ brings the input-output representation~\eqref{eq.temporesponse} into the following form:
\[
	\bv(\bk, \omega)
	\;=\;
	T_{\bv \tilde{\bd}}(\bk, \omega) \, \tilde{\bd}(\bk,\omega)
	\;=\;
	\sum_i
	\sigma_i(\bk,\omega)\, \mathbf{u}_i(\bk,\omega) \inner{\mathbf{w}_i(\bk,\omega)}{\tilde{\bd}(\bk,\omega)},
\]
where $\sigma_i$ is the $i$th singular values of $T_{\bv \tilde{\bd}}(\bk, \omega)$, $\mathbf{u}_i(\bk,\omega)$ is the associated left singular vector, and $\mathbf{w}_i(\bk,\omega)$ is the corresponding right singular vector. The power spectral density (PSD) quantifies the energy of velocity fluctuations $\bv(\bk, \omega)$ across temporal frequencies $\omega$ and spatial wavenumbers $\bk$,
\begin{align*}
	\Pi_{\bv}(\bk,\omega)
	\;=\;
	\trace \left( T_{\bv \tilde{\bd}}(\bk, \omega) \, T^*_{\bv \tilde{\bd}}(\bk, \omega) \right)
	\;=\;
	\trace \left( T_{\bv \bd} (\bk,\omega)\,W\, T^*_{\bv \bd} (\bk,\omega) \right)
	\;=\;
	\trace \left( S_{\bv} (\bk,\omega) \right),
	\end{align*}
and is determined by the sum of squares of the singular values of the frequency response $T_{\bv \tilde{\bd}}(\bk, \omega)$,
	\begin{align*}
	\Pi_{\bv}(\bk,\omega)
	\;=\;
	\sum_i \sigma^2_i(\bk,\omega).
\end{align*}
As described in Section~\ref{sec.stats}, the energy spectrum $E$ in Eq.~\eqref{eq.H2norm} can be obtained by the integration of $\Pi_{\bv}(\bk, \omega)$ over temporal frequency~\cite{jovbamJFM05}, 
	\be
	E (\bk)
	\; = \;
	\dfrac{1}{2 \pi}
	\int_{-\infty}^{\infty}
	\Pi_{\bv}(\bk,\omega)
	\,
	\mrd \omega
	\; = \;
	\dfrac{1}{2 \pi}
	\int_{-\infty}^{\infty}
	\sum_i \sigma^2_i(\bk,\omega)
	\,
	\mrd \omega.
	\non
	\ee
This approach extends standard resolvent analysis~\cite{tretrereddri93,mj-phd04,mcksha10} to stochastically-forced flows and it allows the spatial covariance matrix $W$ of the white-in-time stochastic forcing $\bd$ to be embedded into the analysis. A recent reference~\cite{towschcol18} also establishes relation between spectral decomposition of $S_{\bv} (\bk,\omega)$ and dynamic mode decomposition~\cite{sch10}.

The PSD of the boundary layer flow with $Re_0=232$ subject to near-wall stochastic excitation is shown in Fig.~\ref{fig.temporalresponse}. While locally parallel analysis reveals isolated frequencies at which the PSD peaks, much broader frequency range is important in global analysis. In particular, locally parallel analysis for a flow with (i) $(k_x,\,k_z)=(7 \times 10^{-3},0.32)$ identifies nearly-steady streaks as dominant flow structures (the PSD peaks at $\omega = 0.0063$); and (ii) $(k_x,\,k_z)=(0.19,0.01)$ identifies two peaks at $\omega = 0.08$ and $\omega = 0.19$ which correspond to the TS waves and flow structures in the outer-layer, respectively. On the other hand, the peaks are much less pronounced in the analysis of the spatially-evolving base flow. This suggests that the focus on isolated frequencies in global analysis may not capture the full complexity of the underlying flow structures. In fact, the shapes of spatial profiles associated with principal singular vectors of the frequency response $T_{\bv \tilde{\bd}} (\bk, \omega)$ change for different values of $\omega$. As shown in Fig.~\ref{fig.resolvent_streak}, even though the principal singular values of $T_{\bv \tilde{\bd}}(\bk, \omega)$ for $\omega = 10^{-5}$, $0.01$, and $0.02$ are comparable ($5464$, $4732$, and $3565$, respectively), the corresponding response directions change from streamwise streaks (for steady perturbations) to oblique modes (at larger frequencies). This trend is reminiscent of the various flow structures resulting from the eigenvalue decomposition of the steady-state covariance matrix (cf.\ Section~\ref{sec.global-BL}) and has been also recently observed in spatio-temporal analysis of hypersonic boundary layer flows~\cite{dwisidniccanjovJFM18}.}

\begin{figure}[!ht]
        \begin{tabular}{cccc}
        \subfigure[]{\label{fig.parallel_Pi}}
        &&
        \subfigure[]{\label{fig.parallel_sigma1}}
        &
        \\[-.2cm]
        \hspace{.2cm}
        \begin{tabular}{c}
                \vspace{.4cm}
                \rotatebox{90}{\normalsize $\Pi_{\bv}$}
        \end{tabular}
        &
        \begin{tabular}{c}
                \includegraphics[width=.4\textwidth]{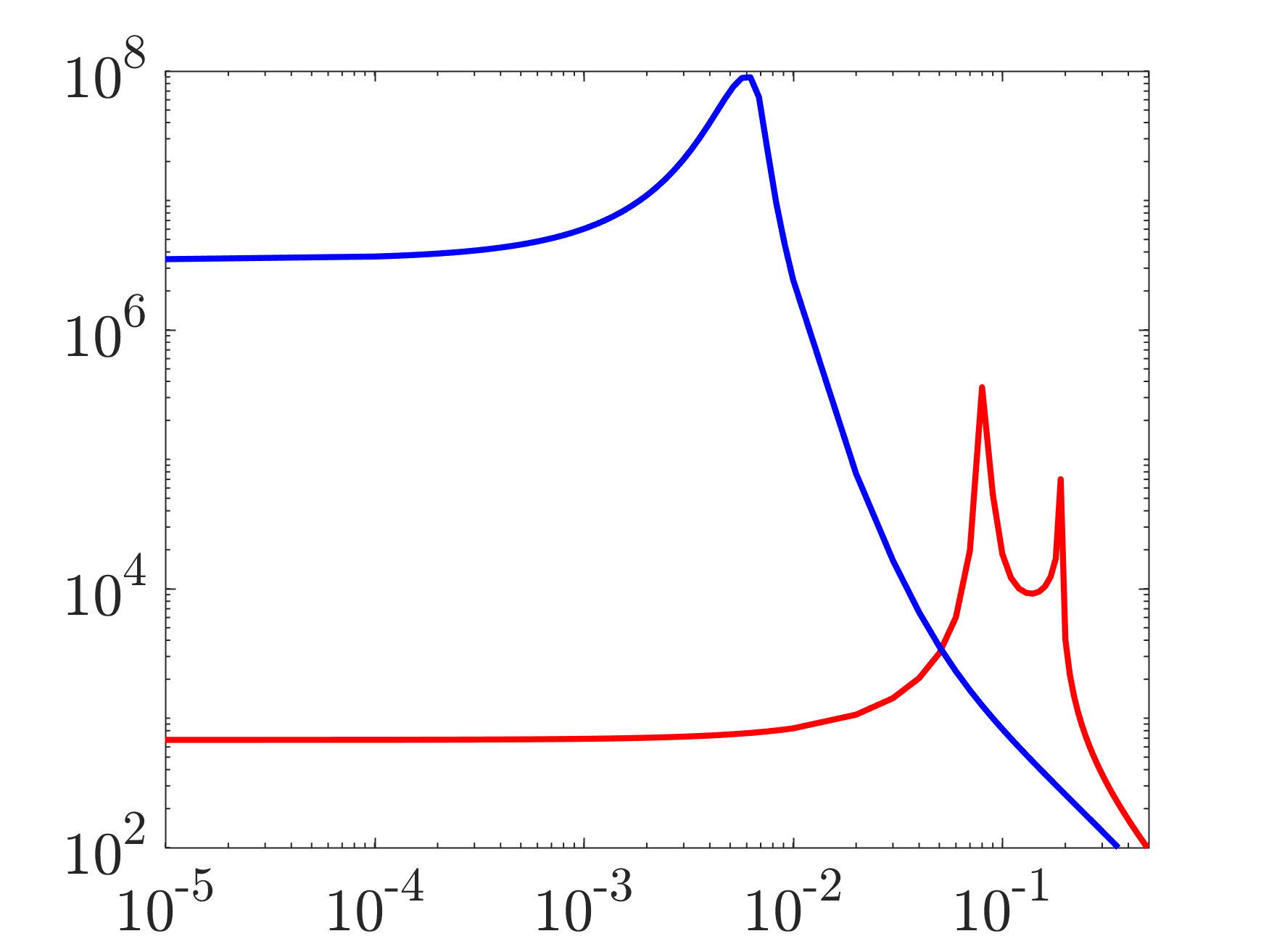}
        \end{tabular}
        &
        \hspace{.2cm}
        \begin{tabular}{c}
                \vspace{.4cm}
                \rotatebox{90}{\normalsize $\Pi_{\bv}$}
        \end{tabular}
        &
        \begin{tabular}{c}
                \includegraphics[width=.4\textwidth]{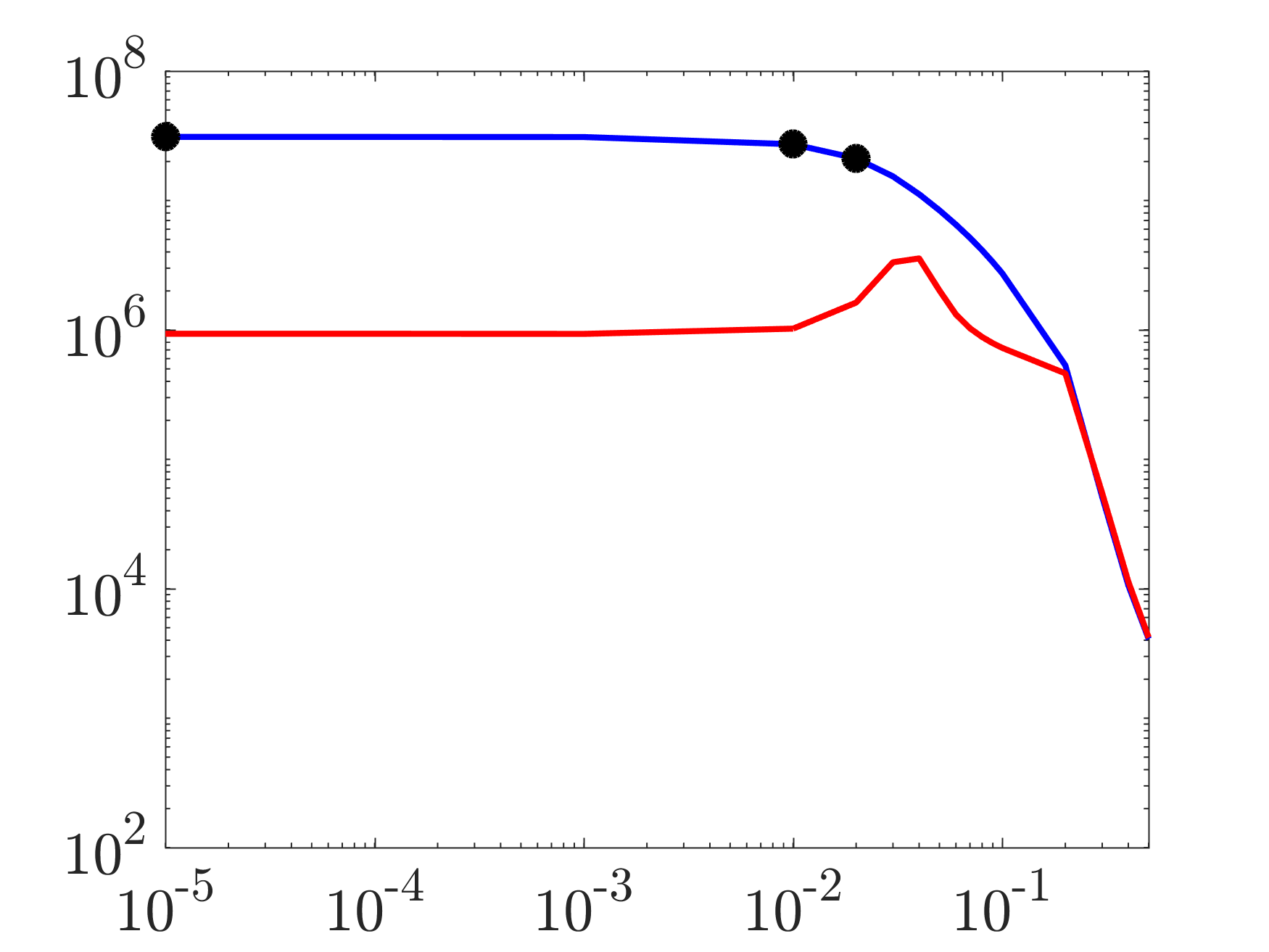}
        \end{tabular}
        \\[-.1cm]
        &
        {\normalsize $\omega$}
        &&
        {\normalsize $\omega$}
        \end{tabular}
        \caption{Power spectral density $\Pi_{\bv}(\bk, \omega)$ as a function of temporal frequency $\omega$ for a boundary layer flow with $Re_0=232$ subject to near-wall white stochastic excitation. (a) Locally parallel analysis with $(k_x,\,k_z)=(7 \times 10^{-3},0.32)$ (blue) and $(k_x,\,k_z)=(0.19,0.01)$ (red) corresponding to streaks and TS waves, respectively. (b) Global flow analysis with $k_z=0.32$ (blue) and $k_z=0.01$ (red). Black dots correspond to the  temporal frequencies of the Fourier modes plotted in Fig.~\ref{fig.resolvent_streak}.}
        \label{fig.temporalresponse}
\end{figure}

\begin{figure}[!ht]
        \begin{tabular}{cccccc}
        \hspace{-.4cm}
        \subfigure[]{\label{fig.resolvent_streakom1e-5}}
        &&
        \hspace{-.2cm}
        \subfigure[]{\label{fig.resolvent_streakom0p01}}
        &&
        \hspace{-.2cm}
        \subfigure[]{\label{fig.resolvent_streakom0p02}}
        &
        \\[-.2cm]
        &
        \hspace{-.5cm}
        \begin{tabular}{c}
                \includegraphics[width=.29\textwidth]{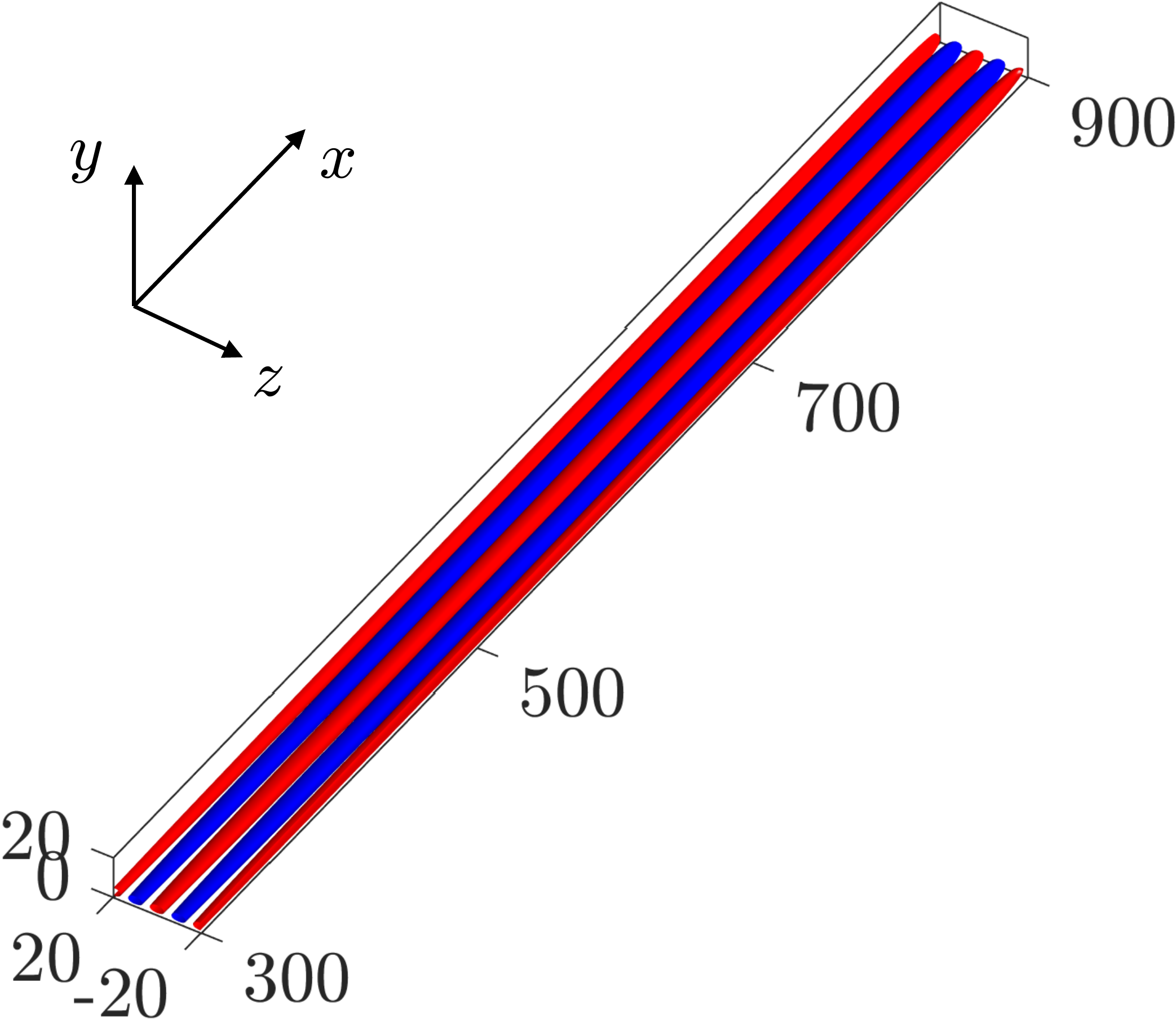}
        \end{tabular}
        &&
        \hspace{-.4cm}
        \begin{tabular}{c}
                \includegraphics[width=.29\textwidth]{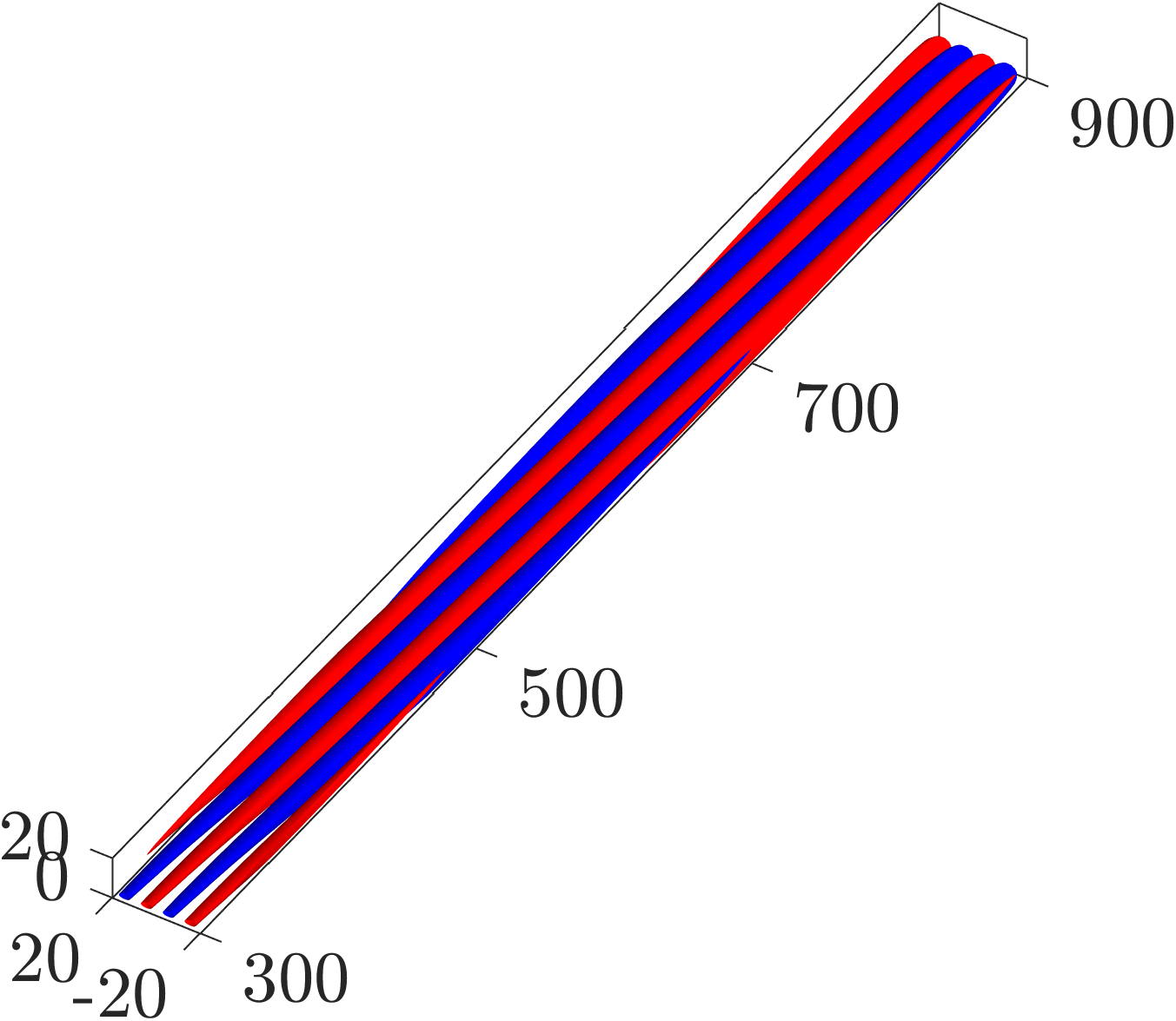}
        \end{tabular}
        &&
        \hspace{-.4cm}
        \begin{tabular}{c}
                \includegraphics[width=.29\textwidth]{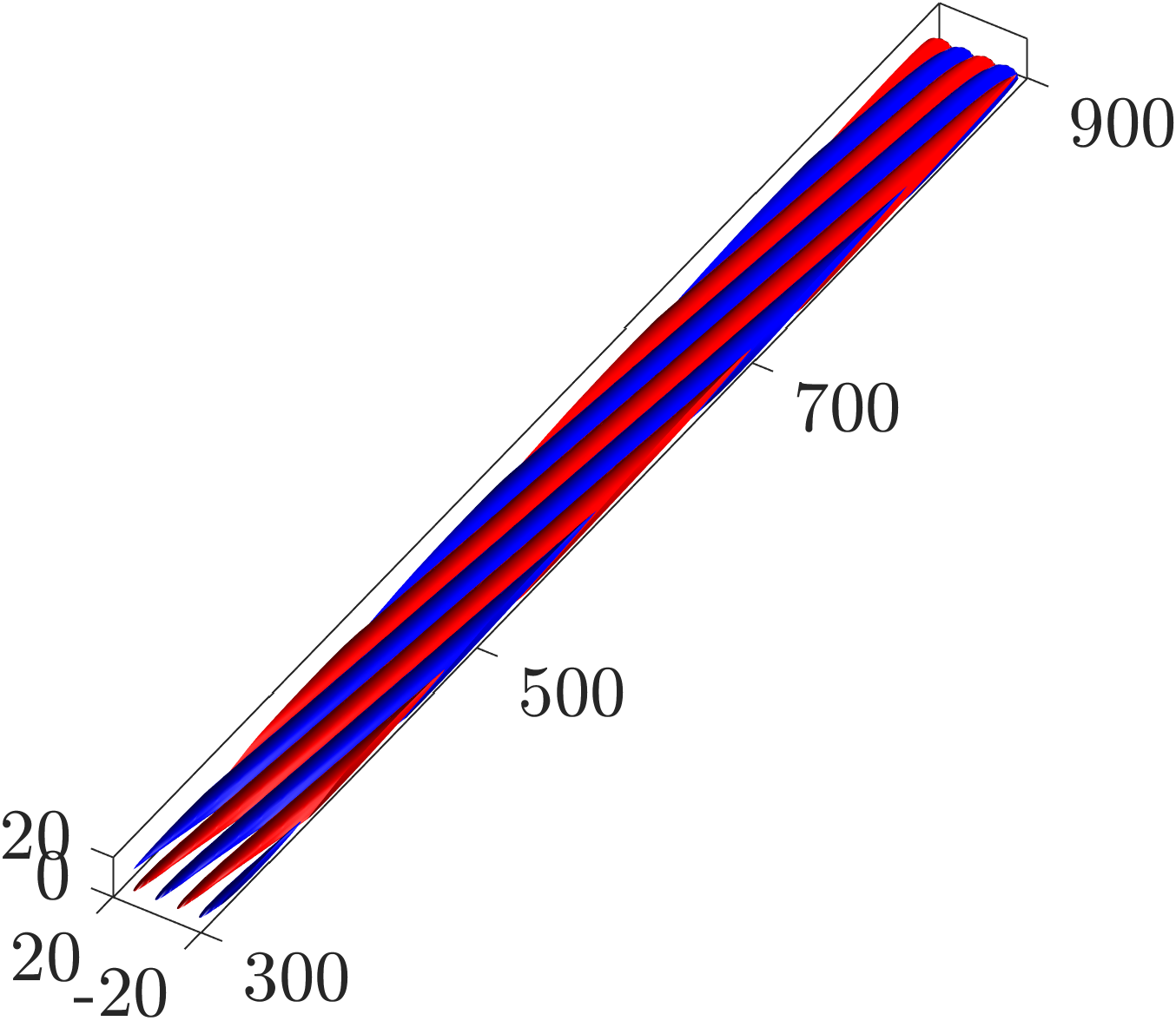}
        \end{tabular}
        \end{tabular}
        \caption{Fourier modes corresponding to the principal response directions of $T_{\bv \bd}(\bk,\omega)$ for a spatially evolving Blasius boundary layer flow with $Re_0=232$ and $k_z=0.32$ subject to near-wall stochastic excitation. (a) $\omega = 10^{-5}$, (b) $\omega = 0.01$, and (c) $\omega = 0.02$.}
        \label{fig.resolvent_streak}
\end{figure}

	\vspace*{-3ex}
\section{Concluding remarks}
\label{sec.conclusion}

In the present study, we have utilized the linearized NS equations to study energy amplification in the Blasius boundary layer flow subject to white-in-time stochastic forcing entering at various wall-normal locations. The evolution of flow fluctuations is captured by two models that arise from locally parallel and global perspectives, and the amplification of persistent stochastic disturbances is studied using the algebraic Lyapunov equation. Both parallel and global flow analyses predict largest amplification of streamwise elongated streaks with similar spanwise wavelength. Moreover, TS wave-like flow structures arise from persistent near-wall stochastic excitation at long spanwise wavelengths. We have shown that as the region of excitation moves away from the wall, energy amplification reduces, which suggests that the near wall region is more sensitive to external disturbances. We have also examined the spatial structure of characteristic eddies that result from stochastic excitation of the boundary layer flow. Our computational experiments demonstrate good agreement between the results obtained from parallel and global flow models. This agreement highlights the efficacy of using parallel flow assumptions in the receptivity analysis of boundary layer flows, especially when it is desired to evaluate the energetic contribution of individual streamwise scales, which are often obscured by the dominant growth of streaks and Tollmien-Schlichting waves in global analysis.

{In contrast to resolvent-mode analysis which quantifies the energy amplification from monochromatic forcing, our stochastic approach incorporates a broad-band forcing model with known spatial correlations that captures the aggregate effect of all time scales. Our Lyapunov-based framework generalizes the concept of receptivity to the amplification of velocity fluctuations from any external source of persistent excitation with known statistical properties. We} note that the ability of the method to capture relevant flow physics relies on the spectral properties of the stochastic forcing that can be used to model the effect of, e.g., free-stream turbulence. In addition to white-in-time stochastic forcing with trivial (identity) spatial covariance operator, we have also investigated energy amplification arising from the streamwise-decaying forcing that corresponds to the spectrum of HIT. Our computations demonstrate close correspondence between these two case studies. The spatio-temporal spectrum of stochastic excitation sources can be further determined in order to provide statistical consistency with the results of numerical simulations or experimental measurements of the boundary layer flow~\cite{zarchejovgeoTAC17,zarjovgeoJFM17}. Implementation of such ideas to leverage statistical data and improve physics-based analysis is a topic for future research.

\vspace*{-3ex}
\begin{acknowledgments}
Part of this work was conducted during the 2016 CTR Summer Program with financial support from Stanford University and NASA Ames Research Center. We thank Prof.\ P.\ Moin for providing us with the opportunity to participate in the CTR Summer Program and Prof. J. W. Nichols for insightful discussions. Financial support from the National Science Foundation under Award CMMI 1739243 and the Air Force Office of Scientific Research under Award FA9550-16-1-0009 and FA9550-18-1-0422 is gratefully acknowledged.
\end{acknowledgments}

\appendix

\section{Operator valued matrices in Eqs.~\eqref{eq.lnse}}
\label{sec.appendix-operators}

Equation~\eqref{eq.lnse} is of the following form:
\begin{eqnarray*}
	\left[
        \begin{array}{c}
        		v_t
        		\\[0.2cm]
        		\eta_t
        \end{array}
        \right]
        &~=~&
        \underbrace{\left[
        \begin{array}{cc}
        \bA_{11} & \bA_{12} \\[0.2cm]
        \bA_{21} & \bA_{22}
        \end{array}
        \right]}_{\bA}
        \,
        \left[
        \begin{array}{c}
        		v
        		\\[0.2cm]
        		\eta
        \end{array}
        \right]
        ~+~
        \underbrace{
        \left[
        \begin{array}{ccc}
        		\bB_{11} & \bB_{12} & \bB_{13}
		\\[0.2cm]
        		\bB_{21} & 0 & \bB_{23}
        \end{array}
        \right]}_{\bB}
        \,
        \left[
        \begin{array}{c}
        		d_u
        		\\[0.2cm]
        		d_v
                \\[0.2cm]
        		d_w
        \end{array}
        \right]
        \\[.15cm]
        \left[
        \begin{array}{c}
        		u
        		\\[0.1cm]
        		v
		\\[0.1cm]
        		w
        \end{array}
        \right]
        &~=~&
        \underbrace{\left[
        \begin{array}{cc}
        \bC_{11} & \bC_{12} \\[0.1cm]
        \bI & 0 \\[0.1cm]
        \bC_{31} & \bC_{32}
        \end{array}
        \right]}_{\bC}
        \,
        \left[
        \begin{array}{c}
        		v
		\\[0.2cm]
        		\eta
        \end{array}
        \right]
\end{eqnarray*}
with operators defined as
\[
\ba{rcl}
\bA_{11}
&=&
\Delta^{-1}\left[\dfrac{1}{Re}\Delta^2 \;-\; U\,\Delta\,\partial_x \;-\; V\,\Delta\,\partial_y \;-\; \partial_y\, V\,\Delta \;-\; 2\,\partial_x\, U\,\partial_{xx} \;-\; \partial_{yy}\, V\,\partial_y \;+\; \partial_{yy} \,U\,\partial_x\right.
\\[0.4cm]
&&
-\left.\partial_{yyy}\, V\;-\;2\left(\partial_{xy}\,U\,\partial_x \;+\; \partial_x\, U\,\partial_{xy}\right)\left(\partial_{xx}\;-\;k_z^2\right)^{-1}\partial_{xy}\right] \;-\; \sigma(x),
\\[0.25cm]
\bA_{12}
&=&
2\, \mri k_z\, \Delta^{-1}\left[\left(\partial_{xy}\,U\,\partial_x\;+\;\partial_x\, U\,\partial_{xy}\right)\left(\partial_{xx}\;-\;k_z^2\right)^{-1}\right],
\\[0.25cm]
\nonumber
\bA_{21}
&=&
\;-\;\mri k_z\, \partial_y\, U,
\quad \quad \quad \quad \quad \quad \quad \quad~~
\bA_{22}\;=\;\dfrac{1}{Re}\Delta\;-\;U\, \partial_x\;-\;V\, \partial_y\;-\;\partial_x\, U \;-\; \sigma(x),
\\[0.5cm]
\bB_{11}
&\!\!=\!\!&
\;-\;\Delta^{-1}\left(f\, \partial_{xy} \;+\;\partial_y\, f\,\partial_x\right),
\quad \quad
\bB_{12}\;=\;\Delta^{-1}\left(f\, \partial_{xx} \;-\; k_z^2\,f\right),
\quad \qquad \!
\Delta
\;=\;
\partial_{xx}\;+\;\partial_{yy}\;-\;k_z^2,
\\[0.5cm]
\bB_{13}
&\!\!=\!\!&
\;-\;\mri k_z\,\Delta^{-1}\left(\partial_{y}\,f\, \;+\; f\,\partial_y\right),
\quad \quad
\bB_{21}\;=\;\;-\;\mri k_z\,f,
\qquad \qquad \qquad \qquad\!\!
\bB_{23} \;=\; \;-\; f\,\partial_x,
\\[0.5cm]
\bC_{11}
&=&
\;-\;\left(\partial_{xx}\;-\;k_z^2\right)^{-1}\partial_{xy},
\quad \quad \quad ~\;\;
\bC_{12}
\;=\;
\mri k_z\left(\partial_{xx}\;-\;k_z^2\right)^{-1},
\\[0.5cm]
\bC_{31}
&=&
\;-\;\mri k_z\left(\partial_{xx}\;-\;k_z^2\right)^{-1}\partial_y,
\quad \quad \quad
\bC_{32}
\;=\;
\;-\;\left(\partial_{xx}\;-\;k_z^2\right)^{-1}\partial_x.
\ea
\]
Here, $\sigma(x)$ {determines} the strength of sponge layers as a function of $x$; see~\cite{ranzarnicjovAIAA17} for additional details. For parallel flows, Fourier transform in $x$ can be used to further parameterize the operators over streamwise wavenumbers; see~\cite{jovbamJFM05} for the expressions of $\bA$, $\bB$, and $\bC$ in such instances.

	\vspace*{-2ex}
\section{Change of variables}
\label{sec.appendix-coc}

The kinetic energy of velocity fluctuations in the linearized NS equations~\eqref{eq.lnse} is defined using the energy norm
\[
	E
	~=\;
	\inner{\bm{\varphi}}{\bm{\varphi}}_e
	\;=\;
	\dfrac{1}{2}
	\int_\Omega \,\bm{\varphi}^*\,\mathbf{Q}\, \bm{\varphi}\, \mrd y
	~=:\;
	\inner{\bm{\varphi}}{\mathbf{Q}\,\bm{\varphi}}
\]
where $\Omega$ is the computational domain, $\inner{\cdot}{\cdot}$ is the $L_2$ inner product and $\mathbf{Q}$ is the operator which determines kinetic energy of the state $\bm{\varphi} = [\,v\,~ \eta\,]^T$ on the appropriate state-space~\cite{redhen93,jovbamJFM05}. With proper discretization of the inhomogeneous directions, the kinetic energy is given by $E=\bm{\varphi}^*\,Q\,\bm{\varphi}$. Here, $Q$ is the discrete representation of operator $\mathbf{Q}$ and is a positive definite matrix. The coordinate transformation $\psi=Q^{1/2} \bm{\varphi}$ can thus be employed to obtain the kinetic energy via the standard Euclidean norm: $E=\bpsi^*\bpsi$ in the new coordinate space. Equation~\eqref{eq.lnse1} results from the application of this change of variables on the discretized state-space matrices $\bar{A}$, $\bar{B}$, and $\bar{C}$
\[
	A \;=\; Q^{1/2}\, \bar{A}\, Q^{-1/2},
	\quad\quad\quad
	B \;=\; Q^{1/2}\, \bar{B}\, I_W^{-1/2},
	\quad\quad\quad
	C \;=\; I_W^{1/2}\, \bar{C}\, Q^{-1/2},
\]
and the discretized input $\bar{\bd}$ and velocity $\bar{\bv}$ vectors
\[
	\bd \;=\; I_W^{1/2}\, \bar{\bd},
	\quad\quad\quad
	\bv \;=\; I_W^{1/2}\, \bar{\bv}.
\]
Here, $I_W$ is a diagonal matrix of integration weights on the set of Chebyshev collocation points.

The operator $\mathbf{Q}$ in the global model is of the form:
\[
	\mathbf{Q}
	~=~
	\left[
	\begin{array}{cc}
	\partial_{xy}^\dagger\, \Theta^\dagger\, \Theta\, \partial_{xy} \;+\; \bI \;+\; k_z^2\, \partial_{y}^\dagger\,\Theta^\dagger\,\Theta\,\partial_y & 0
	\\[0.25cm]
	0
	&
	~~
	k_z^2\, \Theta^\dagger\,\Theta \;+\; \partial_x^\dagger\, \Theta^\dagger\, \Theta\,\partial_x
	\end{array}
	\right]
\]
where $\Theta = (\partial_x^2 - k_z^2)^{-1}$, $\bI$ is the identity operator and $\dagger$ represents the adjoint of an operator. The representation of $\mathbf{Q}$ for parallel flows can be found in~\cite[Appendix A]{jovbamJFM05}.

	\vspace*{-2ex}
\section{Global analysis using the descriptor form}
\label{sec.appendix-descriptor}

The descriptor form of the linearized NS equations around the Blasius boundary layer profile is given by
\begin{align}
	\label{eq.descriptor}
	\ba{rcl}
		{\bF}\,{\dot{\bpsi}}(t)
		& = &
		\bA\, \bpsi(t)
		\;+\;
		\bB\, \bd(t),
		\\[.15cm]
		\bv(t)
		&=&
		\bC\, \bpsi(t),
	\ea
\end{align}
where $\bpsi=[\,u\,~ v\,~ w\,~ p\,]^T$ and
\[
\bF \;=\, \left[
        \begin{array}{cccc}
        \bI\, & \,0\, & \,0\, & \,0\,
        \\[0.2cm]
        \,0\, & \,\bI\, & \,0\, & \,0\,
        \\[0.2cm]
        \,0\, & \,0\, & \,\bI\, & \,0\,
        \\[0.2cm]
        \,0\, & \,0\, & \,0\, & \,0\,
        \end{array}
        \right],
        \quad
	\bA \;=\, \left[
        \begin{array}{cccc}
        \,\bK \,+\, \partial_y\, V\, & \,-\partial_y\, U\, & \,0\, & \,-\partial_x\,
        \\[0.2cm]
        \,0\, & \,\bK \,-\, \partial_y V\, & \,0\, & \,-\partial_y\,
        \\[0.2cm]
        \,0\, & \,0\, & \,\bK\, & \,-\mri k_z\,
        \\[0.2cm]
        \,\partial_x\, & \,\partial_y\, & \,\mri k_z\, & \,0\,
        \end{array}
        \right],
        \quad
	\bB \;=\, \left[
        \begin{array}{ccc}
        \bI\, & \,0\, & \,0\,
        \\[0.2cm]
        \,0\, & \,\bI\, & \,0\,
        \\[0.2cm]
        \,0\, & \,0\, & \,\bI\,
        \\[0.2cm]
        \,0\, & \,0\, & \,0\,
        \end{array}
        \right],
        \quad
	\bC \;=\; \bB^T
\]
where $\bI$ is the identity operator and
\[
	\bK
	~=~
	\frac{1}{Re}\left(\partial_x^2 + \partial_y^2 - k_z^2 \right) \;-\; U\partial_x \;-\; V\partial_y - \sigma(x).
\]
Here, $\sigma(x)$ {determines} the strength of sponge layers as a function of $x$. The width and strength of the sponge layers are selected to guarantee the stability of the generalized dynamics~\eqref{eq.descriptor} in their discretized form, while having minimal influence on velocity fluctuation field. The energy of velocity fluctuations in Eqs.~\eqref{eq.descriptor} can be determined by
\[
	\label{eq.H2norm-descriptor}
	E
	\;=\;
	\trace \left(\bC\, (\bG_c \,+\, \bG_{nc})\, \bC^\dagger\right),	
\]
which is analogous to expression~\eqref{eq.H2norm} for the evolution model with $\bpsi=[\,v\,~ \eta\,]^T$. Here, $\dagger$ represents the adjoint of an operator and $\bG_c$ and $\bG_{nc}$ are the causal and non-causal reachability Gramians that satisfy the following generalized Lyapunov equations:
\begin{align}
	\label{eq.lyap-general}
	\ba{rclcr}
	\bF\,\bG_c\,\bA^\dagger
	&\!\!+\!\!&
	\bA\,\bG_c\,\bF^\dagger
	&=&
	-\bP_l\, \bB\, \bB^\dagger \bP_l^\dagger
	\\[0.15cm]
	\bF\,\bG_{nc}\,\bF^\dagger
	&\!\!-\!\!&
	\bA\,\bG_{nc}\,\bA^\dagger
	&=&
	\bQ_l\, \bB\, \bB^\dagger \bQ_l^\dagger,
	\ea
\end{align}
where $\bP_l$ and $\bQ_l$ are the projection operators that project the state-space into causal and non-causal subspaces; see~\cite[Appendix.E]{rashad-phd12} for additional details.

After proper spatial discretization of the state-space, the procedure for solving the generalized Lyapunov equations~\eqref{eq.lyap-general} consists of the following steps: (i) compute the generalized Schur form of the discretized pair $(A,F)$;  (ii) computing the solution to a system of generalized Sylvester equations; and (iii) solving the generalized Lyapunov equations~\eqref{eq.lyap-general} for Gramian matrices $G_c$ and $G_{nc}$. The Schur decomposition and the solution to the Sylvester equations are required to split the state into slow (causal) and fast (non-causal) parts and to form projection matrices $P_l$ and $Q_l$. For a spatial discretization that involves $n = 4N_x N_y$ states, the overall computational complexity of this procedure is $O(n^3)$, which is significantly higher than the computational complexity of solving the Lyapunov equation~\eqref{eq.standard_lyap} with $n = 2 N_x N_y$. Moreover, since the state-space of the descriptor form has twice the number of states as the evolution model~\eqref{eq.lnse1}, computations based on this representation require more memory. We refer the interested reader to~\cite[Appendix.E]{rashad-phd12} for additional details on computing energy amplification using the descriptor form.

\begin{figure}[!ht]
        \begin{tabular}{cccccc}
        \subfigure[]{\label{fig.descriptorstreakxyz}}
        &&
        \hspace{-.65cm}
        \subfigure[]{\label{fig.descriptorstreakxy}}
        &&
        \hspace{-.65cm}
        \subfigure[]{\label{fig.descriptorstreakyz}}
        &
        \\[-.2cm]
        \begin{tabular}{c}
		\vspace{1.2cm}
		\\{\normalsize $y$}\\\\\\\\\\\hspace{0.3cm}{\normalsize $z$}\\\\
	    \end{tabular}
        \hspace{-.34cm}
        &
        \begin{tabular}{c}
                \includegraphics[width=0.315\textwidth]{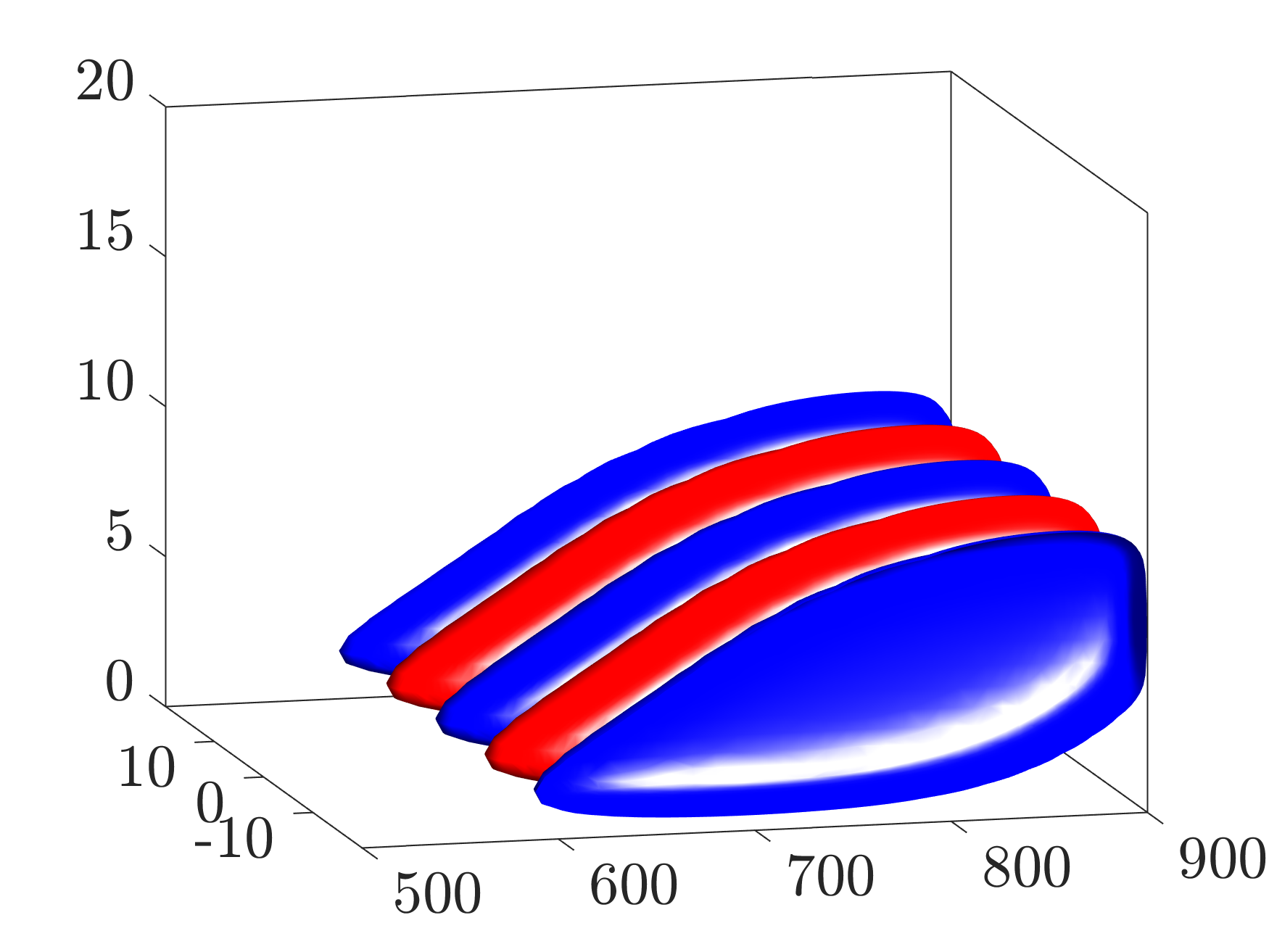}
        \end{tabular}
        &&
        \hspace{-.34cm}
        \begin{tabular}{c}
                \includegraphics[width=0.315\textwidth]{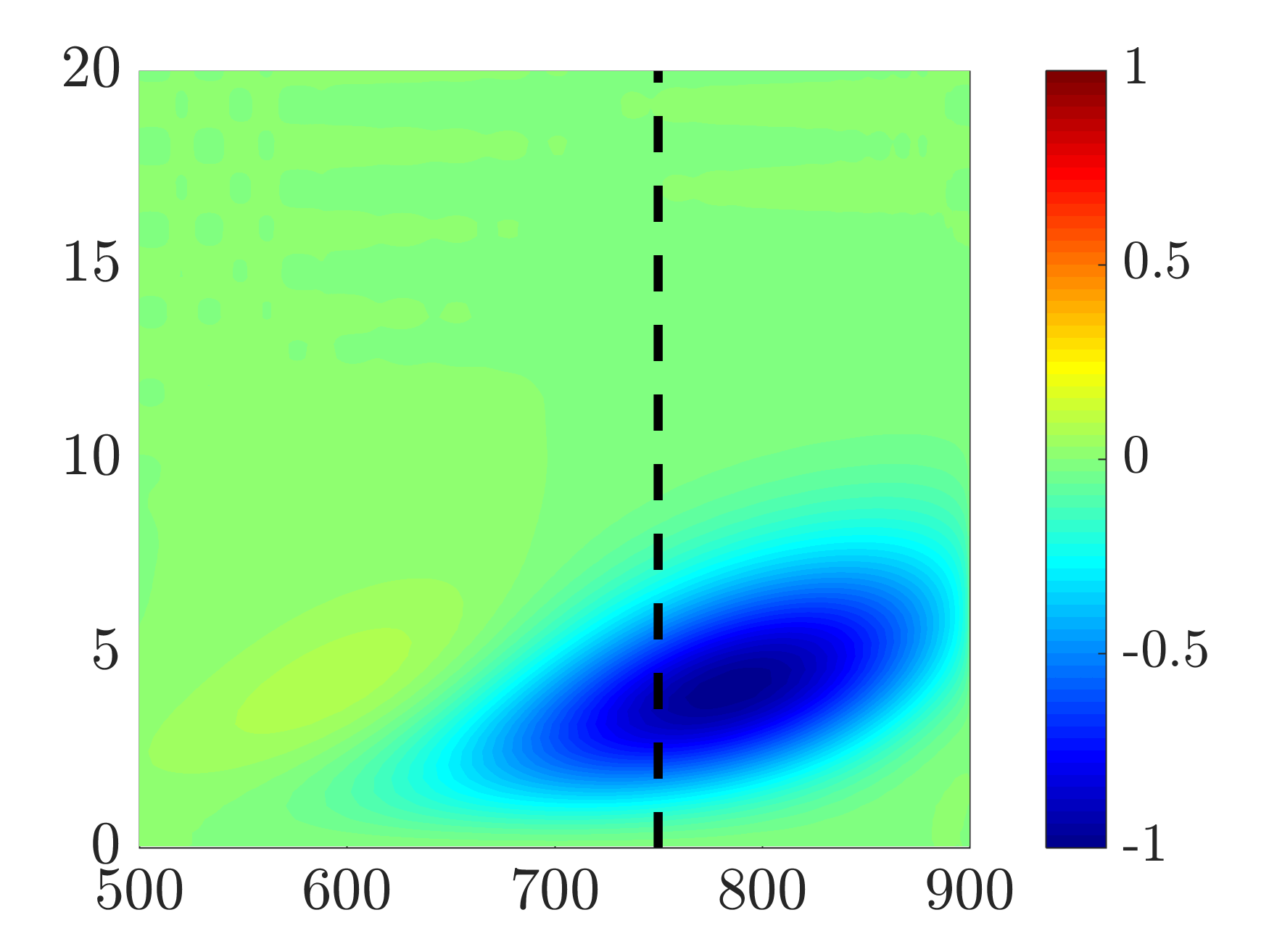}
        \end{tabular}
        &&
        \hspace{-.34cm}
        \begin{tabular}{c}
                \includegraphics[width=0.315\textwidth]{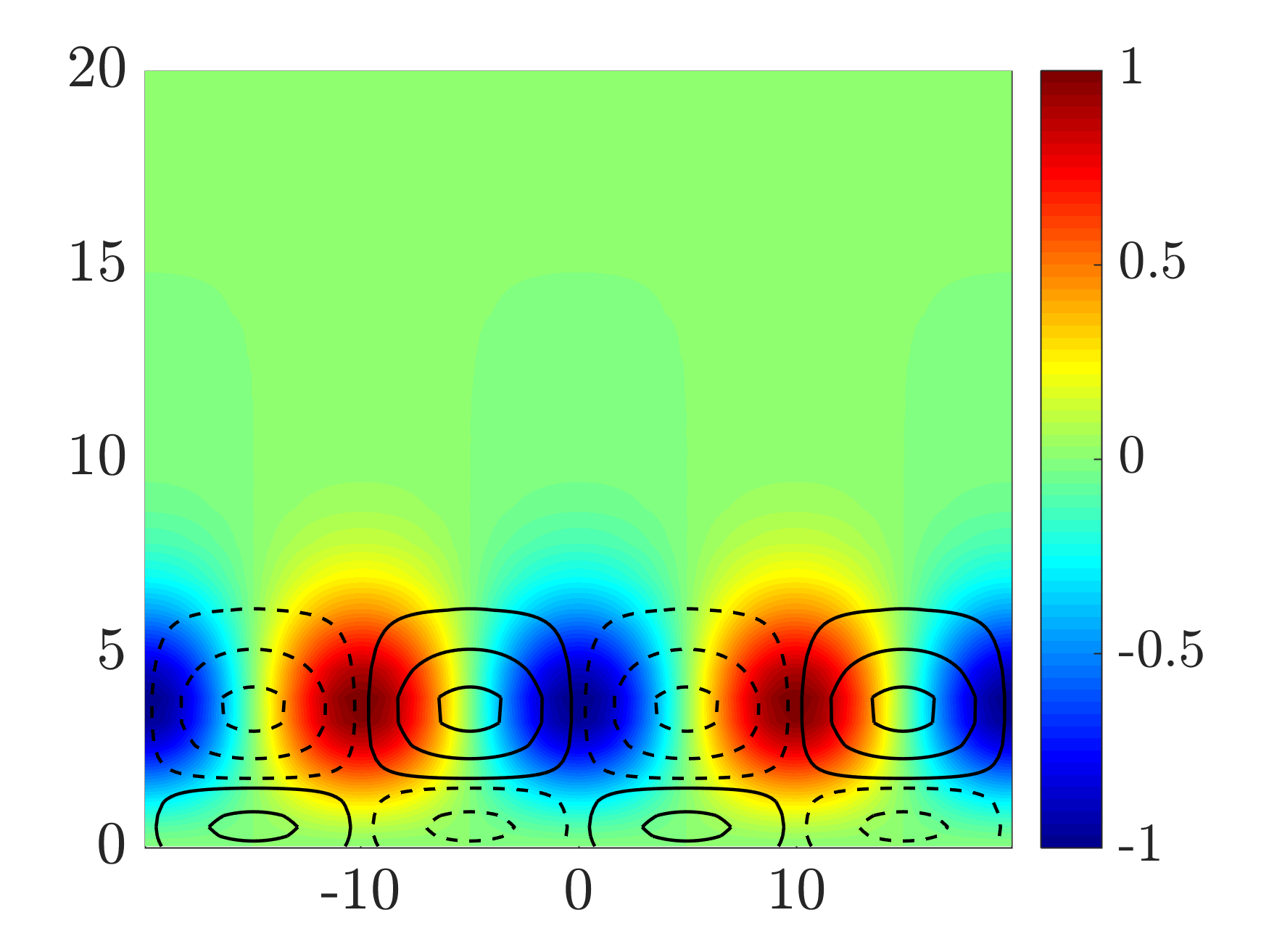}
        \end{tabular}
        \\[-.1cm]
        &
         {\normalsize $x$}
        &&
        \hspace{-.9cm}
        {\normalsize $x$}
        &&
        \hspace{-.8cm}
        {\normalsize $z$}
        \end{tabular}
        \caption{(a) The streamwise component of the principle eigenmode of output covariance matrix $\Phi$ resulting from {near-wall stochastic excitation (case 1 in Table~\ref{tab.forcing-cases}) of the} linearized model~\eqref{eq.descriptor} with $k_z=0.32$ and $Re_0=232$. (b) Streamwise velocity at $z=0$. (c) Slice of streamwise velocity (color plots) and vorticity (contour lines) at $x=750$, which corresponds to the cross-plane slice indicated by the black dashed lines in (b).}
        \label{fig.descriptorstreak}
\end{figure}

In order to demonstrate the close agreement between the outcome of receptivity analysis based on the evolution model of Section~\ref{sec.formulation} and the descriptor form~\eqref{eq.descriptor}, we focus on the energy amplification of flow structures with $k_z=0.32$. Similar to Section~\ref{sec.formulation}, we discretize system~\eqref{eq.descriptor} by applying Fourier transform in $z$ and using a Chebyshev collocation scheme in the wall-normal and streamwise directions. In the wall-normal direction, we enforce homogenous Dirichlet boundary conditions on all velocity components. In the streamwise direction, we use homogeneous Dirichlet boundary conditions at the inflow and spatial extrapolation at the outflow for all velocity components. Moreover, sponge layers are applied at the inflow and outflow to mitigate the influence of boundary conditions on the fluctuation dynamics. As shown in Fig.~\ref{fig.descriptorstreak}, the dominant flow structures that result from near-wall excitation closely resemble the results presented in Fig.~\ref{fig.streamwisevelocity-vorticity-05}.

	\vspace*{-2ex}
\section{Matching the HIT spectrum with stochastically forced linearized NS equations}
\label{sec.appendix-HIT}

We briefly describe how the spectrum of HIT can be matched using stochastically forced linearized NS equations; see~\cite[Appendix C]{rashad-phd12} for additional details. The dynamics of velocity fluctuations $\bv$ around a uniform base flow $\bar{\bu} = [\,1\,~0\,~0\,]^T$ subject to the solenoidal forcing $\bd_s = [\,d_u\,~d_v\,~d_w\,]$ ($\nabla \cdot \bd_s = 0$) are governed by the linearized NS equations
\[
	\bv_t(\bk,\,t)
	~=~
	\bA(\bk)\bv(\bk,\,t) \;+\; \bd_s(\bk,\,t),
\]
where $\bk=[\,k_x \,~k_y \,~k_z\,]^T$ is the spatial wavenumber vector and
\[
	\bA(\bk)
	~=~
	-\left(\mri k_x \;+\; \dfrac{k^2}{Re}\right) \bI_{3\times3},
\]
is the linearized operator. Here, $k^2\,=\,k_x^2\,+\,k_y^2\,+\,k_z^2$ and $\bI_{3\times3}$ is the identity operator. The steady-state covariance of velocity fluctuations $\Phi(\bk)=\lim \limits_{t\to\infty} \left< \bv(\bk,t)\, \bv^*(\bk,t)\right>$ satisfies the following Lyapunov equation
\begin{align}
	\label{eq.lyap-HIT}
	\bA(\bk)\Phi(\bk) \;+\; \Phi(\bk)\bA^*
	~=~
	-{\bf M}(\bk),
\end{align}
where ${\bf M}(\bk)$ denotes the covariance of white-in-time stochastic forcing. The steady-state covariance matrix $\Phi$ corresponding to HIT is given by~\cite{bat53}
\[
	\Phi(\bk)
	~=~
	 \dfrac{E(k)}{4\pi k^2}
	 \left(\bI_{3\times3} \;+\; \dfrac{\bk\,\bk^T}{k^2}\right).
\]
where $E(k)$ is the energy spectrum of the HIT based on the von K\'{a}rm\'{a}n spectrum~\cite{durrei11},
\[
E(k)
~=~
L\,C_{vk}\dfrac{(k\, L)^4}{(1 \,+\, k^2 L^2)^{17/6}}.
\]
Here, $C_{vk}=\dfrac{\Gamma(17/6)}{\Gamma(5/2)\Gamma(1/3)}=0.48$ is a normalization constant in which $\Gamma(\cdot)$ is the gamma function and the integral length-scale $L=1.5$ corresponds to numerical simulations of HIT~\cite{wanchebrawyn96}. The input forcing covariance can be derived by substituting $\Phi(\bk)$ into Eq.~\eqref{eq.lyap-HIT}, which yields
\[
	{\bf M}(\bk)
	~=~
	 \dfrac{E(k)}{2\pi Re}\left(\bI_{3\times3} \;+\; \dfrac{\bk\, \bk^T}{k^2}\right).
\]
After finite dimensional approximation of all operators, the covariance of forcing $\bd_s$, parameterized by $k_z$, is obtained via inverse Fourier transform in $x$ and $y$. The resulting covariance matrix ${\bf M} (k_z)$ includes two-point correlations of the white stochastic forcing in the streamwise and wall-normal directions and it replaces~$W$ in Eq.~\eqref{eq.standard_lyap}.

%
%

\end{document}